\documentclass[12pt]{article}
\pdfoutput=1
\usepackage{graphicx}
\usepackage{epstopdf}
\usepackage{amsmath}
\usepackage{amsfonts}
\usepackage{amssymb}
\usepackage{color}
\usepackage{mathrsfs}

\newcommand{\be}{\begin{equation}}
\newcommand{\ee}{\end{equation}}
\newcommand{\bea}{\begin{eqnarray}}
\newcommand{\eea}{\end{eqnarray}}
\newcommand{\barr}{\begin{array}}
\newcommand{\earr}{\end{array}}

\def\beq{\begin{equation}}
\def\eeq{\end{equation}}
\def\be{\begin{equation}}
\def\ee{\end{equation}}
\def\bea{\begin{eqnarray}}
\def\eea{\end{eqnarray}}
\def\d{{\partial}}

\def\taunl{$g_{NL}\ $}
\def\mpl{M_{\rm Pl}}

\begin{document}

%\begin{titlepage}

\setcounter{page}{1} \baselineskip=15.5pt \thispagestyle{empty}

\begin{flushright}
%hep-th/yymmnnn\\
\end{flushright}
\vfil

\begin{center}

{\Large \bf The Effective Field Theory of Multifield Inflation }
\\[0.7cm]
{\large Leonardo Senatore$^{a,b,c}$,  and Matias Zaldarriaga$^{c}$}
\\[0.7cm]
%\vspace{.7cm}
%\vspace{.3cm}
{\normalsize { \sl $^{a}$ Stanford Institute for Theoretical Physics, Stanford University,\\ Stanford, CA 94305, USA}}\\
\vspace{.3cm}
{\normalsize { \sl $^{b}$ Kavli Institute for Particle Astrophysics and Cosmology,\\Stanford  University and SLAC, Menlo Park, CA 94025, USA}}\\
\vspace{.3cm}
{\normalsize { \sl $^{c}$ School of Natural Sciences, Institute for Advanced Study,\\ Olden Lane, 
Princeton, NJ 08540, USA}}\\
\vspace{.3cm}

\end{center}

\vspace{.8cm}

\hrule \vspace{0.3cm}
{\small  \noindent \textbf{Abstract} \\[0.3cm]
\noindent We generalize the Effective Field Theory of Inflation to include additional light scalar  degrees of freedom that are in their vacuum at the time the modes of interest are crossing the horizon. In order to make the scalars  light in a natural way we consider the case where they are the Goldstone bosons of a global symmetry group or are partially protected by an approximate supersymmetry. We write the most general Lagrangian that couples the  scalar mode associated to the breaking of time translation during inflation to the additional light scalar fields. This Lagrangian is constrained by diffeomorphism invariance and the additional symmetries that keep the new scalars light.
This Lagrangian describes the fluctuations around the time of horizon crossing and it is supplemented with a general parameterization describing  how the additional fluctuating fields can affect cosmological perturbations. We find that multifield inflation can reproduce the non-Gaussianities that can be generated in single field inflation but can also give rise to new kinds of non-Gaussianities. We find several new three-point function shapes. We show that  in multifield inflation it is possible to naturally suppress the three-point function making the four-point function the leading source of detectable non-Gaussianities. We find that under certain circumstances, {\it ie.} if specific shapes of non-Gaussianities are detected in the data,  one could distinguish between single and multifield inflation and sometimes even among the various mechanisms that kept the additional fields light. 
}
 \vspace{0.3cm}
\hrule
\vfil
%\begin{flushleft}
%\today
%March 20, 2008
%\end{flushleft}

%\end{titlepage}

%\newpage
%\tableofcontents
%\newpage

\section{Introduction}

Recent cosmological observations are testing the inflationary paradigm in greater and greater detail. In this situation it is very useful to describe this epoch using the effective field theory approach, {\it i.e.} describe it through its lowest dimension operators compatible with the symmetries. This approach in fact allows for a description of a system in the most general terms, separating in a clear way what is determined by the UV theory and what is instead just the result of the symmetries of the problem. Because this approach concentrates on the theory of the fluctuations it is also the approach that most directly connects to what cosmological observations are actually testing about inflation.

The effective field theory of inflation in the case where there is only one relevant degree of freedom during the inflationary phase was recently developed in \cite{Cheung:2007st,Cheung:2007sv,Senatore:2009gt,Senatore:2009cf,Senatore:2010jy,Creminelli:2006xe,Bartolo:2010bj,Bartolo:2010di}. What makes it possible to describe the theory for the fluctuations  without any assumption about the fundamental degree of freedom that is driving inflation is the fact that the inflationary  period  has to end and give way to the standard FRW cosmology. This simple point implies that time-diffeomorphisms are broken and that therefore there is a Goldstone boson associated with this symmetry breaking. As typical in the case of Goldstone bosons its Lagrangian can be constructed in general terms because  it is very constrained by  symmetries. In practice the Lagrangian for the perturbations in single-clock inflation is constructed by choosing a particular time-slicing where the clock field is taken to be uniform. In this frame, the most general theory is built with the lowest dimension operators invariant under spatial diffeomorphisms, like $g^{00}$, and $K_{\mu\nu}$, the extrinsic curvature of constant time surfaces. Invariance under time-diffeomorphisms is recovered after reintroducing the Goldstone boson $\pi$, which transforms under time diffs.~of parameter $\xi^0(\vec x,t)$ non-linearly as $\pi(\vec x,t)\rightarrow \pi(\vec x,t)+\xi^0(\vec x,t)$. In the high energy limit it turns out that in most cases one can concentrate on the scalar mode $\pi$ which makes the physics appear very transparently. 

This approach to describe inflation has been very successful so far. For example it provided a way to explore in generality the possible signatures of single field inflation, identifying previously missed ones~\footnote{For example, the orthogonal kind of three-point function of the density fluctuations that is currently at about $2\sigma$ level in the WMAP 7year data~\cite{Komatsu:2010fb} was identified in this context~\cite{Senatore:2009gt}.}. Furthermore this approach can be used to translate the constraints on the non-Gaussianities obtained from WMAP data directly onto parameters of the Lagrangian for the fluctuations without any loss of generality \cite{Senatore:2009gt}. In fact it is the Lagrangian for the fluctuations that is directly tested by cosmological observations. This approach of using the experimental data to put constraints on the most generic Lagrangian built with the lowest dimensional operators compatible with the symmetries is  well established in the particle physics community, where it goes under the name of Precision Electroweak Tests \cite{Peskin:1991sw,Barbieri:2004qk}, but it is quite new in the cosmological setting.

The purpose of the present paper is to generalize the construction of the effective field theory of single-clock inflation to include additional light degrees of freedom that might be present during inflation. This is a complicated task that we will develop in a series of two papers. First the Lagrangian for the possible additional light degrees of freedom during inflation will turn out to be much less constrained by the symmetries than the Lagrangian for the Goldstone boson of time translations ($\pi$).  Second, there can be many additional light degrees of freedom involved during inflation: scalar, fermions, vectors. Furthermore it is not necessary that these additional fields are in the vacuum at the moment of horizon crossing. In this first paper we will develop the theory of multifield inflation (or maybe more properly multi-degree-of-freedom inflation) for the case where these additional modes are scalar fields whose fluctuations are in their vacuum during inflation. We will treat the case of fermions, vector fields, and fluctuations not in their adiabatic vacuum in a different paper~\cite{senatore2}~\footnote{Inflationary fluctuations that are not in their vacuum at the time of horizon crossing have been recently considered in~\cite{Chen:2006nt,Holman:2007na,Green:2009ds}.}.

Even for additional scalar degrees of freedom in their vacuum, the construction of the Lagrangian will not be completely straightforward. There are two relevant facts that need to be taken into account. First, the scalar degrees of freedom need to be lighter than the Hubble scale during inflation in order for them to acquire long-wavelength fluctuations. However it is theoretically difficult  to have a naturally light scalar: quantum corrections result in large contributions to its mass.  The only known mechanisms to make a scalar naturally light are either supersymmetry or having the scalar fields be the Goldstone bosons (or the pseudo-Goldstone bosons) of a global symmetry that is spontaneously broken.  We will study both of these possibilities in as much generality as we can. It is of course possible that other additional symmetries or different representations can be found and their study might lead to interesting signatures. What we aim to present here is not just the study of some quite general models, but rather a general formalism that can applied to study inflationary fluctuations in multifield inflation in complete generality.  We further do not address directly the issue of finding UV completions to the effective field theories we present here, but we simply stick to technically natural low energy effective field theories~\footnote{It is possible that some natural low energy effective field theories  do not have a natural UV completion. The question of which natural effective field theories admit a natural UV completion is a very interesting, general and complicated question. The answer is not known in generality, and we do not address this here. We rather stick to the fact that so far in Nature we have found many physical systems that are very well described by low energy natural effective field theories. At the end of the day, even General Relativity is nothing but a natural effective field theory.}.

In the case where the additional scalar degrees of freedom are the Goldstone bosons of a spontaneously broken symmetry, following the standard treatment, it is possible to construct the most generic Lagrangian for the Goldstone bosons with very mild assumptions on the symmetry group that is being spontaneously broken. In comparison to the standard construction in Minkowski space there will be the important novelty that these Goldstone bosons can couple to the Goldstone boson of time translation ({\it i.e.}  $\pi$). This will allow several new terms in the Lagrangian. We will perform the construction both for the case where the symmetry group being spontaneously broken is Abelian and also for the case in which it is non-Abelian, as there will be relevant differences. We will also discuss effects originating from the soft-breaking of these symmetries. 

We will then move on to study the case where the additional light degrees of freedom are approximately supersymmetric. During inflation there is a minimum amount of supersymmetry breaking due to the fact that the vacuum energy is non zero. The additional light fields interact with the field driving inflation at least gravitationally so the effective energy scale of supersymmetry breaking in the sector of the additional fields is at least of order $H$, where $H$ is the 
Hubble constant during inflation. This is too high an energy scale for the scalars to be of interest as they need to be lighter than $H$ to even fluctuate during inflation and  a factor of at least about fifty below $H$ if one wants the fluctuations they induce to be scale invariant at a level consistent with current observational constraints. This would suggest that having an approximately supersymmetric Lagrangian is not enough, however the required fine tuning is not enormous so perhaps one or a few light fields could remain accidentally. We will study this possibility in some detail.

A second fact that needs to be taken into account in order to describe the role of additional light degrees of freedom  is that in cosmology we do not directly observe  fluctuations of the inflaton or the other degrees of freedom. We observe the effect of these fluctuations on the cosmological perturbations such as the CMB anisotropies or galaxy clustering statistics. This implies that it is important to describe how the fluctuations in the additional light degrees of freedom get converted into perturbations of the primordial plasma for example by affecting the duration of inflation or by changing the composition of the plasma. In the case of single field inflation this relationship is very simple: a fluctuation in $\pi$ amounted to a small time-delay in the inflaton trajectory, and therefore in an enhanced expansion of order $H\pi$ at the end of inflation. These are the so-called curvature perturbations and are the only kind of perturbation possible in single-clock inflation because reheating happens in every place in the same way leading locally to the same plasma composition. The only difference between the various regions of the Universe being how much inflation lasted. 

In multifield inflation the situation is much more complicated. A fluctuation in one of the additional scalar fields can be thought of as changing the inflaton trajectory. In order to make contact with with cosmological observations one needs to know how much longer the new trajectory is compared to the unperturbed one and also if the plasma will have a different composition if  reheating is reached from a different trajectory. Contrary to the case of single field inflation, this piece of information is not fully determined by the Lagrangian that describes fluctuations around the time they cross the horizon. How much inflation lasts along each trajectory is an integral effect that depends on the entire trajectory and could even be determined entirely by events at the end of inflation. The same is true for variations of the composition of the plasma after inflation: for example it could be that this is just determined at the time of reheating, independently of the Lagrangian for the fluctuations around horizon crossing. 

Naively, this fact could stop us from developing an effective field theory for multifiled inflation, concluding that a full knowledge of the inflationary Lagrangian in field space is necessary in order to determine the density fluctuations. However we will argue that to a good approximation all of these effects are generated when the relevant modes are out of the horizon and gradients are negligible. This will lead us to conclude that regardless of the mechanism of conversion of the additional scalar field fluctuations into cosmological perturbations it can be parametrized by a relationship between the perturbations in the plasma and the fluctuations of the additional scalar fields at the time of horizon crossing that is {\it local in real space}.   Fluctuations in our universe are not very large and are approximately Gaussian: by Taylor expanding this unknown relationship in small fluctuations we will summarize our ignorance about the conversion mechanism with a set of a few unknown numbers. This is quite similar to the so-called $\delta N$ formalism \cite{Sasaki:1995aw}. Though these parameters can be known only with a detailed knowledge of the model, we are able to identify many interesting signatures that are independent of these unknown numbers provided that the fluctuations are made cosmologically observable.

Since the construction of the effective theory for multifield inflation will be complicated we start by summarizing the main signatures, many of which are unique to the mulfield inflation case.

\subsubsection*{Summary of signatures}

\begin{itemize}
\item Quadratic Lagrangian. It is possible for the fluctuations associated to the additional fields to have  speeds of propagation, or speeds of ``sound",  different from one another and different from one. It is also possible to have dispersion relations of the form $\omega\sim k^2/M$, similar to the Ghost condensate case. Just as in the single field case, modifying the speed of propagation leads to large interaction terms for the fluctuations. If we schematically denote by $\sigma$ one of the additional light scalar fields, then in the limit of small speed of sound or non-linear dispersion relation an interaction of form $\d_i\pi\d_i\sigma\dot\sigma$ becomes large, although still perturbative. Furthermore in the Abelian case it is possible to have a time-kinetic mixing between $\sigma$ and $\pi$ of the form $\dot\pi\dot\sigma$. 

\item Cubic Lagrangian. At cubic level, in the Abelian case operators of the form $\dot\sigma^3$ and $\dot\sigma(\d_i\sigma)^2$  produce the same kind of three-point functions that can be present in single-field inflation. Furthermore there can be operators induced by the fact that the symmetry associated to the Goldstone bosons might have been softly broken at the time. The most interesting operator is $\sigma(\d_\mu\sigma)^2$ and it induces a well defined unique shape that could be detectable and that is usually associated with a comparable-in-size three-point function of the local form and also with a larger-in-size four-point function of local form (as we will discuss). Such a complex signature would be a clear indication of multifield inflation that is not expected to happen in single-clock inflation. It is also possible that operators of the form $\sigma(\d\sigma)^2$, without a Lorentz invariant contraction of the indices, might be detectable. In general the presence of the various operators associated to the explicit soft symmetry breaking depends on the way that the symmetry is explicitly broken. Because of this operators that are naively subleading can become  more easily detectable or even the leading ones. In the non-Abelian case the situation is in general very similar to the Abelian case. However, quite remarkably, for some non-Abelian groups symmetries forbid operators of the form $\dot\sigma^3$ and $\dot\sigma(\d_i\sigma)^2$. In this case, the operators coming from the soft-symmetry breaking, if present, are the most relevant and a three-point function induced by an operator of the form $\sigma(\d\sigma)^2$ could be more easily detectable. As in the Abelian case, such a detection would be generically accompanied by a comparable-in-size three-point function of the local form and also by a larger-in-size four-point function of local form. %A detection of the three-point function induced by the non-Lorentz-invariant operator would be a clear indication of the non-Abelian nature of the group. 
If the soft-symmetry breaking is small, and the nature of the non-Abelian group is such that the other cubic operators are zero, then the four-point function becomes particularly relevant. 
%In the supersymmetric case the detectable three-point function is of the local form. 
In the Abelian case it is also possible to have a dispersion relation of the form $\omega\propto k^2$ with detectable non-Gaussianities induced by the operators $\dot\sigma(\d_i\sigma)^2$ and $(\d_j^2\sigma)(\d_i\sigma)^2$. This is similar to what can happen in single-clock inflation. Both in the Abelian and in the non-Abelian cases with $\omega\propto k^2$ dispersion relation, we can have a detectable three-point function  induced by the operators associated to the explicit soft breaking.
In addition to all of the shapes we have discussed all cases we can have a three-point function of the local type generated at reheating. The local kind is the leading three-point function in the supersymmetric case.

\item Quartic Lagrangian. At quartic level, in the Abelian case with a linear dispersion relation there are three operators  $\dot\sigma^4$, $\dot\sigma^2(\d_i\sigma)^2$ and $(\d_i\sigma)^4$ that induce a four-point function. Notice that only one of these operators, $\dot\sigma^4$, is relevant in single-clock inflation~\cite{Senatore:2010jy}. In general these operators give an effect smaller than the ones from the cubic Lagrangian but we are able identify a set of approximate additional symmetries that can be imposed to make the induced four-point function the leading source of non-Gaussianity. These can be either an approximate symmetry $\sigma\rightarrow-\sigma$, or the requirement that the Lagrangian is approximately Lorentz-invariant~\footnote{In the sense that the scale suppressing operators violating Lorentz invariance is higher that the one suppressing Lorentz-invariant ones.}. In this last case we are left with only the operator $(\d_\mu\sigma)^4$. Notice that this is the first case in which we can have large and detectable non-Gaussianities without having a large violation of Lorentz invariance or of the continuous shift symmetry in the Lagrangian for the fluctuations. There is also another possible local four-point function induced by the operator $\sigma^4$ associated to the small breaking of the Goldstone symmetry. In some region of the parameter space this shape can be the leading one but is degenerate with the four point function that can be created at reheating. Still, this offers a way of generating a detectable four-point function of the local form without at the same time having a comparable or even larger three-point function of the local kind. It is also possible to have a dispersion relation for the modes of the form $\omega\propto k^2$. In this case the leading interacting operators have to come from soft-symmetry breaking, and a local four-point function as generated by the $\sigma^4$ operator, or a four-point function as generated by $\sigma^2(\d_i\sigma)^2$ operator m!
 ight be 
detectable.  Notice that in single field inflation with a dispersion relation of the form $\omega\propto k^2$ there is a larger set of possibilities for operators generating a large four-point function~\cite{Senatore:2010jy}. In the non-Abelian case the discussion is very similar to the Abelian case with the only difference being that there are additional symmetries that  can naturally  suppress the three-point function making the case for searching for a  four-point function even stronger. In the supersymmetric case there is another new shape generated by the operator  $\sigma^2(\d_\mu\sigma)^2$ which could be marginally detectable and that would be accompanied by a larger local-four point function. We are tempted to argue that this shape represents a smoking gun for supersymmetry as an approximate symmetry during inflation. The caveat to this argument is that we find it hard to make statements involving naturalness in the supersymmetric case given that an accidental tuning of order fifty is necessary as a starting point. Finally, in all of these cases (and in the supersymmetric case even without any tuning), it is possible to have a large four-point function of the local kind.

\item Isocurvature fluctuations. In multifield inflation isocurvature fluctuations are possible. The results for the adiabatic fluctuations extends quite simply to this case except for two differences. One novelty is that now  there can be correlations between adiabatic and isocurvature fluctuations. In the Abelian case, we can have a three-point function induced by operators schematically of the form $(\d\sigma)^3$, and also from others schematically of the form $\sigma(\d\sigma)^2$ associated to the symmetry breaking. At the level of the four point function, the leading one is in general an operator of the form $\sigma^4$ that induces a local four-point function. The story with the non-Abelian and the supersymmetric cases proceeds very similarly, with one interesting novelty for the non-Abelian case: it is possible to have a detectable four-point function induced by operators schematically of the form $\sigma^2(\d\sigma^2)$ or $\sigma(\d\sigma)^3$ whose detection would be a clear indication of the nature of the symmetry group being broken. A similar phenomenology holds also in the Abelian and in the non-Abelian case when the dispersion relation is of the form $\omega\sim k^2/M$. Of course, detectability of these isocurvature effects is strongly dependent on the actual size of the isocurvature component which is already constrained to be subleading.

\end{itemize}

\subsubsection*{Distinguishability}

Depending on the details of the situation it is possible that by measuring the departure from Gaussianity of the primordial fluctuations we might be able to distinguish between single and multifield inflation and even to distinguish among the three mechanisms we have identified to keep the additional scalar fluctuations light: an Abelian or non-Abelian global symmetry or supersymmetry. This is possible only in the case of detection of some particular signatures that Nature might be kind enough to imprint on the cosmological perturbations. 
\begin{itemize}
\item There are several observational signatures that would rule  out single field inflation in favor of multifield inflation. In addition to  the observation of isocurvature fluctuations and/or a local type of non Gaussianity~\cite{Cheung:2007sv,Maldacena:2002vr,Creminelli:2004yq}, which have been known as signatures of multi-field inflation for some time, there are  particular shapes of the three and the four-point function that go beyond the local shape and that cannot be generated in single field inflation. A detection of a non-Gaussianity induced by an operator we list other than those listed in table {\ref{tab:singlefield}} would be an indication of the multifield nature of inflation. 
\item If there is a global symmetry, the non-Abelian nature of the group can be inferred by detecting a four-point function in the mixed adiabatic-isocurvature fluctuations induced by operators of the form $\sigma(\d\sigma)^3$, accompanied by the detection of a three point function of the form $(\d\sigma)^3$ in the adiabatic sector, or of the form $\sigma^2(\d\sigma)^2$ without an analogous shape in the purely adiabatic fluctuations. 
%The Abelian nature of the symmetry group can be identified by detecting a four-point function as induced by the operators $\dot\sigma^2(\d_i\sigma)^2$ and $(\d_i\sigma)^4$.
%Also in the case of a dispersion relation $\omega\propto k^2$, a detection of a mixed adiabatic-isocurvature four-point function as induced by the operator $\sigma^2(\d_i\sigma)^2$ would be a clear indication of the non-Abelian nature of the symmetry group. 
\item Supersymmetry could be identified by detection of a subleading four-point function in the adiabatic fluctuations  produced by the  $\sigma^2(\d_\mu\sigma)^2$ operator with the dispersion relation $\omega\sim k$ and subleading with respect to a local four-point function by a factor of order $N_e$. However in the supersymmetric case the caveats we discussed at the end of the paragraph about the quartic Lagrangian in this section apply.
\end{itemize}

In Tables~\ref{tab:multifield} and {\ref{tab:singlefield}} we give a schematic summary of the potential non-Gaussian signals we have been able to identify in this paper, together we a summary of the known results from single-clock inflation with a continuous shift symmetry. The details are discussed in sec.~\ref{sec:signatures}.

The subject of multifield inflation is very large. Some related earlier papers that focused on non-Gaussian 3-point and 4-point functions are for example~\cite{Allen:1987vq,Lyth:2002my,Bernardeau:2002jy,Bernardeau:2002jf, Bernardeau:2007xi,Lyth:2005fi,Bernardeau:2010jp}.

\begin{table}[h!]
\begin{tabular}{| c| c| c| c| c| c|}
\hline
Operator & \multicolumn{2}{c|}{Dispersion}  & Type & Origin & Squeezed L.\\
\hline
    & $w=c_s k$ & $w\propto k^2$ & & & \\
\hline
$\dot\sigma^4\; ,\ \dot\sigma^2(\d_i\sigma)^2\;, (\d_i\sigma)^4$ & X & & Ad., Iso. & Ab., non-Ab. & \\
\hline
$ (\d_\mu\sigma)^4$ & X & & Ad., Iso. & Ab., non-Ab. & \\
\hline
$ \dot\sigma^p(\d_i\d_j\sigma)^{(4-p)}$ &  & X & Ad., Iso. & Ab. & \\
\hline
$ \sigma^4$ & X & X & Ad., Iso. & Ab.$_s$, non-Ab.$_s$, S. & X \\
\hline
$ \dot\sigma\sigma^3$ & X & X & Ad., Iso. & Ab.$_s^\dag$, non-Ab.$_s^\dag$. & X \\
\hline
$ \sigma^2\dot\sigma^2\;,\sigma^2(\d_i\sigma)^2$ & X & X$^\dag{}^\star$ & Ad.$^\dag{}^\star$, Iso. &  non-Ab, Ab.$_s^\dag{}^\star$, non-Ab$._s^{\dag}{}^\star$, & X \\
\hline
$ \sigma^2(\d_\mu\sigma)^2$ & X & &  Ad.$^\dag{}^\star$, Iso. &   non-Ab, Ab.$_s^\dag{}^\star$, non-Ab$._s^{\dag}{}^\star$, S.$^\star$& X \\
\hline
$ \sigma(\d\sigma)^3$ & X &  & Iso. & non-Ab.$_s^\star$. & X \\
\hline
\hline
$\dot\sigma^3\; ,\ \dot\sigma(\d_i\sigma)^2$ & X & & Ad., Iso. & Ab., non-Ab. & \\
\hline
$\dot\sigma(\d_i\sigma)^2\;,\d^2_j\sigma(\d_i\sigma)^2$ &  & X & Ad., Iso. & Ab. & \\
\hline
$\sigma^3$ & X & X & Ad., Iso. & Ab.$_s$, non-Ab.$_s$, S, R & X \\
\hline
$\dot\sigma\sigma^2$ & X & X & Ad., Iso. & Ab.$_s$, non-Ab.$_s$& X \\
\hline
$\sigma\dot\sigma^2\; ,\ \sigma(\d_i\sigma)^2$ & X & X & Ad., Iso. & Ab.$_s^\dag{}^\star$, non-Ab.$_s^\dag{}^\star$ & X \\
\hline
$ \sigma(\d_\mu\sigma)^2$ & X & & Ad., Iso. & Ab.$_s^\dag{}^\star$, non-Ab.$_s^\dag{}^\star$. & X \\

\hline
\end{tabular}
\caption{\small Signatures in Multi-field Inflation. 
In the first column we give the operator generating the non-Gaussian signal: operators quartic in the $\sigma$'s lead to a four-point function, operators cubic in the $\sigma$'s lead to a three-point function. In the second and third columns we explain with which dispersion relation the signal can be generated. In the third we explain if the signal can appear in the Adiabatic (Ad.) or the Isocurvature (Iso.) fluctuations. In the fourth we state the potential origin of the signal.  Here Ab. stands for Abelian; non-Ab. stands for non-Abelian, S stands for supersymmetry, and R stands for generated by non-linearities at reheating. The subscript $_s$ indicates that the term is generated by soft-breaking terms. The symbol $^\dag$ represents that such a signal can be generated in the case the soft symmetry breaking term is such that it forbids some of the lowest dimensional terms. The symbol $^\star$ represents the fact that the signal is in general subleading, but still possibly detectable. In the last column we explicitly mention if the induced signal has a non-vanishing squeezed limit and is therefore detectable also in clustering statistics of collapsed objects. }
\label{tab:multifield}
\end{table}

\begin{table}[h]
\begin{tabular}{| c| c| c| c|}
\hline
Operator & \multicolumn{2}{c|}{Dispersion}    & Squeezed L.\\
\hline
    & $w=c_s k$ & $w\propto k^2$ & \\
\hline
$\dot\pi^4\;$ & X & &  \\
\hline
$ \dot\pi^p(\d_i\d_j\pi)^{(4-p)}$ &  & X &  \\
\hline
\hline
$\dot\pi^3\;,\dot\pi(\d_i\pi)^2$ & X & &  \\
\hline
$\dot\pi(\d_i\pi)^2\;, \d_j^2\pi(\d_i\pi)^2$ &  & X &  \\

\hline
\end{tabular}
\caption{\small Signatures in Single-Clock Inflation with a continuous shift symmetry. The signal associated to the four-point function when the dispersion relation is of the form $\omega\propto k^2$ is quite rich, and we refer to~\cite{Senatore:2010jy} for a discussion about it.}
\label{tab:singlefield}
\end{table}

\section{Effective field theory of single-clock inflation\label{sec:single-field}}

In this section we briefly review the effective action for single-clock inflation. This effective
action was developed in \cite{Cheung:2007st,Creminelli:2006xe} and we refer the reader to those papers for a detailed explanation.
The construction of the effective theory is based on the following consideration. In a quasi de
Sitter background with only one relevant degree of freedom, there is a privileged spatial slicing
given by the physical clock which allows us to smoothly connect to a decelerated hot Big Bang
evolution. The slicing is usually realized by a time evolving scalar $\phi(t)$, but this does not need necessarily to be the case. To describe perturbations
around this solution one can choose a gauge where the privileged slicing coincides with surfaces of
constant $t$, i.e. $\delta\phi(\vec x,t)=0$. In this `unitary' gauge there are no explicit scalar perturbations but only metric
fluctuations. As time diffeomorphisms have been fixed and are not a gauge symmetry anymore,
the graviton now describes three degrees of freedom: the scalar perturbation has been eaten by the
metric. One therefore can build the most generic effective action with operators that are functions
of the metric fluctuations and that are invariant under the linearly-realized time-dependent spatial
diffeomorphisms. As usual with effective field theories, this can be done in a low energy expansion
in fluctuations of the fields and derivatives. We obtain the following Lagrangian~\cite{Cheung:2007st,Creminelli:2006xe}:
 \begin{eqnarray}
\label{eq:actiontad}\nonumber
S_{\rm E.H.\;+\; S.F.} & \!\!\!\!\!\!\!\!\!\!\!\!= \!\!\!\!\!\!\!\!\!& \!\!\!\int  \! d^4 x \; \sqrt{- g} \Big[ \frac12 M_{\rm 
Pl}^2 R + M_{\rm Pl}^2 \dot H
g^{00} - M_{\rm Pl}^2 (3 H^2 + \dot H)+ \\\nonumber 
&&+ \frac{1}{2!}M_2(t)^4(g^{00}+1)^2+\frac{1}{3!}M_3(t)^4 (g^{00}+1)^3+ \\
&& - \frac{\bar M_1(t)^3}{2} (g^{00}+1)\delta K^\mu {}_\mu
-\frac{\bar M_2(t)^2}{2} \delta K^\mu {}_\mu {}^2
-\frac{\bar M_3(t)^2}{2} \delta K^\mu {}_\nu \delta K^\nu {}_\mu + ... \Big]\ ,
\end{eqnarray}
where we denote by $\delta K_{\mu\nu}$ the variation of the extrinsic curvature of constant time surfaces with respect to the unperturbed FRW: $\delta K_{\mu\nu}=K_{\mu\nu}-a^2 H h_{\mu\nu}$ with $h_{\mu\nu}$ being the induced spatial metric, and where $M_{2,3}$ and $\bar M_{1,2,3}$ represent some time-dependent mass scales.

Let us comment briefly on (\ref{eq:actiontad}). The first term is the Eistein-Hilbert term. The first three terms are the only ones that start linearly in the metric fluctuations. The coefficients have been carefully chosen to ensure that when combined the linear terms in the fluctuations cancel. The action must start quadratic in the fluctuations. The terms in the second line start quadratic in the fluctuations and have no derivatives. The terms in third line represent higher derivative terms. Dots represent operators that start at higher order in the perturbations or in derivatives. This is the most general action for single field inflation and in fact it is unique~\cite{Cheung:2007st}.

The unitary gauge Lagrangian describes three degrees of freedom: the two graviton helicities and
a scalar mode. This mode will become explicit after one performs a broken time diffeomorphism
(St\"uckelberg trick) to reintroduce the Goldstone boson which non-linearly realizes this symmetry. In analogy
with the equivalence theorem for the longitudinal components of a massive gauge boson \cite{Cornwall:1974km}, the physics of the Goldstone decouples from the two graviton helicities at high enough energies, equivalently  the mixing can be neglected. The detailed study of \cite{Cheung:2007st} shows that in most situations of interest this is indeed the case and one can neglect the metric fluctuations ~\footnote{Equivalently, the neglected effects are suppressed by slow-roll parameters or by powers of $H/\mpl$. }. 

As anticipated, we reintroduce the Goldstone boson ($\pi$) by performing a broken time-diff., calling the parameter of the transformation $-\pi$, and then declaring $\pi$ to be a field that under time diff.s of the form $t\rightarrow t+\xi^0(x)$ transforms  as
\be
\pi(x)\quad\rightarrow\quad \tilde\pi(\tilde x(x))=\pi(x)-\xi^0(x)\ .
\ee
In this way diff. invariance is restored at all orders. For example the terms containing $g^{00}$ in the Lagrangian  give rise to the following terms:
\be
g^{00}\quad\rightarrow \quad\frac{\d (t+\pi)}{\d x^\mu}\frac{\d (t+\pi)}{\d x^\nu}g^{\mu\nu} \quad\rightarrow\quad g^{00} +2 g^{0\mu} \partial_\mu \pi + (\partial \pi)^2 .
\ee
We refer to~\cite{Cheung:2007st} for details about this procedure.
If we are interested just in effects that are not dominated by the mixing with gravity, then we can neglect the metric perturbations and just keep the $\pi$ fluctuations. In this regime, a term of the form $g^{00}$ in the unitary gauge Lagrangian becomes:
\be
g^{00}\quad\rightarrow\quad -1-2\dot\pi-\dot\pi^2+\frac{1}{a^2}(\d_i\pi)^2\ .
\ee
Further, we can assume that the $\pi$ has an approximate continuous shift symmetry, which becomes exact in the limit in which the space time is exactly de Sitter \cite{Cheung:2007st}. This allows us to neglect terms in $\pi$ without a derivative that are generated by the time dependence of the coefficients in (\ref{eq:actiontad})~\footnote{Notice that this is not always the case. Interesting inflation models both single field and multifield have been recently proposed in which the $\pi$ fluctuations are protected only by an approximate discrete shift symmetry. See for example~\cite{Silverstein:2008sg,McAllister:2008hb,Green:2009ds,Barnaby:2009mc,Flauger:2009ab}.}. 
Implementing the above procedure in the Lagrangian of (\ref{eq:actiontad}), we obtain the rather simple result:
\begin{eqnarray}\label{eq:Spi}
S_{\rm \pi} =\int d^4 x   \sqrt{- g} \left[ -M^2_{\rm Pl}\dot{H} \left(\dot\pi^2-\frac{ (\partial_i \pi)^2}{a^2}\right)
%- M^2_{\rm Pl} \left(3H^2 +\dot{H}\right)+ \right.\\
+2 M^4_2
\left(\dot\pi^2+\dot{\pi}^3-\dot\pi\frac{(\partial_i\pi)^2}{a^2}
\right) -\frac{4}{3} M^4_3 \dot{\pi}^3 +\ldots\right] \ ,
\eea
where for simplicity we have neglected the terms originating from the extrinsic curvature as they are usually important only in a regime where the space time is very close to de-Sitter space \cite{Cheung:2007st}.

We notice that when $M_2$ is different from zero the speed of sound of the fluctuations is different from one. We have the following relationship:
\be\label{eq:M2cs}
M_2^4=-\frac{1-c_s^2}{c_s^2}\frac{\mpl^2\dot H}{2}\ .
\ee
There are two independent cubic self-interactions, $\dot\pi(\d_i\pi)^2$ and $\dot\pi^3$ at this order in derivatives, which can induce detectable non-Gaussianities in the primordial density perturbations. A small speed of sound (i.e.~a large $M_2$) forces large self-interactions of the form $\dot\pi(\d_i\pi)^2$, while the coefficient of the operator $\dot\pi^3$ is not fixed because it also depends on $M_3$. Cosmological data can therefore constrain (or measure) the parameters of the above Lagrangian. This approach has been recently applied to the WMAP data in \cite{Senatore:2009gt}, giving constraints on $M_2$ and $M_3$, as well as on the higher derivative operators that we have omitted in (\ref{eq:Spi}). This is the exact analogous of what happens for data from particle accelerators when the precision electroweak tests of the Standard Model are carried out \cite{Peskin:1991sw,Barbieri:2004qk}.

\section{Additional light scalar fields}

We are now ready to proceed to the construction of the effective action for multifield inflation.
We consider the case where the only light fields during inflation are the inflaton and some scalar fields that we call $\sigma_I$, with $I=1,..,N$ (as mentioned, we will consider other kind of light fields in a subsequent paper \cite{senatore2}). Since quantum corrections typically make it difficult to have light scalar fields  we consider the only two case we are aware of in which a light scalar field can be made technically natural. In the first case the additional fields are the Goldstone bosons, or pseudo-Goldstone bosons, associated with the spontaneous breaking of a (possibly softly broken) global internal symmetry group. If the broken group is Abelian this is equivalent to considering the theory of a scalar with a shift-symmetry. In the second  the additional light fields are approximately supersymmetric.  We start with the case in which the additional scalar fields are the Goldstone bosons arising from the spontaneous breaking of an Abelian group, then we move to the case of a non-Abelian group, and finally we end up this section with the supersymmetric case. 

\subsection{Abelian case ($N$-shift symmetries) \label{sec:Abelian_lagrangian}}

We start with the case in which there are $N$ Goldstone bosons arising from the spontaneous breaking a $U(1)^N$ group. We construct the Lagrangian in unitary gauge. Because of the shift symmetry, each $\sigma_I$ appears with a derivative acting on it. This derivative carries an index and because of the symmetry under time-dependent spatial diff.s of the Lagrangian in unitary gauge, this index has to be contracted in a diff. invariant way or it has to be of the form $g^{0\mu}\d_\mu\sigma_I$. Thus at leading order in derivatives, there are only two operators to construct the $\sigma_I$'s Lagrangian:  $g^{0\mu}\d_\mu\sigma_I$ and  $g^{\mu\nu} \d_\mu\sigma_I \d_\nu \sigma_J$. Keeping in mind that we cannot write tadpole terms for the $\sigma_I$'s, the resulting Lagrangian has the form:
\bea
\label{eq:actiontadmulti}
S_{\rm M.F.} & \!\!\!\!\!\!\!\!\!\!\!\!= \!\!\!\!\!\!\!\!\!& \!\!\!\int  \! d^4 x \; \sqrt{- g} \Big[ \tilde M_1(t)^2{}^{\;I} 
(g^{00}+1)(g^{0\mu}\d_\mu\sigma_I)\\ \nonumber 
&&-e_1(t)^{IJ}(g^{\mu\nu}\d_\mu\sigma_I\d_\nu\sigma_J)+e_2(t)^{IJ}(g^{0\mu}\d_\mu\sigma_I)
(g^{0\mu}\d_\mu\sigma_J)+\\\nonumber 
&&+ e_3(t){}^{IJ} (g^{00}+1)(g^{0\mu}\d_\mu\sigma_I)(g^{0\mu}\d_\mu\sigma_J)+e_4(t){}^{IJ} 
(g^{00}+1)(g^{\mu\nu}\d_\mu\sigma_I\d_\nu\sigma_J)+\\ \nonumber
&&+\tilde M_2(t)^2{}^{\;I} (g^{00}+1)^2(g^{0\mu}\d_\mu\sigma_I)+\\ \nonumber
&&+\tilde M_3(t)^{-2\, , \; IJK}(g^{0\mu}\d_\mu\sigma_I)(g^{0\mu}\d_\mu\sigma_J)(g^{0\mu}\d_\mu
\sigma_K)+\\ \nonumber 
&&+\tilde M_4(t)^{-2\,,\; IJK}(g^{0\mu}\d_\mu\sigma_I)(g^{\mu\nu}\d_\mu\sigma_J\d_\nu\sigma_K)
+... \Big]\ .
\end{eqnarray}
where we have kept only terms up to cubic order in the fluctuations only.
Greek indeces go from $0$ to $3$, capitol latin indexes go from 1 to $N$ and the $\ldots$ represent terms higher 
order in the fluctuations or with higher number of derivative as well as terms suppressed  by the  small breaking of the shift symmetry of the $\sigma_I$'s.  $e_i$, with $i=1,..., 4$, 
are dimensionless time-dependent coefficients, while the $\tilde M_i$'s, with $i=1,\dots,4$, are time-dependent parameters with dimension of mass. It is technically natural to expect the time-dependence of these coefficients to be suppressed by the parameters protecting the approximate shift symmetry of $\pi$,  the slow-roll parameters. Here we stick to this case although examples where this is not satisfied might exist and be interesting.'%The terms suppressed by slow-roll parameters that we just said are represented in the $\dots$ arise when we integrate by parts a derivative, and this acts on the time dependence of these coefficients. As for the terms that violate explicitly the shift-symmetry of the $\sigma_I$'s, we will comment on them explicitly later. 
The terms in the first line start quadratic in the 
field fluctuations, while the terms in the third and fourth lines start at cubic order. Without loss of generality, we can take the kinetic  coefficients $e_1^{IJ}$ to be equal to $\delta^{IJ}$,  
and $e_2$ to be diagonal in field space ($e_2^{IJ}=\tilde e_2^I \delta^{IJ}$). This can always be done with a proper field redefinition.  

Eq.~(\ref{eq:actiontadmulti})  simplifies when we reintroduce the Goldstone boson of time translations $\pi$ and we go in the high energy regime where we can neglect metric fluctuations~\footnote{Metric fluctuations can be treated as in single field inflation. In the ADM formalism, one fixes the gauge, solves the constraint equations for the constrained variables $N,N^i$, and plugs back the solutions in the action. However, the effect from metric perturbations is expected to be very small. If we neglect the term in $\tilde M_1^{I}$, the mixing with gravity starts at non-linear level in the $\sigma$'s, therefore affecting higher order correlation function. This is different to the case of single field inflation where the mixing starts at linear level. Here it becomes irrelevant in the limit $H/\mpl$ very small. The gravitational mixing with the Goldstone boson $\pi$ is instead generically of order of the slow roll parameters. The term in $\tilde M_1^{I}$ induces a linear mixing, and so it could be more relevant, at least in principle. However, as we will see next, in order to avoid ghosts or strong coupling in the theory, we need to have $\tilde M_1^{I}\lesssim (\dot H\mpl^2)^{1/4}$, which reduces the effect of mixing on non-Gaussianities at most of order of the slow roll parameters.}. 
We reintroduce $\pi$ by performing a time diff as in the former section. This time we need the two transformation laws:
\bea
&&g^{0\mu}\quad\rightarrow \quad\frac{\d (t+\pi)}{\d x^\nu}g^{\mu\nu} \quad\rightarrow\quad -\delta^
\mu_0 (1+\dot\pi)+\delta^\mu_i \frac{1}{a^2}\d_i\pi\ ,
\\ \nonumber
&&g^{00}\quad\rightarrow\quad \frac{\d (t+\pi)}{\d x^\nu}\frac{\d (t+\pi)}{\d x^\mu}g^{\mu\nu} 
\quad\rightarrow \quad -1-2\dot\pi-(\d\pi)^2\ ,
\eea
We obtain
\begin{eqnarray}\label{Spimulti} \nonumber
\! \! S_{\rm \sigma\pi}^{(2,3)} & = & \!\!\!\int d^4 x   \sqrt{- g} \left[  
 \tilde M_1^2{}^{I} \left(-2\dot\pi-\dot\pi^2+\frac{(\d_i\pi)^2}{a^2}\right)\left(-(1+\dot\pi)\dot\sigma_I+\frac{1}
{a^2}\d_i\pi\d_i\sigma_I\right)+ \right.
\\ \nonumber 
&&-\d_\mu\sigma_I\d^\mu\sigma_I+\tilde e_2^I\left(-(1+\dot\pi)\dot\sigma_I+\frac{1}{a^2}\d_i\pi\d_i\sigma_I\right)\left(-(1+\dot\pi)\dot\sigma_I+
\frac{1}{a^2}\d_i\pi\d_i\sigma_I\right)+ \\\nonumber 
&&-2\, e_3{}^{IJ} \dot\pi\dot\sigma_I\dot\sigma_J-2\,e_4{}^{IJ} \dot\pi\;\d_\mu\sigma_I
\d^\mu\sigma_J-4\,\tilde M_2^2{}^{I} \dot\pi^2\dot\sigma_I+\\ 
&&-\tilde M_3^{-2\, ,\,IJK}\dot\sigma_I\dot\sigma_J\dot\sigma_K-\tilde M_4^{-2\, , \, IJK}\dot\sigma_I
\;\d_\mu\sigma_J\d^\mu\sigma_K+... \Big]\ ,
\end{eqnarray}
where small latin indexes run from 1 to 3. It is useful to join together this Lagrangian and the one from single field (\ref{eq:Spi}), and to split it into a quadratic and a cubic term. We obtain:
\begin{eqnarray}\label{Spimulti2}
&&\!\!\!\!\!\!\!\!\! S^{(2)} =\\ \nonumber
&&\!\!\!\!\!\!\!\!\!   \int d^4 x   \sqrt{- g} \left[ (2M_2^4-M^2_{\rm Pl}
\dot{H}) \dot\pi^2+\mpl^2\dot H \frac{ (\partial_i \pi)^2}{a^2}
%- M^2_{\rm Pl} \left(3H^2 +\dot{H}\right)+ \right.\\
\right. %\\\nonumber && 
+2 \tilde M_1^2{}^{I} \dot\pi\dot\sigma_I+(1+\tilde e_2^I)\dot\sigma_I\dot\sigma_I+
\frac{\d_i\sigma_I\d_i \sigma_I}{a^2}+... \Big]\ ,
\end{eqnarray}
and
\begin{eqnarray}\label{Spimulti3} 
\! \! S^{(\rm 3)} & \!\!\!\!\!\!\!\!\!\!\!\!= \!\!\!\!\!\!\!\!\!& \!\!\!\int d^4 x   \sqrt{- g} \left[ 
%- M^2_{\rm Pl} \left(3H^2 +\dot{H}\right)+ \right.\\
-2 M^4_2\;
\dot\pi\frac{(\partial_i\pi)^2}{a^2}+ \left(2M_2^4-\frac{4}{3} M^4_3\right) \dot{\pi}^3+ \right. \\
\nonumber
&& -(\tilde M_1^2+4\tilde M_2^2)^I\dot\pi^2\dot\sigma_I-\tilde M_1^2{}^I\frac{(\d_i\pi)^2}{a^2}\dot
\sigma_I-2\tilde M_1^2{}^I\dot\pi\frac{\d_i\pi\d_i\sigma_I}{a^2} \\ \nonumber 
&& 2\left(e_2-e_3+e_4\right)^{IJ}\dot\pi\dot\sigma_I\dot\sigma_J-2e_4^{IJ}\dot\pi\frac{\d_i\sigma_I
\d_i\sigma_J}{a^2}-2 \tilde e_2^I \frac{\d_i\pi\d_i\sigma_I}{a^2}\dot\sigma_I\\ \nonumber 
&&+\left(\tilde M_4^{-2}-\tilde M_3^{-2}\right)^{IJK}\dot\sigma_I\dot\sigma_J\dot\sigma_K-\tilde 
M_4^{-2\, , \, IJK}\dot\sigma_I\frac{\d_i\sigma_J\d_i\sigma_K}{a^2}+\ldots \Big]\ .
\end{eqnarray}
In both equations, $\ldots$ represent higher derivative terms or terms that break the shift symmetry. 
%In the last term we add all the symmetric terms with respect to the $I,J,K$ indexes.
Let us analyze the quadratic and the cubic Lagrangian separately.

\subsubsection*{$\bullet$ Quadratic Lagrangian}

In the $\pi$ Lagrangian the term in $(\delta g^{00})^2$ induces a speed of sounds different from one for the $\pi$ Goldstone boson. Because the Lorentz symmetry is spontaneously broken, a speed of sound equal to one is not protected by any symmetry \cite{Cheung:2007st}. The same is true for the $\sigma_I$ fields. In addition to the standard Lorentz invariant kinetic term for the $\sigma_I$'s the operator proportional to $\tilde e_2$ generates an additional time-kinetic term. This has the effect of changing the speed of sound of the $\sigma_I$'s: each $\sigma_I$ can have a different speed of sound. This generalizes the case of multifield DBI inflation, where it was found that all of the $\sigma_I$ fields and the Goldstone boson $\pi$ have the same speed of sound~\cite{Langlois:2008wt}. This is a restriction that comes from the particular symmetries of the DBI construction but is not general.

The operator proportional to $\tilde M_1^{2\; I}$ generates a kinetic mixing between the $\sigma_I$'s and $\pi$. Again, this mixing is not Lorentz invariant: it is only at the level of the time-kinetic part $\dot\pi\dot\sigma_I$, while the spatial-kinetic part  {\it cannot} mix through terms of the form $\d_i\pi\d^i\sigma_I$. This is an unexpected constraint that follows from  the fact that $\pi$ non-linearly realizes time diff.s. This is analogous to what happened in single field inflation, where the coefficient of the operator $\dot\pi^2$ was not protected by any symmetry, while the coefficient of $(\d_i\pi)^2$ was fixed to be equal to $\dot H\mpl^2$. Unfortunately this mixing term is not important in a wide range of parameter space. In fact, let us consider for simplicity the case in which the mixing is only between $\pi$ and one of the Goldstone bosons $\sigma_I$, that we can call $\sigma$. In that case upon rescaling $\sigma$ and $\pi$ as
\be
\sigma_{\rm res}=(1+\tilde e_2)^{1/2}\sigma\ , \qquad \pi_{\rm res}=\frac{\tilde M_1^2}{(1+\tilde e_2)^{1/2}}\pi\ ,
\ee
and neglecting the time-dependence of the rescaling coefficients, one can put the time-kinetic matrix in the form 
\begin{equation}\label{eq:time-kinetic-mixing}
(\dot \sigma_{\rm res} \quad \dot\pi_{\rm res})\left(
\begin{array}{cc}
1 & 1 \\
1 &(1+e_2) \frac{(2M_2^4-\dot H\mpl^2)}{\tilde M_1^4} 
\end{array} \right)
\left(\begin{array}{c}
\dot\sigma_{\rm res}\\ \dot\pi_{\rm res}\end{array}\right)\equiv
(\dot \sigma_{\rm res} \quad \dot\pi_{\rm res})\left(
\begin{array}{cc}
1 & 1 \\
1 &1+\epsilon_{\rm unmix} 
\end{array} \right)
\left(\begin{array}{c}
\dot\sigma_{\rm res}\\ \dot\pi_{\rm res}\end{array}\right)\ ,
\end{equation}
where $\sigma_{\rm res}$ and $\pi_{\rm res}$ represent the rescaled $\sigma$ and $\pi$ fields, and where we have defined $\epsilon_{\rm unmix}$ as $(1+e_2) (2M_2^4-\dot H\mpl^2)/\tilde M_1^4-1$. We have also neglected numerical coefficients of order one. For $\epsilon_{\rm unmix}<0$, the time-kinetic matrix has a negative eigenvalue. This means that in this limit the theory has a ghost and does not make sense. Clearly in the regime $\epsilon_{\rm unmix}\gg 1$, the effect of the mixing becomes negligible. Therefore, it is only in the region of parameter space where $0\leq\epsilon_{\rm unmix}\lesssim 1$ that the effect of the mixing is important. As we will underline shortly, the kinetic-mixing term is associated with an interaction operator that in the limit $0\leq\epsilon_{\rm unmix}\ll1$ induces a very large level of non-Gaussianity in the density fluctuations.

So far in the Lagrangian in (\ref{eq:actiontadmulti}) we have neglected to write down operators of the form
\be\label{eq:ghost-dispertion-operator}
\bar{\bar M}^{-2\ IJ}\d^2\sigma_I \d^2\sigma_J\ ,
\ee
where two derivatives act on the $\sigma_I$ fields. The reason for this is that generically these operators are important only at energy scales of order of the cutoff of the theory and therefore negligible at energies of order $H$. However there is a regime where the operator which contains  only spatial derivatives can be important. This happens because in inflation we are not interested in arbitrary low energy fluctuations but rather in energies of order $H$~\cite{Cheung:2007sv,Senatore:2009cf}. In this case (for simplicity we neglect the internal indexes $I$) taking the limit of very large $\tilde e_2$ it is possible that for frequencies of order $H$ the spatial-kinetic term of the form $(\d^2_i\sigma)^2/\bar{\bar{M}}^2$ dominates with respect to the standard $(\d_i\sigma)^2$ term. This happens for $\tilde e_2\gtrsim \bar{\bar M}^2/H^2$. In this limit the dispertion relation is of the form $\omega^2\sim k^4/\bar{\bar M}^2$, which is the same kind of very non-relativistic dispersion relation found in the case of ghost inflation~\cite{Cheung:2007sv,ArkaniHamed:2003uz}. We see that this can also happen even in the case of multifield inflation.

\subsubsection*{$\bullet$ Cubic Lagrangian}

The cubic Lagrangian is quite complicated. All the interactions of the form $\pi^3,\; \pi^2
\sigma,\;\pi\sigma^2$ and $\sigma^3$ subject to the constraint that a derivative must act on each fluctuations and the operator must be rotational invariant are present. This leaves us with a number of cubic operators equal to
\be
2+3N+2 \frac{N (N+1)}{2}+N^2+\frac{N(N+1)(N+2)}{6}+N\frac{N(N+1)}{2}=\frac{6+13N+9N^2+2N^3}{3} \ ,
\ee
where $N$ is the number of the $\sigma$ fields (there are $10$ operators even for a single additional scalar). 

We notice that the coefficients of some of these operators are {\it univocally determined} in terms of some the operators that appear in the quadratic Lagrangian. This is so because in unitary gauge we are allowed to write down operators that are not time diff. invariant. Full diff. invariance is recovered once we reintroduce the Goldstone boson $\pi$, but it does so only at non-linear level. This means that if at a given order in the fluctuations we have a time-diff violating term, upon reinsertion of the $\pi$ this term will induce  operators containing at least one extra $\pi$ fluctuation.

For example for  single field inflaton the coefficient of the operator $\dot\pi(\d_i\pi)^2$ was uniquivocally fixed in terms of the speed of sound and the canonical normalization of the quadratic Lagrangian:
\be
M^4_2\dot\pi^2 \qquad \rightarrow \qquad M^4_2 \dot\pi(\d_i\pi)^2\ .
\ee
The same happens here for the non diff. invariant kinetic terms involving the $\sigma_I$ fields. For example, the operator $\tilde e_2^I (g^{0\mu}\d_\mu\sigma_I)^2$ which changes the speed of sound of the fluctuations induces the following interaction of the form $\pi\sigma^2$:
\be
(-1+\tilde e_2^I)\dot\sigma_I\dot\sigma_I \qquad\rightarrow\qquad \tilde e_2^I\frac{\d_i\pi\d_i\sigma_I}{a^2}\dot
\sigma_I\ ,
\ee
whose coefficient cannot be altered by any other operator. Notice instead that the coefficient of the operator $\dot\pi\dot\sigma^2$ is not fixed by $\tilde e_2$ uniquely. 

Similarly, for the time-kinetic mixing $\dot\pi\dot\sigma_I$, we have
\be\label{eq:operator_mix}
 \tilde M_1^{2\; I} \dot\pi\dot\sigma_I\qquad\rightarrow\qquad -\tilde M_1^{2\; I}\left(\frac{(\d_i\pi)^2}{a^2}\dot
\sigma_I+2\dot\pi\frac{\d_i\pi\d_i\sigma_I}{a^2}\right)\ ,
\ee
 and again the coefficient of these two $\pi^2\sigma$ operators cannot be changed once $\tilde M_1^2$ is fixed. Notice that in the limit $\epsilon_{\rm unmix}\ll1$ one of the fields becomes very strongly coupled. If $\epsilon_{\rm unmix}$ is very small, $\pi$ and $\sigma$ are approximately forty-five degrees mixed. Upon diagonalization, one of the diagonal fields has a very small coefficient in front of its kinetic term proportional to $\epsilon_{\rm unmix}$. After canonical normalization, the operators in (\ref{eq:operator_mix}) represent an interaction for this field that scales as $1/\epsilon_{\rm unmix}^{3/2}$ and the theory becomes strongly coupled in this limit. 

\subsubsection*{$\bullet$ Quartic Lagrangian}

It is interesting to write down the quartic Lagrangian for the $\sigma$ fields. 
In unitary gauge, there are only three operators which involve four $\sigma_I$ legs that are compatible with the approximate shift symmetry of the $\sigma_I$'s. The 
Lagrangian reads:
\begin{eqnarray}\label{Smulti4}
\! \! S^{(\rm 4)}_\sigma &= & \!\!\!\int d^4 x   \sqrt{- g} \left[ \tilde{\tilde M}_1^{-4\,,\, IJKL} \left(g^{{\mu
\nu}}\d_\mu\sigma_I\d_\mu\sigma_J\right) \left(g^{{\mu\nu}}\d_\mu\sigma_K\d_\mu\sigma_L\right)+ 
\right.\\ \nonumber 
&&\left.\tilde{\tilde M}^{-4\,,\, IJKL}_2 \left(g^{{0\mu}}\d_\mu\sigma_I\right)  \left(g^{{0\mu}}\d_\mu
\sigma_J\right) \left(g^{{\mu\nu}}\d_\mu\sigma_K\d_\mu\sigma_L\right)+\right.\\ \nonumber
&&\left.\tilde{\tilde M}^{-4\,,\, IJKL}_3 \left(g^{{0\mu}}\d_\mu\sigma_I\right)  \left(g^{{0\mu}}\d_\mu
\sigma_J\right)\left(g^{{0\mu}}\d_\mu\sigma_K\right)  \left(g^{{0\mu}}\d_\mu\sigma_L\right)
 \right]\ .
\end{eqnarray}

Reinsertion of the $\pi$ will induce some operators that are quintic in the fluctuations, of the form $\pi\sigma^4$. We neglect them here. After setting to zero the metric fluctuations, we are left with:

\begin{eqnarray}\label{Spimulti4}
&&S^{(\rm 4)}_{\sigma,\pi} = \int d^4 x   \sqrt{- g} \left[ \left(\tilde{\tilde M}_1^{-4}- \tilde{\tilde 
M}_2^{-4}+\tilde{\tilde M}_3^{-4}  \right)^{IJKL}\dot\sigma_I\dot\sigma_J\dot\sigma_K\dot\sigma_L+\right.
\\ \nonumber 
&&\left.\left(-2\tilde{\tilde M}_1^{-4}+\tilde{\tilde M}_2^{-4}  \right)^{IJKL}\dot\sigma_I\dot\sigma_J (\d_i
\sigma_K\d_i\sigma_L)+\tilde{\tilde M}_1^{-4\,,\,IJKL} (\d_i\sigma_I\d_i\sigma_J)(\d_i\sigma_K\d_i\sigma_L)\right]\ ,
\end{eqnarray}
%where in the last term we add all the permutations of the internal indexes ($I,J,K,L$). 
Notice that the combination proportional to $\tilde {\tilde M}_1$ is the only one that is Lorentz invariant. We will come back to this point later.  We stress that this is not the full Lagrangian at quartic order in the fluctuations. We are just considering the terms that are quartic in the $\sigma_I$'s. In principle, there are also terms like $\pi\sigma^3,\;\pi^2\sigma^2\,\,\ldots$ that we have neglected. We did this because it will turn out that this quartic Lagrangian in $\sigma^4$ can generically be important for observations inducing a large and detactable four point function. The other terms are less important. But in order to explain how this can happen we need to explain how the $\pi$ and the $\sigma_I$ fluctuations are related to observables like the curvature perturbations. We will do this in the next section.

\subsubsection*{$\bullet$ Soft-breaking Lagrangian}

As we will argue later, the $U(1)^N$ symmetry of the Goldstone bosons need not only be spontaneously broken, but also {\it it has}  to be explicitly broken for otherwise a $\sigma_I$ fluctuation would have no effect on the curvature perturbations of the universe. This breaking can be concentrated just around the reheating time, but there might be a soft breaking of the symmetry at the time the modes cross the horizon. We will concentrate on this possible breaking in this section and come back to the breaking at the time of reheating in the next section. For simplicity we will concentrate on a single $U(1)$ symmetry, the generalization to the $U(1)^N$ case is straightforward.

Let us imagine that the $U(1)$ symmetry is explicitly broken by a term that transforms under the $U(1)$ with charge one, and whose size is controlled by a parameter $\mu^4\ll M^4$, with $M$ being the typical scale suppressing the higher dimension operators in (\ref{eq:actiontadmulti}). This implies that the soft-symmetry breaking is signaled by the appearance in the Lagrangian of terms like $e^{i\sigma}$ which have charge one under the $U(1)$. By treating $\mu^4$ as a spurion transforming under the $U(1)$ with charge minus one, it is possible to construct the potential, which reads:
\be\label{eq:potential}
V(\sigma)=-a_1 \mu^4\cos\left(\frac{\sigma}{M}+\theta\right)+b_1\frac{\mu^8}{M^4}\cos\left(2\frac{\sigma}{M}\right)+b_2\frac{\mu^8}{M^4}\sin\left(2\frac{\sigma}{M}\right)+\ldots \ .
\ee
The terms proportional to $b_{1,2}$ are higher order in $\mu/M\ll1$, and dots represent even higher order terms that are negligible. The effect of the phase $\theta$ is to make $\sigma=0$ not a minimum of the potential. At leading order in $\mu/M$ the minimum is at $\sigma\simeq M \theta$. Expanding around the minimum, with a suitable redefinition of $\sigma$:
\be
\sigma=M \theta+\tilde\sigma\ ,
\ee
we obtain a leading potential given by 
\be\label{eq:potential-even}
V(\tilde\sigma)_{\rm leading}=-a_1 \mu^4\cos\left({\tilde\sigma \over M}\right)\simeq a_1 \mu^4\left(1-\frac{1}{2}\frac{\tilde\sigma^2}{M^2}+\frac{1}{24}\frac{\tilde \sigma^4}{M^4}+\ldots\right)\ .
\ee
For the $\sigma$ field to have sizable fluctuations, we need to require its mass to be much smaller than $H$. This implies the condition
\be\label{eq:constraint}
\mu^4\ll H^2 M^2\ ,
\ee
with $H\ll M$ from imposing the cutoff of the theory be larger than $H$. Notice that in the presence of a mass term $m$, fluctuations of the $\sigma$ fields keep evolving outside of the horizon and decrease their amplitude according to the equation $3H \dot\sigma_I+m^2\sigma_I=0$. Since modes that exited the horizon at earlier times have longer time to evolve, this effect induces a tilt in the two-point function  of order $m^2/H^2$. Therefore we need to impose $\mu^4/(M^2 H^2)\lesssim n_s-1 \ll 1$. The first inequality comes from the fact that the tilt $n_s-1$ receives additional contributions from the scale dependence of $H$ and therefore the actual tilt $n_s-1$ can be larger than $\mu^4/(M^2 H^2)$~\footnote{As we anticipated, we are here interested in the case where the additional scalar fields are directly observables and so must have quasi scale invariant fluctuations. We do not study here the case, for example as treated in~\cite{Chen:2009we}, in which the additional fields have mass of order $H$, do not have scale invariant fluctuations, but nevertheless lead to physical effects because back-reacting on $\pi$. We will study these models in a subsequent paper.}.  The potential in (\ref{eq:potential}) is even under $\tilde\sigma\rightarrow -\tilde\sigma$. The leading contributions breaking this symmetry come at subleading order in $\mu/M$, and are of the form
\be\label{eq:potential-odd}
V(\tilde\sigma)_{\rm sub-leading}\sim b_1 \frac{\mu^8}{M^4}\sin\left(2\frac{\tilde\sigma}{M}\right)\sim - b_1 \mu^8 \frac{\tilde\sigma^3}{M^7}\ ,
\ee
where we have neglected an irrelevant tadpole term that is canceled when we impose the minimization at higher order in $\mu/M$.

The other leading interactions coming from the breaking of the $U(1)$ symmetry are given by the operators of the form
\bea\label{eq:interactions_mixed}\label{eq:operator-peculiar}
\frac{\mu^4}{M^2}\dot{\tilde \sigma}\left(\cos\left(\frac{\tilde\sigma}{M}\right)-1\right)\qquad&\rightarrow&\qquad \frac{\mu^4}{M^4}\dot{\tilde\sigma}\tilde\sigma^2\ \qquad \rightarrow\qquad \frac{\mu^4 H}{M^4}\tilde\sigma^3, \\ \nonumber
\frac{\mu^4}{M^4}(\d \tilde\sigma)^2\sin\left(\frac{\tilde\sigma}{M}\right)\qquad&\rightarrow&\qquad \frac{\mu^4}{M^5}(\d \tilde\sigma)^2\tilde\sigma\ , \\ \nonumber
\frac{\mu^4}{M^4}(\d \tilde\sigma)^2\left(\cos\left(\frac{\tilde\sigma}{M}\right)-1\right)\qquad&\rightarrow&\qquad \frac{\mu^4}{M^6}(\d \tilde\sigma)^2\tilde\sigma^2\ , \\ \nonumber
%&&\frac{\mu^4}{M^6}(\d \tilde\sigma)^2\dot{\tilde\sigma} \sin\left(\frac{\tilde\sigma}{M}\right)\qquad\rightarrow\qquad \frac{\mu^4}{M^7}(\d \tilde\sigma)^2\dot{\tilde\sigma}\tilde\sigma\ . 
\eea
where here $(\d\tilde\sigma)^2$ stays for $(\d_i{\tilde\sigma})^2$, $\dot{\tilde\sigma}^2$ or a combination of the two. As we have highlighted, in an expanding space the first operator is not a total derivative. Upon integration by parts it picks up a factor of $H$ and becomes of the form $\tilde\sigma^3$.
%We will later see that the constraint (\ref{eq:constraint}) will make all the above interactions in (\ref{eq:potential-even}), (\ref{eq:potential-odd}) and (\ref{eq:interactions_mixed}) irrelevant from the observational point of view.

Our assumption of having only one spurion of charge one is technically natural, and it is at the basis of the suppression of the $\sigma^3$ interaction. This will have important observational consequences. However, it also possible to add a second spurion $\tilde \mu^4$ with charge two and with size comparable to $\mu^4$. In this case terms of the form 
\be
\tilde \mu^4\; \sin\left(2\frac{\sigma}{M}\right)
\ee
are allowed. Upon minimization of the potential, terms cubic in $\tilde\sigma$ would survive with coefficients of order $\tilde\mu^4/M^3$:
\be\label{eq:cubic-spurion-two}
\tilde \mu^4\; \sin\left(2\frac{\sigma}{M}\right) \qquad\supset\qquad  \frac{\tilde \mu^4}{M^3}\tilde\sigma^3\ .
\ee

We finally notice that in unitary gauge the parameters $\mu^4,\tilde\mu^4,\ldots\;,$ can depend explicitly on time $t$ (for example with a linear dependence). In this case we could even consider terms where in unitary gauge we act with a time-derivative on $\mu^4,\tilde\mu^4,\ldots\;$. For terms linear in $\mu^4$ this time derivative could be integrated away by parts but this is not possible to do in general at higher order in $\mu^4,\tilde\mu^4,\ldots\;$. Upon reinsertion of $\pi$ the time dependence becomes dependence on the combination $t+\pi$ while every time-derivative in unitary gauge now becomes a space-time derivative as for example $\d^0=g^{0\rho}\d_\rho$. This means that when $\mu^4,\tilde\mu^4,\ldots\;,$ are time-dependent, we will have additional couplings between $\pi$ and $\sigma$. Similar considerations apply to the case for example when we multiply an operator in $\sigma$ with factors of $\delta g^{00}$, the leading being a mixing term $\delta g^{00}\sigma$. The importance of these additional $\pi-\sigma$ couplings  is quite modest as they are already  proportional to $\mu^4,\tilde\mu^4,\ldots\;$. Further, as we will explain later, they can be of some importance only when both $\pi$ and $\sigma$ are relevant for the curvature perturbations at a comparable level. This is in some sense a  tuned case whose phenomenology lies beyond the scope of the present paper.

\subsection{Non-Abelian case\label{sec:non-Abelian-case}}

In the former section we have considered the simpler case of the Golsdstone bosons $\sigma_I$'s originating from  the spontaneous breaking of a $U(1)^N$ group. We  now consider the analogous non-Abelian  case. While the generalization is  straightforward we will obtain some new interesting signatures that are specific of the non-Abelian nature of the symmetry group. The novelty here arises from the fact that if the broken group is non-Abelian the Goldstone bosons need not to be protected by a shift symmetry to be light. Therefore we will not be forced to have a derivative acting on each of the $\sigma$ fields. At cubic level for example in the Abelian case we simply had operators of the form $\dot\sigma^3$ and $\dot\sigma(\d_i\sigma)^2$.  The constraint of having a derivative acting on each of the $\sigma$ restricted the possible terms we could write. In the non-Abelian case we might have operators of the form $\sigma\dot\sigma^2$, $\sigma(\d_i\sigma)^2$ and $\sigma^2\dot\sigma$. In the Abelian case we had these operators only after considering terms coming from the soft-symmetry breaking, and their form was therefore strongly constrained. In the non-Abelian case these operators can appear already in the limit of exact spontaneously broken symmetry, though the symmetry might impose further constraints that we have to explore.

 Let us therefore consider a global symmetry group $G$ that is spontaneously broken to a subgroup $H$. For simplicity, we take $G$ to be compact.  The theory for the resulting Goldstone bosons together with a historical account of the developments is nicely explained in \cite{Weinberg:1996kr} which we follow here. The only generalization we will have to make here is to couple these Goldstone bosons to the Goldstone boson of time translations ($\pi$).

Let us set up some notation. We label the generators of $H$ by $t_i$, while those of $G$ and not of $H$  by $x_a$. The Lie algebra takes the following form
\bea
&&[t_i,t_j]=i C_{ijk } t_k\\ \nonumber
&&[t_i,x_a]=i C_{iab} x_b\\ \nonumber 
&&[x_a,x_b]=i C_{abi} t_i+i C_{abc} x_c\ .
\eea
We notice that the broken generators $x_a$'s transform under the same representation of $H$. The Goldstone bosons $\sigma_a$'s are, apart from a normalization, a parametrization of the right coset $G/H$ that we can chose as
\be
\gamma(\{\sigma_a(x)\})={\rm Exp}\left[i \sigma_a(x) x_a\right]\ .
\ee
The Lagrangian for the Gosldstone bosons is the most general Lagrangian built in terms of the operator $D_{a\mu}$, which is defined as
\be
\gamma^{-1}(x)\d_\mu\gamma(x)=i x_a D_{a\mu}(x)+i t_i E_{i\mu}(x)\ . 
\ee
At leading order in the fields and in the derivatives, $D_{a\mu}$ and $E_{i \nu} $ are given by
\bea\label{eq:defD}
&&D_{a\mu}=\d_\mu\sigma_a+\frac{1}{2}C_{abc}\sigma_b\d_\mu\sigma_c+\frac{1}{6}\left(C_{cde} C_{bea}+C_{cdi}C_{bia}\right)\sigma_b\sigma_c\d_\mu\sigma_d+{\cal{O}}(\sigma^3\d_\mu\sigma)\\ \nonumber
&& E_{i\mu}=\frac{1}{2}C_{abi}\sigma_a\d_\mu \sigma_b+\frac{1}{6}C_{acd}C_{bdi}\sigma_a\sigma_b\d_\mu\sigma_c+{\cal{O}}(\sigma^3\d_\mu\sigma)\ .
\eea
The Lagrangian can also include higher derivative operators which have to be built out of covariant derivatives defined as:
\be
\left({\mathscr{D}}_\nu D_{ \mu}\right)_a=\d_\nu D_{a \mu}+C_{iab} E_{i \nu} D_{b\mu}\ .
\ee

While the $\sigma_a$ transform linearly only under $H$ they do not transform linearly under generic $G$ transformation. Instead $D_{a\mu}$ transforms linearly under all $G$ transformations. In particular if we define:
\be\label{dmu}
D_\mu\equiv D_{a\mu} x_a\ ,
\ee
it transform under an element $g\in G$ as
\be
D_\mu'=h\left(\sigma(x),g)\right)D_\mu h\left(\sigma(x),g)\right)^{-1}\ ,
\ee
where $h\left(\sigma(x),g)\right)$ is an element of $H$ that depends on both $\sigma$ and $g$ \footnote{Using the notation of \cite{Weinberg:1996kr}, if $g\in G$ we have
\be
D'_{a\mu}(x)={\mathscr{D}}_{ab}\left(\sigma(x),g\right) D_{b\mu}(x)\ .
\ee
where ${\mathscr{D}}_{ab}$ are the matrixes of the adjoint representation of $G$:
\be
h x_b h^{-1}={\mathscr{D}}_{ba} x_a\ ,
\ee
where $h$ is an element of $H$, that realize a reducible representation of $H$. This suggests to conveniently group the operators $D_{a\mu}$ into the matrixes $D_\mu$ of equation (\ref{dmu}).}.

The Lagrangian for the Goldstone bosons is obtained by constructing the most generic Lagrangian with the operators $D_{a\mu}$ and its covariant derivatives that is invariant under linear transformation of the unbroken $H$ group and the remaining symmetries of the problem. This automatically ensures invariance under non-linear transformation of the full group $G$. Notice that symmetries dictate that each operator containing Goldstone bosons must contain at least one derivative acting on one of the Goldstone bosons. Each term in $D_{\mu}$ contains at least one derivative, and therefore expanding in $D_{\mu}$ corresponds to a low energy expansion. The Lagrangian we construct differs from the standard one by the additional couplings to the Goldstone boson of time-translations. By now we have learned how to couple to this Goldstone boson: we write the action in unitary gauge where we restrict the spacetime symmetries to be simply time-dependent spatial diffeomorphisms. The resulting Lagrangian reads

\begin{eqnarray}\label{Spimulti-non-Abelian} \nonumber
S_{\sigma} = \int d^4 x   \sqrt{- g} &&{\rm Tr}\left[F_1^2 D_{\mu }D^{\mu}+ F_2^2 D^{0}D^{0}+F_3^3(g^{00}+1)D^0+\right.\\ \nonumber
  &&\left. F_4^2(g^{00}+1) D_{\mu}D^{\mu}+F_5^2(g^{00}+1) D^{0}D^{0}+ \right.\\
&&  \left. \bar F_1 D_{\mu}D^\mu D^0+\bar F_2 D^0 D^0 D^0+\ldots \right] \ ,
  %\\ \nonumber &&\left.F_4\epsilon^{abc}D^{0}_ {a}D^{\mu}_{ b}D_{\mu c}+F_4\epsilon^{abc}D^{0}_{a}D^{0}_bD^{0}_ {c}\right]\ ,
\end{eqnarray}
where $F_{1,2,3,4,5},\, \bar F_{1,2}$ are time-dependent numbers with the dimension of mass that we expect to be of comparable size. The time-dependence of the coefficients breaks the shift symmetry of $\pi$ and is therefore expected to be small, proportional to the slow-roll parameters.  We have also assumed for simplicity that the generators $x_a$ are hermitian (this is true for all the Cartan's classical Lie groups) and that the $x_a$ transform in an irreducible representation of $H$, the generalization being straightforward.
%Here we have assumed that all the generators $x_a$ lie in a  representation of the group $H$. The generalization is  straightforward and does not lead to any  interesting additional effects.  Dots represent higher order or higher derivative terms. 
The $\pi$ gets reinserted in the usual way. In particular:
\be
D^{0}_{ a}\qquad\rightarrow\qquad \frac{\d(t+\pi)}{\d x^\mu}D^{\mu}_{ a}\ .
\ee
Upon reinsertion of the field $\pi$, neglecting metric perturbations, we obtain:
\begin{eqnarray}\label{Spimulti-non-Abelian-pi} \nonumber
S_{\pi\sigma} = \int d^4 x   \sqrt{- g} &&{\rm Tr}\left[F_1^2 D_{\mu }D^{\mu}+ F_2^2 D^{0}D^{0}+ 2 F_2^2\d_\mu\pi D^{\mu}D^{0}-2F_3^3\dot\pi D^0+F_3^3(\d_\mu\pi)^2D^0+\right.\\ &&\left. -2 F_4^2 \dot\pi D_{\mu}D^{\mu}-2 F_5^2\dot\pi D^{0}D^{0}+\right. %\\ \nonumber && 
\left. \bar F_1 D_{\mu}D^\mu D^0+\bar F_2 D^0 D^0 D^0+\ldots \right] 
%&&\left.F_4\epsilon^{abc}D^{0}_ {a}D^{\mu}_{ b}D_{\mu c}+F_4\epsilon^{abc}D^{0}_{a}D^{0}_bD^{0}_ {c}\right]\ ,
\end{eqnarray}
where we have kept terms in $\pi$ only up to those of the form $\pi\sigma^2$ and $\pi^2\sigma$ because we will later see that those are the relevant ones.

\subsubsection*{$\bullet$ Quadratic Lagrangian}

Using (\ref{eq:defD}) we can see that the first two terms in the first line of eq. (\ref{Spimulti-non-Abelian-pi}) give rise to the kinetic term for the $\sigma_a$'s. The first contains the standard Lorentz invariant term while the second one contains  time-kinetic term. This implies that the speed of sound of the Goldstone fluctuations can be different from one:
\be
c_s^2\sim \frac{F^2_2}{\left(F^2_1+F^2_2\right)}\ .
\ee
As usual, having a speed of sound different from one introduces an interaction with $\pi$ at higher order. This comes from the fact that $\pi$ non-linearly realizes time-diffs. When the speed of sound becomes very small, it is possible that higher-derivative terms, schematically of the form
\be
{\rm Tr}\left[{\mathscr{D}}_\nu D^{\mu}{\mathscr{D}}^\nu D_{\mu}\right]
\ee
and others with similar index contractions, dominate the spatial part of the dispersion relation at horizon crossing, and induce a dispersion relation of the form $\omega^2\propto k^4$ as it happens in the Abelian case and in the single-clock case.

The fourth term in the first line denotes a kinetic mixing between $\sigma_a$ and $\pi$. As in the Abelian case, such a term in the quadratic Lagrangian induces interactions between $\pi$ and $\sigma$. In order for this term not to be zero, it needs that ${\rm Tr}[x_a]\neq 0$. This in practice signals the presence of an Abelian commuting subgroup being spontaneously broken: a kinetic mixing between $\pi$ and $\sigma$ is only possible for Goldstone bosons associated to the breaking of an Abelian group.

%Notice that contrary to the Abelian case, there is a kinetic mixing term between the Goldstone bosons and $\pi$ only in the case one of Goldstone bosons of the non-Abelian group is in a singlet representation. In that case, we could add to the above Lagrangian a term of the form $\delta g^{00}D^0$, which is invariant and give rise to a time-kinetic mixing between $\pi$ and the Goldstone boson. 
%: unless the group is Abelian, a unitary-gauge term of the form $\delta g^{00}D^0_a$ is not invariant.

\subsubsection*{$\bullet$ Cubic Lagrangian}

The cubic Lagrangian can be quite constrained in the non-Abelian case, the details depending on the particular group being broken. First, the fact that the symmetries do not require a derivative acting on each of the fields might allow for interactions of the form $\sigma(\d\sigma)^2$ that would be the leading ones in a derivative expansion. However, the antisymmetry of the structure constants makes these terms vanish. Among the terms with a derivative acting on each of the $\sigma$'s the ones proportional to $\bar F_{1}$ and $\bar F_2$ give rise to the same kind of interactions $\dot\sigma(\d_i\sigma)^2$ and $\dot\sigma^3$ that can be produced in the Abelian case. However, the coefficients of these interactions are proportional to ${\rm Tr}[x_a x_b x_c]$ which {\it can vanish} for some groups (for example for $SU(2)$). Further, there are {\it no} interactions of the form $\pi^2\sigma$ in the absence of an Abelian commuting subgroup. The only remaining cubic interactions are of the form $\pi\sigma^2$, and only a small subset of these are present. In fact, while it would not be naively required by symmetries, the $\sigma$'s appearing in the interactions  always have a derivative acting on them. In summary, in the absence of an Abelian commuting subgroup, we have the following schematic cubic interactions:
\be\label{eq:non_Abelian_three_point}
\dot\sigma(\d_i\sigma)^2\ , \quad\dot\sigma^3\ , \qquad\dot\pi\dot\sigma\dot\sigma\ , \qquad \dot\pi(\d_i\sigma)^2\ , \qquad \d_i\pi\d_i\sigma \dot\sigma \ ,
\ee
Just as in the Abelian case the coefficient of $ \d_i\pi\d_i\sigma_a \dot\sigma_a $ is uniquely fixed once the speed of sound of the $\sigma$'s is fixed. Furthermore the first two of these interactions vanish in the case ${\rm Tr}[x_a x_b x_c]=0$. This fact will have important consequences.

%In the case of a non-Abelian group, the interactions are highly constrained. Not all the interactions that were allowed in the Abelian case are now allowed. First of all there is {\it no} cubic interaction purely in the $\sigma_a$'s. This can be noticed by expanding the first line of the action (\ref{Spimulti-non-Abelian-pi}) up to cubic order and by noticing that the structure constants $C_{abc}$ are completely antisymmetric. Further, there are {\it no} interactions of the form $\pi^2\sigma_a$. The only cubic interactions are of the form $\pi\sigma_a^2$, and only a small subset of these are present. In fact, while it would not be naively required by symmetries, the two $\sigma_a$'s appearing in the interactions have always a derivative acting on them. In summary, we have the following cubic interactions:
%\be\label{eq:non_Abelian_three_point}
%\dot\pi\dot\sigma_a\dot\sigma_a\ , \qquad \dot\pi(\d_i\sigma_a)^2\ , \qquad \d_i\pi\d_i\sigma_a \dot\sigma_a \ .
%\ee
%Similarly to what happened in the Abelian case the coefficient of $ \d_i\pi\d_i\sigma_a \dot\sigma_a $ is uniquely fixed once the speed of sound of the $\sigma$'s is fixed.

\subsubsection*{$\bullet$ Quartic Lagrangian}

We first concentrate on the part of the Lagrangian that is purely quartic in the $\sigma$'s. It is possible to write down operators of the form ${\rm Tr}[D^0D^0D^0D^0],\ {\rm Tr}[D^\mu D_{ \mu }D^0D^0]$, and ${\rm Tr}[D^\mu D_{\mu}D^\nu D_{\nu}]$ that give rise to the same quartic interactions we found in the Abelian case. However in some cases there are lower-dimensional quartic interactions. The lowest dimensional ones come from expanding up to quartic order the terms in the first line in (\ref{Spimulti-non-Abelian-pi}). These are schematically of the form 
\bea\label{eq:non_Abelian_quartic_interactions}
F_1^2\left(C_{ebc}C_{eda}+C_{daf} C_{bfc}+C_{dai}C_{bic} \right)& \times &(\d_\mu\sigma_a)\sigma_b(\d_\mu\sigma_c)\sigma_d\ ,%\qquad C_{abc}C_{alm}\; \sigma_b\d_0\sigma_c\;\sigma_l\d_0\sigma_m\ , 
\\  \nonumber
%\left(C_{cde} C_{bea}+C_{cdi}C_{bia}\right)    \;  \d^\mu\sigma_a\; \sigma_b \sigma_c\d_\mu\sigma_d    \ ,\qquad 
F_2^2\left(C_{ebc}C_{eda}+C_{daf} C_{bfc}+C_{dai}C_{bic} \right) &\times &\dot\sigma_a\sigma_b\dot\sigma_c\sigma_d\  .
%\left(C_{lba}C_{lcd}+C_{cde} C_{bea}+C_{cdi}C_{bia}\right)   && \;  (\d_0\sigma_a) \sigma_b \sigma_c(\d_0\sigma_d)\ .
\eea
where we have neglected numerical factors. 
These quartic interactions are expected to dominate because they have two less derivatives than the interactions one gets from the operators $\sim D_{ \mu}{}^4$ which would have a derivative acting on each $\sigma$. The quartic interactions distinguish between Abelian and non-Abelian nature of the broken group. Notice also that in some well known groups some of the structure constants are zero and therefore the interactions can be further constrained. For example for $SU(2)\times SU(2)$ broken to the diagonal subgroup $C_{abc}$ is zero. As we will explain in detail later the total antisymmetry of the structure constants makes these interactions irrelevant for purely adiabatic fluctuations. The next-to-lowest dimensional quartic interactions come from quartic terms arising from expanding  the operators proportional to $\bar F_{1,2}$ in (\ref{Spimulti-non-Abelian-pi}). They are schematically of the form
\be\label{eq:non-Abelian-four-mixed}
{\rm Tr}[x_a x_b x_c]\; C_{cde}\;\d \sigma_a \d\sigma_b \d\sigma_d \sigma_e \ ,
\ee
with two possible combinations of space and time derivatives compatible with rotational invariance. It is important to notice that these operators are present only when lower dimensional cubic ones of the form $(\d\sigma)^3$ are present, which will make these operators always subleading from the observational point of view. We will however see that they can  still be quite interesting.

We will find that there exist symmetries that can forbid the leading cubic (and quartic) interactions, making  interactions of the form $\sim D_{a \mu}{}^4$ the leading ones. These come from the Lagrangian
\begin{eqnarray}\label{Spimulti-non-Abelian-4}
S^{(4)}_{\sigma} = \int d^4 x   \sqrt{- g} &&{\rm Tr}\left[ \tilde{\tilde c}_{11}D_{ \mu }D^{\mu}D_{\nu }D^{\nu}+ \tilde{\tilde c}_{12}D_{\mu }D_{ \nu }D^{\mu}D^{\nu}+\right.\\ \nonumber
&&\left. \tilde{\tilde c}_{21} D^{0}D^{0}D_{ \nu }D^{\nu}+ \tilde{\tilde c}_{22} D^{0}D_{ \nu }D^{0}D^{\nu}+\right.\\ \nonumber &&\left. \tilde{\tilde c}_{31} D^{0}D^{0}D^{0}D^{0}\right]\\ \nonumber  
\simeq\int d^4 x   \sqrt{- g}\ \ {\rm Tr}\left[x_a x_b x_c x_d \right]&&\left[ \tilde{\tilde c}_{11}\d_\mu\sigma_a\d^\mu\sigma_b \d_\nu\sigma_c\d^\nu\sigma_d+ \tilde{\tilde c}_{12}\d_\mu\sigma_a \d_\nu\sigma_b\d^\mu\sigma_c\d^\nu\sigma_d+\right.\\ \nonumber&&\left. \tilde{\tilde c}_{21} \dot\sigma_a\dot\sigma_b \d_\nu\sigma_c\d^\nu\sigma_d+\tilde{\tilde c}_{22} \dot\sigma_a\d_\nu\sigma_b\dot\sigma_c \d^\nu\sigma_d+\right.% \\ \nonumber &&
\left.\tilde{\tilde c}_{31}   \dot\sigma_a\dot\sigma_b \dot\sigma_c\dot\sigma_d\right]\ .
\end{eqnarray}
where $\tilde{\tilde c}_{11,12,21,22,31}$ are time-dependent dimensionless numbers of order unity, and where in the second passage we have kept only terms of quartic order in the $\sigma_a$'s. Notice that, depending on the properties of the group, it is possible that the operators multiplying $ \tilde{\tilde c}_{11}$ and $ \tilde{\tilde c}_{12}$, as well as $ \tilde{\tilde c}_{21}$ and $ \tilde{\tilde c}_{22}$, are the same, effectively reducing the number of free parameters.

\subsubsection*{$\bullet$ Soft-breaking Lagrangian}

It is quite difficult to discuss in full generality the soft-breaking of a non-Abelian group. In order to do this one has to specify the spurionic representation of the terms explicitly breaking the symmetry. While in the $U(1)$  case  in practice there is only one representation, for a non-Abelian group there are many, and the Lagrangian differs in the different cases. It is however quite straightforward to realize that all the operators that were associated to the explicitly braking in the $U(1)$ case, as for example the $\sigma(\d\sigma)^2$ operators, can be obtained also in this case. The actual presence (or absence) of each breaking term however depends on the spurionic properties of the terms explicitly breaking the symmetry.  

%There is however the possibility of obtaining, at least in principle, new operators of the form:
%\be\label{eq:operator-peculiar}
%\frac{\mu^4}{F^4}\dot\sigma_a \sigma_b\sigma_c\ ,\qquad\frac{\mu^4}{F^5}\dot\sigma_a \sigma_b\sigma_c\sigma_d\ ,
%\ee
% where in the non-Abelian case the spurionic symmetry-breaking parameter $\mu^4$ is a matrix that we can imagine transforms appropriately. Notice that in the $U(1)$ case these operators are total derivatives: there need to be more than one kind of fluctuation in the operator for it not to be a total derivative, and this forces the broken group to be either $U(1)^N$ or non-Abelian. However in the $U(1)^N$ case the coefficient of these operators need to be proportional to the spurion associated to at least two of the broken $U(1)$'s, making the importance of this operator in the $U(1)^N$ case quite small~\footnote{In an inflating spacetime, these operators would not be total derivative as, by integration by part, the derivative would hit the scale factor. In this case they would however reduce to a sum of $\sin(\sigma/M)$ and $\cos(\sigma/M)$ that we discussed before.}.

In the case the symmetry is softly broken, the Goldstone bosons have a potential and, depending on the initial conditions at the beginning of inflation, they can have a time-dependent vacuum expectation value (vev). Since a time-dependent vev amounts to spontaneously breaking time-diffs, this means that $\pi$ will be mixed with those Goldstone bosons $\sigma_a$ that acquire a vev. Let us take for simplicity the case in which one of the Goldstone bosons $\sigma_a$ acquires a vev and let us say that this is $\sigma_1$, as one can always rotate the generators so that this is the case. 

In  order to construct the Lagrangian for the fluctuations, it is best to proceed in the following way. We can just take the Lagrangian we have constructed and give a time-dependent vev to $\sigma_1$. Since the potential is small we can identify two regimes of interests.  The first in when the higher-dimensional terms schematically of the form $D_\mu^3,\ldots,\;$ are assumed to be negligible. In this case the equation of motion for the vev, $\langle\sigma_1\rangle$, will be
\be\label{eq:goldstone-inflaton}
\ddot{\langle\sigma_{1,c}\rangle}+3H \dot{\langle\sigma_{1,c}\rangle}-\frac{\mu^4}{F}\sin\left(\frac{\langle\sigma_{1,c}\rangle}{F}\right)=0\ .
\ee
where the subscript $_c$ denotes the canonically normalized field.
Here for simplicity we have taken all the $F$'s as being the same. Notice that since the inflaton can be given by a combination of different fields (in principle not even scalar fields), in addition to $\sigma_1$, the Hubble constant $H$ is here not univocally determined in terms of $\mu^4$. Clearly, there are two possible sub-regimes for the solution of (\ref{eq:goldstone-inflaton}), depending mainly of the value of $H$. One possibility is that the field oscillates around the minimum. In this case the oscillations get damped exponentially fast in time. We therefore neglect this possibility. A second regime is when the field is slowly rolling down the hill. In this case, we have
\be\label{eq:time-dependent-vev}
|\dot{\langle\sigma_{1,c}\rangle}|=\frac{\mu^4}{3HF}|\sin\left(\frac{\langle\sigma_{1,c}\rangle}{F}\right)|\leq  \frac{\mu^4}{3HF}\ .
\ee
This estimate ensures that the higher dimensional operators $D_\mu^3,\ldots,\;$ and higher derivative operators are negligible.  In order to construct the Lagrangian for the fluctuations, we reinsert $\pi$ by performing the field redefinition
\be\label{eq:subs}
\sigma_1(\vec x,t)=\langle\sigma_1\rangle(t+\pi(\vec x,t))+\delta\sigma_{tr,1}(\vec x,t)\ ,
\ee
where $\delta\sigma_{1,tr}$ represents the scalar field fluctuation remaining after the definition $\langle\sigma_1\rangle(t+\pi)$. This happens when the field $\sigma_1$ is not the only field that develops a time-dependent vev. Since the number of light fields is an input in the Lagrangian, the presence of an additional clock-field appears as a free choice. By substituting (\ref{eq:subs}) into the Lagrangian and expanding in fluctuations, there will be terms independent of the fluctuations or just dependent on $\pi$. These should be discarded, as they can just be moved to the part of the Lagrangian that has to do with the clock field $\pi$ and whose coefficients were generic to start with (in fact, there could be additional fields that have a vev and contribute to it). The remaining terms involve directly the $\sigma_a,\ a\neq 1$ fluctuations together with additional couplings involving $\pi$ or $\delta\sigma_{tr,1}$. If the $\pi$ field happens to be almost entirely composed of the $\sigma_1$ field, then the component $\delta\sigma_{tr,1}$ is irrelevant and should be discarded. If instead $\sigma_{1}$ does not contribute relevantly to $\pi$, then the field $\delta\sigma_{tr,1}$ appears in the Lagrangian as the other $\sigma_a,\; a\neq1$ components. Intermediate cases are somewhat tuned and we do not consider them here. Similarly if $\pi$ is dominated by the $\sigma_1$ component, then its Lagrangian will be fully determined by the coefficients related to $\langle\sigma_1(t)\rangle$, implying a connection between the coefficients of the $\pi$ Lagrangian and the Goldstone Lagrangian. However in this case the $\pi$ Lagrangian is not very interesting, representing a standard slow roll inflation model. Instead in the opposite case in which $\sigma_1$ does not contribute much to $\pi$, then those terms are quite irrelevant for the $\pi$ Lagrangian. The vev of $\sigma_1$ leads to additional interactions. The most relevant ones that are not present in the absence of symmetry breaking and without the time-depende!
 nt vev a
re schematically of the form:
\be\label{3-pt-iso-non-Abelian}
F\,{\rm Tr}[x_1 x_a x_b]D^{1,0}D_\mu^aD^{b,\mu}\qquad\supset\qquad F\,{\rm Tr}[x_1 x_a x_b] \langle\dot\sigma_1\rangle C_{aef}\sigma^e\d_\mu\sigma^f \d^\mu\sigma^b\ ,
\ee
where the second index in $D^{1,0}$ is a spacetime index. Similar combinations of the derivatives on the $\sigma^{f,b}$ fields are also possible. At quartic level, we have operators of the form
\bea\label{4-pt-iso-non-Abelian}
&&F\,{\rm Tr}[x_1 x_a x_b]D^{1,0}D_\mu^aD^{b,\mu}\ \quad\supset\quad\ F\,{\rm Tr}[x_1 x_a x_b] \langle\dot\sigma_1\rangle\left( C_{cde}C_{fea}+ C_{cdi}C_{fia}\right)\sigma^f\sigma^c\d_\mu\sigma^d \d^\mu\sigma^b\ , \nonumber \\
&&{\rm Tr}[x_1 x_a x_b x_c]D^{1,\mu}D_\mu^aD^{b,\nu}D^{c}_\nu\ \quad\supset\quad\  {\rm Tr}[x_1 x_a x_b x_c] \langle\dot\sigma_1\rangle C_{aef} \sigma^f\d_\nu\sigma^e\d_\nu\sigma^b \d^\nu\sigma^c\ , 
\eea
and similar index contractions. Notice that the same situation we are describing here in the non-Abelian case can happen also in the Abelian case. In that case however the field redefinition in (\ref{eq:subs}) is trivial, and we therefore did not treat that case explicitly in the former subsection.

The second regime of interests for the possible background solution of $\langle\dot\sigma_1\rangle$ is when the speed is very large: $\langle\dot\sigma_1\rangle\sim F^2$. In this case one should consider the whole set of higher dimensional operators, for example a generic function $P(\langle\dot\sigma_1\rangle^2)$. In the absence of a large gradient in the potential energy as in our case, the only solution that does not redshifts away very quickly is that for which $P'=0$. If we were dealing with a single scalar field with a shift symmetry, this solution would represent the ghost condensate~\cite{Cheung:2007sv,ArkaniHamed:2003uy,ArkaniHamed:2003uz} and so we would be tempted to call this case as a `non-Abelian' ghost condensate. One should consider generic functions of ${\rm Tr}[D_\mu D^\mu],\, {\rm Tr}[D_\mu D^\mu D_\nu D^\nu],\ldots,$ suppressed by a common scale $F$ with generic order one coefficients. However, it seems quite generic, for example by looking already at the particular function $P(X)=F^4\left[\alpha(X/F^2-1)^2+\beta(X/F^2-1)^4+\ldots\right]$, with $X={\rm Tr}[D_\mu D^\mu]$ and $\alpha,\beta,\ldots$ numerical coefficients, that if $\dot\sigma_1$ gets a vev of order $F$, the additional Goldstone bosons receive a mass of order $F$ and can therefore be integrated out of the theory. In this case we are back to the standard single-field ghost-condensate which is included in the single field Lagrangian~\footnote{It is straightforward to generalize this to the case in which multiple degrees of freedom  contribute to $\pi$, just as did for the case where the vev of $\dot\sigma$ is small (see eq.~(\ref{eq:time-dependent-vev})).}. It is conceivable that by choosing a particular symmetry group and/or by adding additional symmetries, it might be possible to find cases in which the additional Goldstone bosons do not get a mass. The search for such a possibility lies beyond the scope of the present paper.

\subsection{Supersymmetric case}

In the former two subsections we have justified the presence of additional naturally-light  scalar fields by assuming that they were the Goldstone bosons associated to the spontaneous braking of some internal symmetry. However it is well known that supersymmetry offers another reason for having a naturally light scalar field. In our case the simple fact that the universe is inflating spontaneously breaks supersymmetry. In the limit in which we neglect the effects coming from supergravity, we can imagine that the sector of the additional fields $\sigma$ is decoupled from the inflaton ({\it} i.e. the $\pi$ field). In this way supersymmetry breaking is transmitted to the sector of the $\sigma$ fields at leading order only through the effect that the curvature has on the loops \cite{Senatore:2009cf}. Let us consider for simplicity a simple massless chiral multiplet consisting of a scalar doublet $\sigma$ and a Weyl fermion $\psi_\sigma$ with a superpotential and a Kahler potential of the form 
\be
W=\lambda \Sigma^3\ ,\qquad K=\Sigma^\dag\Sigma
\ee
where $\Sigma$ is the chiral multiplet. If we compute loop corrections to the mass of the $\sigma$ field, we have that in Minkoswky space the loop from the quartic self-interaction cancels with the loop with the fermion contribution so that the mass remains zero. In Minkowsky space supersymmetry is not broken. However when we put this same theory in an inflationary spacetime and we try to compute loop corrections to the mass we realize that because of the inflationary background the scalar and the fermion propagators are altered generically in different ways for modes with frequency $\omega\lesssim H$. On the other hand for frequencies $\omega\gg H$, the modes live effectively in Minkowsky space and so they are unchanged~\cite{Senatore:2009cf}. This means that loop corrections to the mass will cancel for $\omega\gg H$ but not for $\omega\lesssim H$. This means that the natural value of the mass of $\sigma$  is $
m\sim \delta m\sim \lambda\, H\ $. Even though supersymmetry is broken at energies of order $H$, the effective susy breaking scale for the $\sigma$ field is $\lambda\, H$, which for $\lambda\ll1$ is much smaller than $H$. This conclusion gets partially altered when we reinsert back the supergravity effects. Among the various other corrections, now the scalar potential gets multiplied by Exp$[K/\mpl^2]$. This induces a tree level mass for the $\sigma$ of order $H^2$. This is bad news for our multifield inflationary model, as we need a mass much smaller than Hubble. However, it is still true that loops are cut off at the scale of order $H$ and therefore we can imagine to perform the  mild tuning to make this tree level mass much smaller than Hubble.  In practice we require it to be of order $(n_s-1)^{1/2}H$ where $n_s$ is the tilt of power spectrum. Loops will not renormalize this mass at a relevant level. We conclude that imposing supersymmetry in the sector of the $\sigma$ field is a good tool to make the scalar fields naturally light if the sector is not directly coupled to the inflaton at the cost of assuming a tuning of order $n_s-1$ for the tree level mass.

The leading Lagrangian neglecting supergravity corrections reads:
\begin{eqnarray}\label{Spimulti-susy} \nonumber
S_{\sigma} &=& \int d^4 x   \sqrt{- g} \left[\d_\mu\sigma \d^\mu\sigma^*+m^2 \sigma\sigma^*+3m_s \lambda (\sigma+\sigma^*)\sigma\sigma^*+9\lambda^2 (\sigma\sigma^*)^2\right. \\ 
&&\left. +\frac{(\sigma+\sigma^*)}{M_{s,1}}\d_\mu\sigma\d^\mu\sigma^*+\frac{(\sigma+\sigma^*)^2}{M_{s,2}^2}\d_\mu\sigma\d^\mu\sigma^*\right]\ ,
\end{eqnarray}
where we have added a mass term to the superpotential and two of the leading  non renormalizable (or irrelevant) terms to the Kahler potential: $(\sigma+\sigma^*)^3/M_{s,1}+(\sigma+\sigma^*{}^2)^4/M_{s,2}^2$~\footnote{Of course here we should include all the interactions allowed by symmetries at the same order of $1/M$, and they exceeds the two ones we have written. However, as we will see, here we will be interested in just the dimensionality of the operators in order to extract the observational consequences of all of them, and the two we have considered will be enough.}.  In the above equation, $m_s$ is the supersymmetric mass, of order $H^2$ in order to cancel the $H^2$ mass from supergravity effects.

If we now consider supergravity corrections, their importance depends on the specific inflationary model. For example, in the particular case of small field inflation, the leading ones come from the vacuum energy of order $H^2\mpl^2$. It is straightforward to check that in this case the supergravity corrections to the interaction terms in the superpotential are negligible.  Instead they are relevant for the interactions associated to the Kahler potential. Symbolically, the leading ones are given by
\be\label{eq:sugra-kahler}
\frac{H^2}{M_{s,1}}\left(\sigma+\sigma^*\right)^3\ , \qquad \frac{H^2}{M_{s,2}^2}\left(\sigma+\sigma^*\right)^4\ .
\ee
By adding all the possible Kahler terms of the right dimensionality, it is possible to tune away these two terms. This would still make the theory technically natural, as loop corrections generate those terms at a much smaller value than their naive tree level value.

Summarizing:  in the supersymmetric case we can have the following interactions in addition to the ones in (\ref{eq:sugra-kahler}). In the superpotential we have a cubic and a quartic interaction of order:
\be\label{eq:susy-interactions}
\lambda m_s\, \sigma^2\sigma^\star+\; h.c. \ , \qquad \lambda^2\sigma^\star{}^2\sigma^2\ .
\ee
From the Kahler potential we have derivative interactions of the form:
\be
\frac{(\sigma+\sigma^\star)}{M_s} \d_\mu\sigma\d^\mu\sigma^\star+\; h.c.\ ,\qquad \frac{(\sigma+\sigma^\star)^2}{M_{s,2}^2}\d_\mu\sigma\d^\mu\sigma^* \ .
\ee
The susy breaking effects generate a cubic term with coefficient of order $\lambda^2 H$, which is smaller than the tree level one coming from the supersymmetric mass $m_s\sim H$. This one is of order $\lambda H$.
% for values of $m$ smaller than $\lambda H$ agrees at the level of order of magnitude with the one from (\ref{eq:susy-interactions}) taking $m\sim \lambda H$.
Notice that, contrary to the Abelian and non-Abelian cases, here interactions without derivative acting on any of the $\sigma$ fields are only marginally suppressed by the smallness of $\lambda H$. They are partially protected by supersymmetry. We will see that this will be important from the observational point of view. Finally, we stress that all of the above interactions are Lorentz invariant. This is so because we have avoided introducing a direct coupling of the $\sigma$ to the $\pi$ Goldstone boson as this would lead to a supersymmetry breaking in this sector at a much higher scale. 

Finally, in order to avoid the problem of having to cancel the order $H^2$ supergravity induced mass, it is possible to impose a shift-symmetry on the imaginary part of the chiral field $\sigma$ by choosing for example a Kahler of the form
\be
K=(\Sigma+\Sigma^\dag)^2\ ,
\ee 
and by forbidding a mass term in the superpotential. In this case the imaginary part of the scalar component ${\rm Im}(\sigma)$ does not get a mass from supergravity corrections. If the shift symmetry of ${\rm Im}(\sigma)$ is softly broken the leading interactions involving only ${\rm Im}(\sigma)$ come from the superpotential term. They are of the form $\lambda^2{\rm Im}(\sigma)^4$. A cubic term proportional to ${\rm Im}(\sigma)^3$ is not generated. Notice that supersymmetry plays an important role in this case. The non-renormalization of the superpotential allows to break the shift-symmetry in a more general way than in the non-supersymmetric case.
 
This concludes the construction of the Effective Lagrangian for the fluctuations around the time of horizon crossing. As anticipated, in order to be able to deduce the observational consequences of this effective Lagrangian, we need to see how the fields that enter in it are related to what we observe in the sky. We are now ready to do this.

\section{Relating to curvature perturbations\label{sec:curvature-conversion}}

When the quasi de Sitter phase ends, the universe eventually undergoes reheating and reaches thermal equilibrium~\footnote{In some models, like the curvaton \cite{Lyth:2002my}, this might occur after a phase of radiation or matter domination.}. We will assume for the moment that the final composition of the universe is independent of value of the $\sigma$ fields which 
means that the only difference between the various universes is the size of the scale factor. This is what the curvature perturbation $\zeta$ parametrizes. Isocurvature perturbations correspond to the case in which the final composition is spatially dependent. We will come back to this possibility later in this subsection.
 
In single field inflation, there is a very simple relationship between the Goldstone boson $\pi$ and 
the curvature perturbation $\zeta$. This is due to the fact that in single field inflation there is only one 
classical trajectory, and a $\pi$ fluctuation corresponds to a time delay on this unique trajectory. On 
the other hand, a $\zeta$ fluctuation corresponds to how much the universe expanded by the 
end of inflation, on surfaces where the physical clock (or the temperature of the universe) is uniform. 
Because of the uniqueness of the trajectory, it is therefore easy to relate $\pi$ and $\zeta$. At 
second order, when the mode is outside the horizon $\zeta$ is given by \cite{Cheung:2007sv}:
\begin{eqnarray}\label{eq:zeta_pi}
\zeta &=& -H \pi +H \pi \dot{\pi}
  + \frac{1}{2} \dot{H} \pi^2 
\end{eqnarray}
The linear piece of this relationship is easy to interpret. In the presence of a $\pi$ 
fluctuation that acts as a time-delay $\delta t\sim\pi$ ($\pi$ has in fact units of time), the universe 
undergoes an enhanced expansion by $\zeta\sim H\delta t\sim H\pi$.

\subsubsection*{$\bullet$ Abelian case}

\begin{figure}
\begin{center}
\includegraphics[width=10cm]{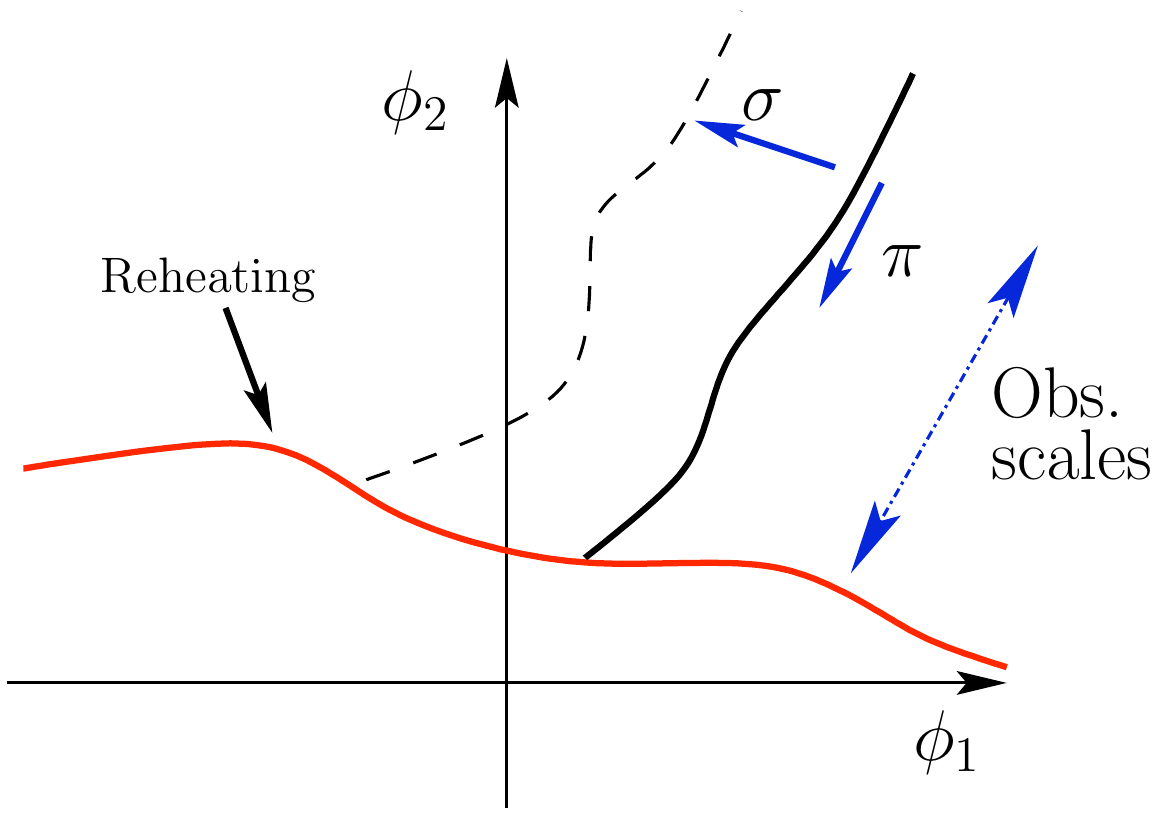}
\caption{\label{fig:multifield_potential} \small Representation of a typical multifield potential. Modes of interest for observation cross the horizon about sixty $e$-foldings before the end of inflation. Therefore, effects coming from the evolution of the filds after horizon crossing can be treated locally in real space. Our effective theory is more general than this example, as it does not assume that the inflaton is a scalar field. This example is however interesting in helping in visualizing the different scales in the problem.}
\end{center}
\end{figure}

In multifield inflation the situation is much more complicated. The novel feature of multifield inflation 
is that there are multiple classical trajectories, associated to the various fluctuations of the $
\sigma_I$ fields. The problem is that it is not easy to relate how a fluctuation in a 
$\sigma_I$ direction occurring about sixty $e$-foldings before the end of inflation translates into how much the universe will have expanded by the end of inflation. At linear level in single 
field inflation this was simply given by  $H\pi$ because the only possible fluctuation mode
was moving along the same (unperturbed) trajectory. In general, the relationship between $
\zeta$ and a $\sigma_I$ fluctuation will be  a complicated function of the entire trajectory, from the 
time a mode crosses the horizon to the reheating time (see Fig.~\ref{fig:multifield_potential}). This however will not prevent us from developing an effective treatment.

The question we should answer is how much more the universe expands (or equivalently inflation lasts) if we 
change trajectory by an amount $\sigma_I$.
Since we are interested in modes that cross the horizon about sixty $e$-foldings before the end of 
inflation, the $k$-mode associated with that fluctuation is outside the horizon for all of those sixty $e$-folding minus a number of order one that corresponds to the time during which the mode 
crosses the horizon. Let us first concentrate on those first order one $e$-foldings and convince 
ourselves that no large effect is generated during that period. This is so because the $\sigma_I$ fields have an approximate shift symmetry during inflation. This suppresses how much 
the potential can change along the $\sigma_I$ direction before the end of inflation. 

The only possibilities for a large effect can come from a cumulative effect during all the sixty $e$-foldings or due to a large breaking of the shift symmetry at the time of reheating. In fact we expect that at this time all the shift symmetries, both of $\pi$ and of the $\sigma_I$ fields can be broken. Thus we concentrate on the remaining sixty $e$-foldings during which the modes lie
outside of the horizon. In the presence of a $\sigma_I$ 
fluctuation larger than the horizon the universe in each region of space evolves to a very good approximation as if the $
\sigma_I$ field had no gradients and it was instead uniform in that region.  This means that at linear level no matter how complicated the relationship between $\sigma_I$ and $\zeta$ is it must be a relationship local in real space 
and therefore has to have the form
\be\label{eq:linear_relationship}
\zeta(x)\simeq\left.\frac{\d\zeta}{\d\sigma_I}\right|_0\sigma_I(x)\ ,
\ee
where $\left.\d\zeta/\d\sigma_I\right|_0$ is a parameters of dimension of mass$^{-1}$ 
which represents the Taylor expansion of the relationship between $\zeta$ and $\sigma_I$ around the unperturbed trajectory ($\sigma_I=0$). Notice that we have neglected any dependence on $\dot\sigma_I$ at horizon crossing. Note that $\dot\sigma_I$ might not equal to zero at that time because of the softly broken shift symmetry. However once a mode $\sigma_I$ is out of the horizon it is on an attractor solution: $\dot\sigma_I$ is just a function of $\sigma_I$. We are therefore not loosing any information by neglecting terms in $\dot\sigma_I$ in (\ref{eq:linear_relationship}).

This can be extended also to non-linear level. Since from observations we know that the 
non-Gaussianities cannot be very large we can compute them perturbatively as corrections to the 
Gaussian evolution. We concentrate on the three and the four-point function at tree-level. In the case of the three 
point-function  two of the fluctuations correspond to 
observable modes and are computed using the Gaussian evolution while the third one is evaluated on 
the first correction to the Gaussian evolution, a term of second order in the Gaussian 
fluctuations. When one takes the expectation value each of the two momenta of the second 
order corrections has to be equal to one of the two momenta of the perturbations which are taken to 
be Gaussian. Since their momenta are out of the horizon this 
means that all the momenta of interest, even the ones that enter in the second order fluctuations
are momenta that cross the horizon around sixty $e$-foldings before the end of inflation. Similar conclusions apply for the four point function.

Putting together all of the above reasoning we conclude that even at non-linear level, the 
relationship between a $\sigma_I$ fluctuation and the curvature perturbation $\zeta$  will be to a 
good approximation a non-linear function that is {\it local in real space}  and that we can Taylor 
expand
\be
\zeta(x)=\left.\frac{\d\zeta}{\d\sigma_I}\right|_0\sigma_I(x)+\frac{1}{2!}\left.\frac{\d^2\zeta}{\d
\sigma_I\sigma_J}\right|_0 \sigma_I\sigma_J+\frac{1}{3!}\left.\frac{\d^3\zeta}{\d\sigma_I\d\sigma_J\d\sigma_K}\right|_0 
\sigma_I(x)\sigma_J(x)\sigma_K(x)+...\ .
\ee 
Here as in (\ref{eq:linear_relationship}) $\left.\frac{\d\zeta(x)}{\d\sigma^n}\right|_0$ are the Taylor expansion parameters with dimension of mass$^{-n}$, $n$ being the number of $\sigma_I$ fluctuations, and should be evaluated on the unperturbed trajectory ($\sigma_I=0$).

By including also the contribution from $\pi$, we obtain the generic expression
\bea\label{eq:zeta_relation}
\zeta(x)&=&\left.\frac{\d\zeta}{\d\pi}\right|_0\pi(x)+\left.\frac{\d\zeta}{\d\sigma_I}\right|
_0\sigma_I(x)+\\ \nonumber 
&&\frac{1}{2!}\left.\frac{\d^2\zeta}{\d\pi^2}\right|_0 \pi(x)^2+\frac{1}{2!}\left.\frac{\d^2\zeta}{\d
\pi\d\sigma_I}\right|_0 \pi(x)\sigma_I(x)+\frac{1}{2!}\left.\frac{\d^2\zeta}{\d\sigma_I\d
\sigma_J}\right|_0 \sigma_I(x)\sigma_J(x)+\\ \nonumber 
&&\frac{1}{3!}\left.\frac{\d^3\zeta}{\d\sigma_I\d\sigma_J\d\sigma_K}\right|_0 \sigma_I(x)
\sigma_J(x)\sigma_K(x)+...\ .
\eea
The terms in the first line represent the linear relationship between $\zeta$ and $\pi$ and $\sigma_I
$, and they are important to give the overall normalization of the power spectrum. The terms in the 
second line are instead important for the three-point function, while the terms in the fourth line (of 
which we included only the term that we find can be more important observationally) are important in the four-point function.

Notice that the terms in the above Taylor expansion that contain only $\pi$ can be read off from 
eq.~(\ref{eq:zeta_pi}): their overall mass scale is given by Hubble. They are fixed by symmetries and are not model dependent. The 
coefficients proportional to $\sigma_I$ are instead model dependent, and their magnitude represents how much a $
\sigma_I$ fluctuation affects the overall duration of inflation. When these coefficients are large enough so that the $\sigma_I$ fluctuations are as or more important for $\zeta$ than the $\pi$ fluctuations, then we are in a truly multifield inflationary model.  

In the second line the term mixed in $\pi$ and $\sigma_I$ deserves a special discussion. It represents how much 
a $\sigma_I$ fluctuation affects the subsequent $\pi$ fluctuation (or viceversa). Clearly, because of 
the approximate shift symmetry around horizon crossing of the $\sigma_I$ fields, this effect is 
suppressed by the breaking of the shift symmetry either of the $\pi$ field or of the $\sigma_I$ field, 
and can therefore be neglected.

In eq.~(\ref{eq:zeta_relation}), the field fluctuations should be thought of as evaluated at the time 
when the modes of interest have just crossed the horizon. Neither the $\pi$ fluctuations nor the $
\sigma_I$'s are constant outside of the horizon. However one can show that for $\sigma_I=0$, the combination in  (\ref{eq:zeta_pi}) is constant at non linear level when gradients are negligible \cite{Cheung:2007sv,Maldacena:2002vr}. The situation is 
a bit more complicated for the $\sigma_I$ fluctuations. They are not constant, and they get some non-linear corrections from the evolution outside of the horizon. We can deal with this by treating the fluctuations of $\sigma_I$ 
after horizon crossing as initial conditions that will determine how much the universe will expand 
given such an initial condition. In addition  because of the non-linear evolution 
outside of the horizon, the initial distribution of the $\sigma_I$ fields will receive a non-Guassian 
correction (i.e. even an initially Gaussian distribution 
becomes non-Gaussian). However  as noted in \cite{Zaldarriaga:2003my} since this evolution will occur outside of the horizon and can be treated  perturbatively the final form of the correction  will be local in real 
space:
\be\label{eq:correction}
\sigma^{\rm corr.}_I(x)=\sigma_I(x)+f_{{\rm NL,\; corr.};\; I}^{{\rm loc.};\, JK}\left( \sigma_J(x)\sigma_K(x)-\langle\sigma_J(x)\sigma_K(x)\rangle + \cdots
\right)\ .
\ee
Naively, this correction is expected to be small because even though it comes from the build up 
during the sixty $e$-foldings of time of evolution outside of the horizon, the non-derivative self 
couplings of the $\sigma_I$ fields are suppressed by the parameter controlling the breaking of the shift symmetry for 
most of the duration of inflation and therefore the final effect is suppressed. However in the last 
part of the evolution the shift symmetry of the $\sigma_I$  and $\pi$ fields may 
be violently broken which could make the effect more relevant. 

Because of the local structure of the induced non-Gaussian correction, implementing 
eq.~(\ref{eq:correction})  amounts just to a redefinition of the coefficients  that involve derivatives 
with respect to $\sigma_I$. Since we do not know them the result in (\ref{eq:zeta_relation}) is unchanged. Notice that this redefinition does not effect the overall suppression of the term in $\pi\sigma_I$ in (\ref{eq:zeta_relation}).

There is one final subtlety to notice about the former treatment. Once we look at the $\pi$ and $\sigma_I$ fluctuations in Fourier space, the various modes will have crossed the horizon at slightly different times (see Fig.~\ref{fig:multifield_potential}). This means that we have to allow the various Taylor coefficients as well as the amplitude of the modes to have a slight dependence on the time of horizon crossing. This will allow us to recover information about deviations from scale invariance. This is very similar to what happens for single field inflation \cite{Cheung:2007st}, and the situation is treated analogously. The only difference is that while  in single field inflation the tilt was just a function of the parameters of the unperturbed trajectory ($H,\;\dot H,$ and $\ddot H$) and of the speed of sound of the $\pi$ fluctuations here the tilt is allowed to be a more general function.

Our relationship (\ref{eq:zeta_relation}), even after the redefinition (\ref{eq:correction}),  is quite similar to what is usually referred to as the $\delta N$ formalism \cite{Sasaki:1995aw}. However, the redefinition in (\ref{eq:correction}) shows that our parameterization, at least for the quadratic terms, is different: for example, the coefficient of the $\sigma_I\sigma_J$ term is not equal to the second derivative of the overall expansion with respect to the $\sigma_I$ fields. However this difference only enters at non-linear level. Furthermore we stress again that this approach neglects the contribution to the expansion coming from the order one $e$-foldings during which the modes cross the horizon, and that we argue is negligible at leading order. The literature making use of the $\delta N$ formalism is very vast. For example it has been applied in many models of multifield inflation to calculate the cumulative effect on the two-point function of the different inflationary trajectory (see for example \cite{Sasaki:1995aw}). After a generalization to treat non-linear effects the formalism was applied to higher order statistics (see for example \cite{Lyth:2005fi,Bernardeau:2002jy,Bernardeau:2010jp}). There are also models where the effects on the overall expansion originates mainly from variation in the reheating process \cite{Zaldarriaga:2003my} or from a new late phase of matter domination as in the curvaton model \cite{Lyth:2002my}.

\subsubsection*{$\bullet$ Non-Abelian and Supersymmetric cases}

The relationship between $\zeta$ and the Goldstone's boson $\sigma_I$'s in the non-Abelian case proceed as in the Abelian case except that in the non-Abelian case the Goldstone bosons transform under some representation of the unbroken group $H$. Let us consider the linear term in $\sigma_I$ from eq.~(\ref{eq:zeta_relation}). First of all in order for there to be a linear term the group $H$ needs to have been broken by the time of rehating. This is  a reasonable assumption, as the full group $G$ might have been softly broken already at the beginning of inflation. Let us consider the case in which the Goldstone bosons transform in an irreducible representation of the unbroken group $H$ as we did in  (\ref{Spimulti-non-Abelian}). In this case, we could perform the following transformation. Given $h\in H$, we perform
\be\label{eq:redefinition}
\left.\frac{\d\zeta}{\d\sigma_I}\right|_0\sigma_I(x)=\left.\frac{\d\zeta}{\d\sigma_K}\right|_0 {\mathscr {D}}(h)^{-1}_{KI} {\mathscr {D}}(h)_{IJ} \sigma_J(x)=\widetilde{\left.\frac{\d\zeta}{\d\sigma_1}\right|_0}\sigma_1'\ ,
\ee
 where we have chosen the particular transformation such that 
 \be
 \left.\frac{\d\zeta}{\d\sigma_K}\right|_0 {\mathscr {D}}(h)^{-1}_{KI} =\widetilde{\left.\frac{\d\zeta}{\d\sigma_1}\right|_0}\delta_{I1}\ , 
\ee 
and we have redefined
 \be\label{eq:conversion_non_Abelian}
\sigma_I'= {\mathscr {D}}(h)_{IJ} \sigma_J(x)\ .
 \ee
Independently of the fact that the group $H$ might be strongly broken by the end of inflation the  transformation $\sigma\rightarrow \sigma'$ leaves invariant the Lagrangian in (\ref{Spimulti-non-Abelian}), with the exception of the soft-breaking that might be present at the time when the modes cross the horizon which we keep general in any case. This means that at the level of the linear dependence of $\zeta$ on the  $\sigma_I$ fields  we can take $\zeta$ to depend on a single representative mode for each irreducible representation of the group $H$. The quadratic and higher order terms in the relationship (\ref{eq:zeta_relation}) are unaffected by the transformation in (\ref{eq:redefinition}) and are therefore generic. This will be particularly relevant for the generation of non-Gaussianities.

The case where the additional light fields are approximately supersymmetric proceeds in exactly the same way with the only different that now $\zeta$ has to depend both on the real and the imaginary part of the $\sigma$ fields. Notice that the fact that the $\sigma$ fields have an influence on the curvature perturbations means that there is a coupling with the inflaton that breaks supersymmetry directly at some lower energy also in the $\sigma$ sector.

In summary: curvature perturbations are given by (\ref{eq:zeta_relation}) where the fields are 
evaluated at a time just after the observable modes have crossed the horizon during inflation. While the coefficients of the terms in $\pi$ are at most of order $H$, the coefficients of the terms involving only powers of the $\sigma_I$'s are instead free. These coefficients effectively parametrize our ignorance about the full evolution of 
the fields during the sixty $e$-foldings of inflation and the epoch of reheating. As we let them vary, we interpolate from effectively single field inflation to a truly multifield inflationary model. Contrary to the case of single field inflation, this treatment cannot be made indefinitely exact. The error is associated with the fact that eq.~(\ref{eq:zeta_relation}) does not properly account for those order one $e$-foldings during which the $\sigma_I$ fields are crossing the horizon. There is an irreducible error associated with this effect that we expect to be of order $1/N_e$, where $N_e\sim60$ is the number of $e$-foldings that inflation lasts since the scales we observe left the horizon. In reality we expect the error to be even smaller because of  the $\sigma_I$ fields are approximately massless and that should make only the last $e$-foldings of evolution relevant for (\ref{eq:zeta_relation}). Since the term in $\pi \sigma_I$ is expected to be small, and since as we will explain later we expect the term in $\sigma_I\sigma_J\sigma_K$ also to be  irrelevant, we are left with a number of unknown coefficients equal to
\bea
N+\frac{N(N+1)}{2}=\frac{N(N+3)}{2}\ , &&\qquad {\rm\ Abelian\ case}\ , \\ \nonumber
\sum_r (1+\frac{N_r(N_r+1)}{2})=\sum_r\frac{1}{2}(N_r^2+N_r+2)\ , &&\qquad {\rm\ non-Abelian\ case}\ ,\ \\ \nonumber 
2N+N(N+1)=N(N+3)\ , &&\qquad {\rm\ supersymmetric\ case}\ , 
\eea
where for the non-Abelian case the sum of $r$ runs over the number of reducible representations of $H$ each one with dimension $N_r$, and for the supersymmetric case we have assumed that there are $N$ chiral multiplets unrelated by any symmetry.

It is worth to explicitly point out that in order for the additional fields to play a relevant role for the cosmological perturbations, their internal symmetry (either the global one or the supersymmetric one) needs to be explicitly broken. As we have seen, there are small breaking effects already in the Lagrangian that describes the fluctuations around horizon crossing, but there can be additional effects that become relevant around the time of reheating or even later. The actual way the symmetry is broken at this stage only has one more relevant observable consequence, determining explicitly the coefficients of the terms in (\ref{eq:zeta_relation}). Determining their size would require a complete description of the model both during inflation and during reheating and it probably faces, in each specific model, additional constraints coming from the explicit breaking of the internal symmetries in each model. Such a description seems to us difficult to achieve in more specific and yet still general terms than the ones we have provided here. It lies beyond the scope of the present paper. 

\subsection{Isocurvature fluctuations\label{sec:isocurvaute}}

It is possible that the composition of the plasma after reheating depends on the particular inflaton trajectory, i.e. on the value of the $\sigma_I$'s. Fluctuations of the composition of the plasma on surfaces of constant value of the physical clock are known as isocurvature fluctuations. 

Given the discussion of the former subsection, it is  straightforward to offer an effective parameterization for the isocurvature fluctuations. The composition of the plasma can be parameterized by the various ratios of the densities\footnote{ There are other isocurvature perturbations related to velocities and higher order moments of the distribution function of the various components of the primordial plasma \cite{Bucher:1999re}. They can be included in our treatment in a straightforward way but for clarity in the main text we focus explicitly on perturbations in the plasma composition.}. Let us denote by $n_i$ the number density of some species $i$
\be\label{eq:composition}
\delta\left(\frac{n_{i}}{n_\gamma}\right)\ .
\ee
where the index $i$ runs over baryonic, neutrino and all the possibile other species.
%as well as isocurvature velocity or even isocurvature fluctuations associated to higher moments of the distribution function of these components.  
The number density of photons is denoted by ${n_\gamma}$. The composition (\ref{eq:composition}) is expected to be a function of some coupling constant and decay channels that could depend on the inflaton trajectory. Since  the interesting modes are all well outside the horizon these functions will be  local in real space and thus  can be Taylor expanded around $\sigma_I=0$ as we did for (\ref{eq:zeta_relation}). The additional information we need is that the perturbations to the composition of the plasma have to go to zero in the limit in which the $\sigma_I$'s do: a simple $\pi$ fluctuation is just a time delay of the same unperturbed solution and therefore reheating in the presence of only a $\pi$ fluctuation happens in the same way as in the unperturbed case. We obtain:
\bea\label{eq:isocurvature}\nonumber
&&\delta\left(n_{i}/n_\gamma\right)(x)=\left.\frac{\d \left(n_{i}/n_\gamma \right)}{\d\sigma_I}\right|_0\sigma_I(x)+\frac{1}{2}\left.\frac{\d \left(n_{i}/n_\gamma \right)}{\d\sigma_I\d\sigma_J}\right|_0\sigma_I(x)\sigma_J(x)+\left.\frac{\d \left(n_{i}/n_\gamma \right)}{\d\pi\d\sigma_I}\right|_0\pi(x)\sigma_I(x)\ .\\
\eea
The same arguments we used in the former section imply that the coefficients of the terms in $\pi\sigma_I$ are suppressed by the breaking of the shift symmetries and is therefore small.

In the non-Abelian case we can perform a rotation such that $\zeta$ depends only on one particular $\sigma_I$. However this need not be the same one as the one for the adiabatic fluctuations. This will be important for the non-Gaussianities.

We are left with a number of model dependent isocurvature parameters equal to 
\bea
n\times\left(N+\frac{N(N+1)}{2}\right)=\frac{n}{2}N(N+3)\ , &&\qquad {\rm\ Abelian\ case}\ , \\
\sum_r n\times (1+\frac{N_r(N_r+1)}{2})=\sum_r\frac{n}{2}(N_r^2+N_r+2)\ , &&\qquad {\rm\ non-Abelian\ case}\ , \\ \nonumber
n\times\left(2N+N(N+1)\right)\,N(N+3)\ , &&\qquad {\rm\ supersymmetric\ case}\ , \\
\eea
where $n$ is the number of isocurvature fluctuations we wish to consider.
These are the parameters constrained by experiments.

\section{Signatures\label{sec:signatures}}

\subsection{Detectable four-point function from multi-field inflation\label{sec:four-point-multifield}}

We now begin the analysis of the signatures of multifield inflation in light of the effective field theory we have developed. We start with the four-point function.
In Ref.~\cite{Senatore:2010jy} it was shown that in single field inflation it is possible  to have a large and detectable four-point function without having at the same time a large and detectable three-point function. This was due to a parity symmetry $\pi\rightarrow-\pi$ that turned out to be an accidental approximate symmetry once we introduced a large coefficient for the operator $(\delta g^{00})^4$. This had the consequence of allowing a large coefficient for the operator $\dot\pi^4$ without at the same time a large coefficient for the $(\d\pi)^3$ operators. One can therefore have a large four-point function without a large three point function. In the case of a linear dispersion relation $\omega\propto k$ only one shape for the four-point function is possible, while in the case of a non-linear dispersion relation $\omega\propto k^2$ many possible shapes are allowed~\cite{Senatore:2010jy} (see Table~\ref{tab:singlefield}).

In this section we are going to show that in multifield inflation it is possible to have a larger class of four-point functions that can be observationally relevant.  In fact contrary to the case of single field inflation it is possible to impose symmetries that shut down all the cubic self-interactions and also those quartic in $\pi$ without shutting down the $\sigma$'s quartic self-interactions~\footnote{In this section we will neglect to specify to indexes of the $\sigma$ fields expect for the cases where it is required for clarity.}. This makes the four-point function the leading source of non-Gaussianity in these models. In the case there is a large cubic self-interaction, there is a non-vanishing four-point function. However, the signal in the three-point function is much larger~\cite{Senatore:2010jy}.

There are at least two symmetries that shut down the interactions that involve a $\pi$ field. One is of course to impose a $\pi$-parity $\pi\rightarrow-\pi$ to be a good approximate symmetry. This shuts down all the cubic interactions, including the ones in $\pi\sigma^2$. Notice that the operator $\dot\pi^4$ that was studied in Ref.~\cite{Senatore:2010jy} becomes more and more suppressed in the limit in which we make the $\pi\rightarrow-\pi$ symmetry more exact, as it is originated by the operator $(\delta g^{00})^4$ which does not respect this symmetry. However the parity symmetry for $\pi$  can never be exact. There is a minimum violation originating from the time dependence of the Hubble constant. Upon reinsertion of $\pi$ this induces terms with all powers in $\pi$. This implies  that the breaking is suppressed by slow roll parameters which in turn means that the resulting cubic operators induce a negligible three-point function. For the $U(1)^N$ case, by adding an approximate $\sigma_I\rightarrow-\sigma_I$ parity symmetry (that for example can become strongly broken only at the time of reheating), all the cubic interactions are suppressed. In the non-Abelian case the $\sigma_I^3$ are automatically suppressed. 

Another way to shut down all the cubic interactions is by imposing that Lorentz invariance be a good approximate symmetry for the fluctuations. In fact at least in single field inflation with an approximate continuous shift symmetry for the Goldstone boson, the only way one can have a large three-point function is for the $\pi$ Lagrangian to be very non-Lorentz invariant. For example, we saw that large non-Gaussianities were obtained in the limit of a very small speed of sound or in the case the $\dot\pi^3$ term was made large. In multifield inflation we can instead require that the scale suppressing the Lorentz-violating operators be much higher than the one suppressing the Lorentz-invariant ones. This effectively makes Lorentz symmetry an approximate symmetry for the flucutations. This symmetry removes all the $\pi$ self-interactions, and it also removes all the remaining cubic interactions in $\sigma^3$, in $\pi^2\sigma$ and $\pi\sigma^2$.

Neither of the symmetries just mentioned shuts down interactions of the form $(\d\sigma)^4$ which can lead to large and detectable four-point functions. Before proceeding to study them we point out that  in order for the non-Gaussianity induced by the $\sigma$'s to be relevant, the $\sigma$'s need to play a relevant role in determining an observable quantity. For example, this is the case if the influence of $\pi$ on $\zeta$ is much smaller than the one from the $\sigma$'s: 
\be
\label{eq:dominance_condition}
\frac{H^2}{\mpl^2\,\epsilon\, c_s^{(\zeta)}}\ll \left(\left.\frac{\d\zeta}{\d\sigma}\right|_0\right)^2 \frac{H^2}{c_s^{(\sigma)}}\ ,
\ee
where $c_s^{(\zeta, \sigma)}$ is the speed of sound of $\zeta$ and $\sigma$ fluctuations respectively. Here for simplicity we have considered the case in which all the fields have dispersion relations of the form $\omega\sim c_s k$, a generalization to the case where some of the fields have a dispersion relation of the form $\omega\sim k^2/M$ is straightforward (see~\cite{Cheung:2007sv,Senatore:2009gt} for the size of the fluctuations in this case).  
%Here and in the rest of this section for simplicity we suppress the internal indexes $I$, which is straightforward to reintroduce.

There is at this point a very  straightforward way in which the $\sigma$ fields could induce a large $\zeta$ four-point function, while inducing a negligible three-point function: this is by imagining that the coefficient $\d^3\zeta/\d\sigma^3$ in (\ref{eq:zeta_relation}) is much more important than the coefficient $\d^2\zeta/\d\sigma^2$. The term in $\d^3\zeta/\d\sigma^3$ induces a four-point function that we call local and  we paremetrized by $g_{NL}^{\rm loc.}$. The name local comes from the fact that the four-point function in this case is induced by a local-in-space non-linear relationship between $\zeta$ and $\sigma_I$. Its size is of the order of
\be
g_{NL}^{\rm loc.}\zeta^6\sim \langle\zeta^4\rangle\sim \left.\frac{\d^3\zeta}{\d\sigma^3}\right|_0\frac{1}{\left(\left.\frac{\d\zeta}{\d\sigma}\right|_0\right)^3} \zeta^6\qquad \Rightarrow \qquad 
g_{NL}^{\rm loc.}\sim \left.\frac{\d^3\zeta}{\d\sigma^3}\right|_0\frac{1}{\left(\left.\frac{\d\zeta}{\d\sigma}\right|_0\right)^3} \ .
\ee
Analogously, $\d^2\zeta/\d\sigma^2$ induces a local three point function parametrized by $f_{NL}^{\rm loc.}$ of the form
\be\label{eq:fnlloc}
f_{NL}^{\rm loc.}\zeta^4\sim \langle\zeta^3\rangle\sim \left.\frac{\d^2\zeta}{\d\sigma^2}\right|_0\frac{1}{\left(\left.\frac{\d\zeta}{\d\sigma}\right|_0\right)^2} \zeta^4\qquad \Rightarrow \qquad f_{NL}^{\rm loc.}\sim \left.\frac{\d^2\zeta}{\d\sigma^2}\right|_0\frac{1}{\left(\left.\frac{\d\zeta}{\d\sigma}\right|_0\right)^2} \ .
\ee
In order for the four-point function induced in this way to be observationally larger than the three point function, we need that
\be\label{eq:condition}
g_{NL}\zeta\gtrsim f_{NL}\qquad\Rightarrow \qquad \left.\frac{\d^3\zeta}{\d\sigma^3}\right|_0\gtrsim \frac{1}{\zeta} \left.\frac{\d^2\zeta}{\d\sigma^2}\right|_0\left.\frac{\d\zeta}{\d\sigma}\right|_0\ .
\ee
Now, the derivatives $\d^n\zeta/\d\sigma^n$ represent how much the expansion of the universe depends on the various trajectories, parametrized by the values of $\sigma$. Their presence is associated with the breaking of the shift symmetry along the $\sigma$ directions that can happen around the time of reheating or as a cumulative effect during the approximately sixty $e$-foldings of inflation. The important point here is that once the linear term is generated with a scale of order
\be
\left.\frac{\d\zeta}{\d\sigma}\right|_0\sim \frac{1}{M_c}\ ,
\ee
where $M_c$ is a mass scale associated with the conversion mechanism,  we expect all the higher derivatives to scale as 
\be
\left.\frac{\d^n\zeta}{\d\sigma^n}\right|_0\sim\frac{1}{\alpha}\frac{1}{M_c}\left.\frac{\d^{n-1}\zeta}{\d\sigma^{n-1}}\right|_0
\ee
where $\alpha$ is a numerical coefficient representing the efficiency of the conversion mechanism. The dependence on $\alpha$ can arise when $\zeta$ is a local function of $\sigma$ given by the following form
\be
\zeta=\alpha\left(c_1\left(\frac{\sigma}{M_b}\right)+c_2\left(\frac{\sigma}{M_b}\right)^2+c_3\left(\frac{\sigma}{M_b}\right)^3+\ldots\right)\ ,
\ee
where $c_{1,2,3}$ are order one coefficients. $M_b$ represents a scale associated to the breaking of the symmetry protecting the lightness of the $\sigma$ fields at the scale of reheating, and we expect all powers of $\sigma$ to be suppressed by the same scale $M_b$. The overall conversion of the $\sigma$ fluctuations into $\zeta$ can be controlled by an additional parameter $\alpha$, that we can imagine as representing the inefficiency of the conversion mechanism, for $\alpha\lesssim 1$, or the efficiency in the opposite case (see~\cite{Zaldarriaga:2003my} for a first realization of this possibility).
Given the current constraint on the local $f_{NL}$, we have that $\alpha$ is larger than order $10^{-1}$-$10^{- 2}$. This means that for the relationship in (\ref{eq:condition}) to be satisfied, a tuning at least of about $\alpha/\zeta\sim10^{3}$ is necessary. That is to say we do not expect the four-point function generated this way at the time of reheating to be more relevant than the three-point one. This leads us to consider other ways in which a four-point function can be made larger than a three-point function. Let us start with the Abelian case.

\subsubsection*{$\bullet$ Abelian case}

One way to make the quartic interactions the leading ones is by imposing that Lorentz invariance be an approximate symmetry of the Lagrangian for the fluctuations, in the sense that the scale suppressing the operators that violate Lorentz invariance is made parametrically higher that the scale suppressing the operators that respect it. This makes irrelevant all the $\pi$ self-interactions, the mixing term in $\pi\sigma$, all the cubic interactions involving any power of the $\sigma$'s (with one derivative acting on each one) and the $\pi$'s, and of all the quartic interactions in (\ref{Spimulti4}) but the unique combination  that is Lorentz invariant:
\be\label{eq:quartic-Lorenzt}
\tilde{\tilde M}^4\d_\mu\sigma\d^\mu\sigma\d_\nu\sigma\d^\nu\sigma\ ,
\ee
where for simplicity we are suppressing all the internal indexes. In order for the four-point function produced by this operator  to be relevant, it is not enough for the $\sigma_I$'s to be non-Gaussian, they also need to play a relevant role in determining an observable quantity. For example, we could ask that they dominate $\zeta$, which requires eq.~(\ref{eq:dominance_condition}) to be satisfied (in this case with $c_s=1$).  The interaction in (\ref{eq:quartic-Lorenzt}) induces a $g_{NL}$ of the form
\be
g_{NL}\zeta^2\sim\left.\frac{{{\cal L}}_4}{{\cal L}_2}\right|_{E\sim H}\sim\frac{H^2\sigma^2}{\tilde{\tilde M}^4}\sim
\frac{H^2}{\tilde{\tilde M}^4}\frac{1}{\left.\frac{\d\zeta}{\d\sigma}\right|_0^2}\zeta^2\qquad\Rightarrow
\qquad g_{NL}\sim 10^{10} \frac{H^4}{\tilde{\tilde M}^4}\ ,
\ee
where we have used that 
\be\label{eq:zeta_norm1}
\zeta\sim 10^{-5}\sim \left.\frac{\d\zeta}{\d\sigma_I}\right|_0 H\ .
\ee
The resulting value of $g_{NL}$ can be very large. The cutoff of the theory is of order $\tilde{\tilde M}$, and therefore \taunl is just bounded to be smaller than $10^{10}$. In particular $\left.{{\cal L}}_4/{\cal L}_2\right|_{E\sim H}$ can be larger than $10^{-4}$ and thus detectable. In determining the thresholds for detection, we will estimate the size of the constraints as~\cite{Creminelli:2006gc}
\be
\Delta [f_{NL} \zeta] \sim \frac{1}{N_{\rm pix}^{1/2}} \ ,\qquad \Delta[g_{NL} \zeta^2]\sim \frac{1}{N_{\rm pix}^{1/2}} \ ,
\ee
where $N_{\rm pix}$ is the number of signal dominated modes. Given the current experimental bounds we will impose the thresholds for detection by imposing the above ratios to be tentatively larger than $\sim 10^{-4}$. Given the fact that we will neglect all the numerical factors, we will interpret these limits loosely. We leave a detailed study of the quantitative predictions and of the detectability thresholds, along the lines of what done in~\cite{Senatore:2009gt} for single-clock inflation, to future work.

At the level of the cubic operators, the requirement of having quasi-Lorentz invariance has got rid of the cubic operators with one derivative acting on each fluctuation. This symmetry does not forbid operators with one more derivative, such as $\Box\sigma(\d_\mu\sigma)^2/M^3$. If these operators were to contribute to the three-point function, their signal would be larger than the one on the four-point function. However, it is quite straightforward to realize that operators involving $\Box\sigma$ do not contribute to the three-point function at tree-level, as they are proportional to the linear equations of motion, by which $\Box\sigma=0$. Operators of the form $\nabla_\mu\d_\nu\sigma\d^\nu\sigma\d^\mu\sigma$ can be put in the same form by an integration by parts.  In an expansion in derivatives, the next cubic operators allowed by this symmetry contain six derivatives and are therefore subleading.

We would like to stress that imposing the Lagrangian to be quasi Lorentz invariant has reduced the large number of interaction operators  to just one. There is therefore only one shape for the four-point function in this limit, which is a smoking gun for multifield inflation and strong evidence for approximate Lorentz symmetry of the Lagrangian of the inflationary perturbations~\footnote{For the UV oriented reader, it should be particularly easy to UV complete models with these kinds of symmetries, as the Lagrangian for the $\pi$ sector can be taken as the one of standard slow roll inflation. In general it is interesting to try to find particular UV completions of the models described by the Effective Field Theory of Inflation. But this is not necesary to have a consistent and predictive theory.}. 

Notice that the quartic operator in (\ref{eq:potential-even}), schematically of the form $\sigma^4$, that is present in the case of explicit symmetry breaking, is not eliminated by imposing an approximate Lorentz invariant, and his contribution can be important. Comparing with the leading term we are considering we obtain:
\bea\label{eq:est1}
&&\left.\frac{{\cal L}^{(4)}_{\rm soft-breaking}}{{\cal L}^{(4)}}\right|_{E\sim H}N_e\sim \frac{\mu^4\sigma^4/{\Lambda_{U,S}}^4}{(\d_\mu\sigma)^4/M^4}N_e\sim N_e\frac{\mu^4}{ H^4}\frac{M^4}{\Lambda_{U,S}^4}\ . %\\ \nonumber
%&&\frac{L^{(4)}_{\rm soft-breaking}}{L^{(4)}}\sim \frac{\mu^4(\d_\mu\sigma)^2\dot\sigma\sigma/M^7}{(\d_\mu\sigma)^4/M^4}\sim \frac{\mu^4}{M^3 H}\ll 1, 
\eea
Here we have normalized the $\sigma$ terms in the symmetry breaking potential with a scale $\Lambda_{U,S}$ that represents the unitarity bound coming from the interactions associated to the symmetry breaking potential. The scales $\Lambda_{U,S}$ and $M$ are independent ($M$ is the the unitarity bound coming from interactions compatible with the shift-symmetry).   Because of the discrete shift-symmetry $\sigma\rightarrow \sigma+2 \pi \Lambda_{U,S}$, the interactions suppressed by $M$ do not renormalize $\Lambda_{U,S}$ and because of the continuous shift symmetry they do not renormalize the scale $\mu$ either. On the other hand, interactions in $\mu^4$ generate derivative interactions suppressed by $M$ and this leads to the inequality $M^4\lesssim \Lambda_{U,S}^8/\mu^4$. This leaves room for interesting phenomenology as we will see.

In (\ref{eq:est1})  the factor of $N_e$ comes from the fact that the operator  $\sigma^4$ keeps operating after modes have crossed the horizon.  According to our prescription for the treatment of the conversion of $\sigma$ fluctuations into density perturbations  described in sec.~\ref{sec:curvature-conversion},  the non-Gaussianities induced when the modes are outside of the horizon are incorporated in the coefficients of eq.~(\ref{eq:zeta_relation}), which are kept generic. Here however we explicitly include this effect in the estimate just to stress that, in the absence of cancellations from the terms coming from reheating, we expect the generated non-Gaussianity to be proportional to $N_e$. 

For $1\lesssim \mu^4 N_e M^4/(H^4 \Lambda_{U,S}^4)\lesssim (n_s-1) N_e M^4/(H^2 \Lambda_{U,S}^2)$, this four point function  is the leading signal. It induces a $g_{NL}$ of the local kind of the order of
\be\label{eq:est2}
g_{NL}\zeta^2\sim\left.\frac{{{\cal L}}_4}{{\cal L}_2}\right|_{E\sim H}N_e\sim\frac{\mu^4}{\Lambda_{U,S}^4}N_e\ ,
\ee
which can be detectable. A first analysis of this kind of four-point function in the WMAP data has been recently performed in~\cite{Smidt:2010sv}. In principle, the cubic operator in (\ref{eq:cubic-spurion-two}) of the form $\tilde\mu^4\sigma^3/\Lambda_{U,S}^3$ is allowed by this symmetry and would give a larger effect on the three-point function. However, we saw that there are several symmetries that can forbid such an operator. On the other hand, the cubic operator in  (\ref{eq:interactions_mixed}) schematically of the form $\sigma(\d_\mu\sigma)^2$ is negligible. By performing a similar comparison, we obtain~\footnote{\label{footnote:mass scales}Notice that in principle the soft-breaking operators like the one we consider here could appear in the form $\mu^4\sin(\sigma/\Lambda_{U,S})(\d\sigma)^2/\Lambda_U^4$, as $(\d\sigma)^2$ is compatible with the shift symmetry. For simplicity we consider only the case in which all terms are suppressed by $\Lambda_{U,S}$. It is easy to check that our estimates would not change relevantly. The same applies to similar operators in the rest of the section.}
\be\label{eq:cubic-broken-ratio}
\left.\frac{{\cal L}^{(3)}_{\rm soft-breaking}}{{\cal L}^{(4)}_{\rm soft-breaking}}\right|_{E\sim H}\frac{1}{N_e}\sim \frac{\mu^4(\d_\mu\sigma)^2\sigma/\Lambda_{U,S}^5}{\mu^4 \sigma^4/\Lambda_{U,S}^4}\frac{1}{N_e}\sim \frac{H}{\Lambda_{U,S}}\frac{1}{N_e}\ll 1\ .
\ee
This could be only marginally detectable if we happen to see a local four-point function close to its current upper limit.
We therefore conclude that we can have a detectable four-point function with the  shape induced by the quartic operator $\sigma^4$. This shape is of the local form and is degenerate with what can be produced at reheating, but here we notice that this mechanism allows for having a large four-point function of the local form without at the same time having a local three-point function. In fact, as we just argued, we expect that non-linearities at reheating will induce non-Gaussianities dominated by the three-point function. %At subleading level we might detect a three-point function as induced by operators of the form $\sigma(\d\sigma)^2$ and, only marginally, operators of the form $\sigma^2(\d\sigma)^2$. Given the more stringent constraints on the local three-point function, these subleading operators are undetectable if the $\sigma^3$ operator is present.

There is another way in which we can obtain a large four-point function without having a detectable three-point function. We can impose the parity symmetry 
\be
\sigma_I\rightarrow -\sigma_I\ ,
\ee
as an approximate symmetry in the Lagrangian. This symmetry is expected to be broken at the time of reheating. This symmetry removes all the cubic operators in $\sigma$ as well as the mixing $\pi\sigma$, and leaves only the cubic operators of 
the form $\pi\sigma\sigma$ and the quartic of the form $\pi\pi\sigma\sigma$ and $\sigma\sigma
\sigma\sigma$. Now, if the influence of $\pi$ on $\zeta$ happens to be much smaller than the one of the $\sigma
$'s (see eq.~(\ref{eq:dominance_condition}))
%, which is the case if:
%\be
%\frac{H^2}{\mpl^2\epsilon\, c_s^{(\zeta)}}\ll \left(\left.\frac{\d\zeta}{\d\sigma}\right|_0\right)^2 \frac{H^2}{c_s^{(\sigma)}}\ ,
%\ee
then the operators containing $\pi$ are negligible and the leading interaction terms are of the form $(\d\sigma)^4$ as contained in (\ref{Spimulti4}). Alternatively, we can impose the additional
\be
\pi\quad\rightarrow\quad -\pi \ , 
\ee
approximate symmetry in the Lagrangian, and we are left again only with the $(\d\sigma)^4$ operators.
Because of the kind of symmetry we are imposing, the speed of sound of the $\sigma$ fields need not be equal to one nor equal to one another. For simplicity we will take all the sound speeds to be equal, the generalization being straightforward. In this case the speed of sound is given by 
\be
c_s^2\sim\frac{1}{1+\tilde e_2}\ .
\ee
where we have suppressed the $^I$ index in $\tilde e_2^I$ because we are taking all of the speeds of sound equal.
If all the coefficients $\tilde{\tilde M}_i$ for the various operators are the same, in the presence of a speed of sound different from one different operators would contribute differently to the non-Gaussianities, their contribution differing by powers of $1/c_s$ according to the number of spatial derivatives versus time derivatives they have. However unless the coefficients of the various operators are constrained to be in some particular relationship by some symmetry  it is reasonable to argue that there is a natural scaling relationship among the coefficients $\tilde{\tilde M}_i$ such that they differ by powers of $1/c_s$. The natural relationship is such that  all of these operators become strongly coupled at the same energy scale, and therefore because they all have the same scaling dimensions, they have comparable effects at energy scales of order $H$ that are relevant for inflationary perturbations. By natural relationship among the various coefficients we mean that if the coefficient of one of the operators does not respect this scaling relationship, then loop corrections will tend to renormalize with power law divergencies the various operators. This renormalization becomes an order one effect if the coefficients respect the natural scaling relationship. This statement is the exact analogous of the one we usually make for relativistic theories, when we say that all the various operators are suppressed by the same mass scale. In the presence of a speed of sound different from one, additional powers of $c_s$ need to be added.

It is straightforward to find this natural scaling among the various coefficients.  Since we are dealing only with the self-interactions involving the $\sigma$-fields, there is no way for them to realize that their speed of sound is different from one: this means that we can redefine the spatial coordinates and the fields in such a way as to make the $\sigma$ fields have an effective speed of sound equal to one. Let us therefore redefine the spatial coordinates in the following way
\be
\vec x\quad\rightarrow\quad \vec{\tilde x}=\frac{\vec x}{c_s}\ ,
\ee
and canonically normalize the $\sigma$ field as
\be
\sigma\quad\rightarrow\quad\sigma_c=c_s^{1/2} \sigma\ .
\ee
In this way the quadratic Lagrangian becomes
\begin{eqnarray}
 \int dt\,d^3 x  \; \sqrt{- g}\; \frac{1}{c_s^2}\left(\dot\sigma^2-c_s^2(\d_i\sigma)^2\right)\qquad\rightarrow  \qquad
 \int dt\,d^3 \tilde x   \; \sqrt{- g}\;\left(\dot\sigma_c^2-(\tilde\d_i\sigma_c)^2\right)\ ,
\eea
where $\tilde\d_i=\d_{\tilde x_i}$. The effective speed of sound in this rescaled coordinates has become equal to one. The quartic Lagrangian becomes
\begin{eqnarray}\label{eq:rescaled-quartic}
&& \int dt\,d^3 x \;  \sqrt{- g}\; \left[\frac{1}{\tilde{\tilde M}_a^4}\dot\sigma^4+\frac{1}{\tilde{\tilde M}_b^4}\dot\sigma^2\d_i\sigma\d_i\sigma+\frac{1}{\tilde{\tilde M}_c^4}\d_i\sigma\d_i\sigma\d_j\sigma\d_j\sigma\right]\qquad\rightarrow \\ \nonumber 
&& \rightarrow \qquad
 \int dt\,d^3 \tilde x \;  \sqrt{- g}\; \left[\frac{c_s}{\tilde{\tilde M}_a^4}\dot\sigma_c^4+\frac{1}{\tilde{\tilde M}_b^4c_s}\dot\sigma^2\tilde\d_i\sigma\tilde\d_i\sigma+\frac{1}{\tilde{\tilde M}_c^4c_s^3}\tilde\d_i\sigma\tilde\d_i\sigma\tilde\d_j\sigma\tilde\d_j\sigma\right]\ ,
 \eea
 where in the first passage we have simply defined the $\tilde{\tilde M}_{a,b,c}$ in terms of the $\tilde{\tilde M}_{1,2,3}$ of equation~(\ref{Spimulti4}). Since we did not rescale time, we can read off the unitarity bound directly by inspection of the mass scale suppressing the operators in the rescaled Lagrangian in (\ref{eq:rescaled-quartic}). In particular, notice that if the  $\tilde{\tilde M}_{a,b,c}$ satisfy the following approximate relationship $\Lambda_U^4\sim\tilde{\tilde M}_c c_s^3\sim \tilde{\tilde M}_b c_s\sim \tilde{\tilde M}_a/c_s$, then all of these operators become strongly coupled at the same unitarity bound scale $\Lambda_U$. It is arguable that loop corrections indeed force this relationship in a natural way. This line of reasoning was developed initially in \cite{Senatore:2009gt,Senatore:2010jy}.  Unless explicitly specified in the rest of the paper we will assume this scaling in $c_s$.  
 As a result the $g_{NL}$ that can be generated by these quartic interactions is of order
\be
g_{NL}\zeta^2\sim\left.\frac{{{\cal L}}_4}{{\cal L}_2}\right|_{E\sim H}\sim\frac{H^2\sigma_c^2}{\Lambda_U^4}\sim
\frac{H^4}{\Lambda_U^4}\qquad\Rightarrow
\qquad g_{NL}\sim 10^{10} \frac{H^4}{\Lambda_U^4}\ ,
\ee
where we have used (\ref{eq:zeta_norm1}). The induced $g_{NL}$ could be detectable. 
In this case there are three different operators that can induce a large four point function,
\be
\dot\sigma_I\dot\sigma_J\dot\sigma_K\dot\sigma_L\ , \qquad \dot\sigma_I\dot\sigma_J\d_i\sigma_K\d_i\sigma_L\ , \qquad \d_i\sigma_I \d_i\sigma_J \d_i\sigma_K \d_i\sigma_L\ ,
\ee
whose coefficients can be read off in (\ref{Spimulti4}) and are basically independent. Because these are derivative operators, the shape of the induced non-Gaussianity is of the equilateral 
kind, where all the sides of the quadrilateral in momentum space are comparable. Notice that only the shape generated by the operator $\dot\sigma^4$ is equal to the one that can be generated in single-field inflation~\cite{Senatore:2010jy}. The remaining three shapes can only be generated in multi-field inflation. 

The $\sigma^4$ interaction coming from the soft symmetry-breaking terms can lead to a detectable local four-point function. It gives:
\be\label{eq:4-point-1}
g_{NL}\zeta^2\sim\left.\frac{{{\cal L}}_4}{{\cal L}_2}\right|_{E\sim H}\sim\frac{\mu^4 N_e}{\Lambda_{U,S}^4}\lesssim (n_s-1)N_e\left(\frac{H}{\Lambda_{U,S}}\right)^2\ .
\ee
If we happen to detect this four point function close to its current experimental bound, it is possible that we will have the chance to detect also subleading operators, such as $\dot\sigma\sigma^3$, which is suppressed with respect to the leading term by a factor of $H/\Lambda_{U,S}$ and which gives rise to the same local four-point function. % and, potentially marginally, also operators of the form $\sigma^2(\d\sigma)^2$, which are suppressed by a factor of $H^2/(\Lambda_U^2N_e)$ and that give rise to a four-point function that is peaked both on equilateral and on squeezed configurations.
It is interesting to point out that in the inflationary case it is possible that symmetry is explicitly broken in a way that goes to zero in the limit the time-traslations are recovered as an exact symmetry~\footnote{For example if the clock field were to be a standard scalar field, it is possible that the symmetry breaking is proportional to the inflaton time-derivative.}. In this case the lowest dimension operators associated to the soft symmetry breaking might be absent, subleading operators with derivatives acting on the $\sigma$'s might become the leading ones, and operators that are not detectable if the lower dimension operators are present, such as those of the form $\mu^4\sigma^2(\d\sigma)^2/\Lambda_{U,S}^6$, can indeed become detectable.
Notice also that if the leading operator has only one derivative acting on the $\sigma$'s, then the upper limit on $\mu^4$ that is determined by the induced mass of the $\sigma$'s goes from  $\mu^4\lesssim (n_s-1)H^2\Lambda^2_{U,S}$ to $\mu^4\lesssim (n_s-1)H\Lambda^3_{U,S}$. If instead the leading operator has two derivatives, then we have $\mu^4\ll \Lambda_{U,S}^4$. For simplicity in doing our estimates we shall consider the first upper bound $\mu^4\lesssim (n_s-1)H^2\Lambda_{U,S}^2$, but it is worth to keep in mind that these other possibilities are present.
 
So far in this section we have concentrated on the case in which the dispersion relation of the $\sigma$ fields is of the form $\omega\propto k$. As we mentioned in sec.~\ref{sec:Abelian_lagrangian}, the dispersion relation at horizon crossing can be of the form $\omega^2\simeq k^4/(\tilde e_2 \bar{\bar M}^2)$ if the parameter $\tilde e_2$ satisfies $\tilde e_2\gtrsim  \bar{\bar M}^2/H^2$. This very non-relativistic dispersion relation is the same as the Ghost Condensate~\cite{ArkaniHamed:2003uy}, and similarly to that case~\cite{ArkaniHamed:2003uz,Cheung:2007st,Senatore:2009gt}, we can have large non-Gaussianities even in this regime. The discussion proceeds in a very similar way to the case of a linear dispersion relations. While in this case we cannot impose that the theory be quasi Lorentz invariant in order to have a large four-point function without a detectable three-point function but we we can still impose an approximate $\sigma\rightarrow-\sigma$ symmetry. Now the scaling dimensions of the operators are very non-relativistic. If we rescale $t\rightarrow s^{-1} t$, in order to keep the kinetic term invariant we need to rescale the spatial coordinates as $x^i\rightarrow s^{-1/2} x^i$. Invariance of the quadratic Lagrangian than implies that we need to rescale $\sigma$ as
\be
\sigma\quad\rightarrow\quad s^{1/4}\; \sigma\ .
\ee
The leading quartic operators has scaling dimension $1/2$ and is of the form $(\d_i\sigma)^4$. Remarkably, it is unique. This operator induces  a four-point function of the order of 
\be\label{eq:4-point-k2}
g_{NL}\zeta^2\sim\left.\frac{{{\cal L}}_4}{{\cal L}_2}\right|_{E\sim H}\sim\frac{{\bar{\bar M}}^2\sigma^2}{\tilde{\tilde M}_1^4\tilde e_2}\sim\frac{H^{1/2}\bar{\bar M}^{7/2}}{\tilde{\tilde M}_1^4\tilde e_2^{1/4}}\sim \left(\frac{H}{\Lambda_U}\right)^{1/2}
\qquad\Rightarrow
\qquad g_{NL}\sim 10^{10} \left(\frac{H}{\Lambda_U}\right)^{1/2}\ ,
\ee
where the unitarity bound $\Lambda_U$ is given by $\Lambda_U\sim\tilde e_2^{1/2} \tilde{\tilde M}_1^{8}/\bar{\bar M}^{7}%\sim \tilde e_2^{1/2}\bar{\bar M}^{7}
$ and where we have used that at energy scales of order $H$ the canonically normalized $\sigma_c$ goes as $\sigma_c\sim\tilde e_2^{3/8} \bar{\bar M}(H/\bar{\bar M})^{1/4}%\sim \Lambda_U(H/\Lambda_U)^{1/4}
$ (see~\cite{Senatore:2010jy} for how this scaling is derived). Contrary to the case of single-clock inflation, here there is in general no symmetry protecting the generation of a standard spatial kinetic term $(\d_i\sigma)^2$ from a loop constructed by contracting two fluctuations of the $(\d_i\sigma)^4$ operator. Imposing that such a generation is still subleading with respect to the tree-level value of $(\d_i\sigma)^2$ restricts $\tilde{\tilde M}_1\lesssim\bar{\bar M}/ \tilde e_2^{1/8}$. Since the leading interactions are proportional to ${\tilde {\tilde M}}_1$, imposing such an inequality is radiatively stable. 
Notice however than since $\tilde e_2\gtrsim \bar{\bar M}^2/H^2$, this constraint implies $\Lambda_U\lesssim H$. Given the estimate in (\ref{eq:4-point-k2}), this tells us that the theory with a $(\d_i\sigma)^4$ interaction has to be strongly coupled at energies scales of order $H$, and therefore not viable.

There are two ways to proceed at this point. The first is to impose a symmetry that ensures that the operator $(\d_i\sigma)^4$ is small enough not to renormalize the $(\d_i\sigma)^2$ term. This can be done by remembering that as shown in~\cite{Senatore:2010jy}, in single field inflation with the dispersion relation $\omega\propto k^2$ it is possible to impose an approximate $\pi\rightarrow-\pi$ symmetry that makes quartic operators of the form $\dot\pi^p (\d_i\d_j\pi)^{4-p}$ the leading interaction operators. This is done by keeping  the contribution from the operators $\dot\pi^3,\, \dot\pi(\d_i\pi)^2$ and $(\d_i\pi)^4$ small and by imposing their coefficients have a particular relationship among themselves so that their renormalization to the kinetic term $(\d_i\pi)^2$ cancel each other. This is the result of a symmetry. The renormalization of $(\d_i\pi)^2$ needs to be proportional to terms that breaks the shift-symmetry of $\pi$ and it is therefore small. For similar reasons the operators $\dot\pi^3,\, \dot\pi(\d_i\pi)^2$ and $(\d_i\pi)^4$ do not get renormalized by higher dimensional ones. If we impose that the $\sigma$ Lagrangian has the same structure as the $\pi$ Lagrangian then we can have the same signatures. In particular we can have a leading four-point function of the form $\dot\sigma^p (\d_i\d_j\sigma)^{4-p}$  with size of order
\be\label{eq:4-point-k3}
g_{NL}\zeta^2\sim\left.\frac{{{\cal L}}_4}{{\cal L}_2}\right|_{E\sim H}\sim\frac{H^{5/2}\bar{\bar M}^{3/2}}{\tilde{\tilde M}_1^4\tilde e_2^{5/4}}\sim \left(\frac{H}{\Lambda_U}\right)^{5/2}
\qquad\Rightarrow
\qquad g_{NL}\sim 10^{10} \left(\frac{H}{\Lambda_U}\right)^{1/2}\ ,
\ee
where we have used that he unitarity bound $\Lambda_U$ is given by $\Lambda_U\sim \tilde e_2^{1/2}\tilde{\tilde M}_1^{8/5}/\bar{\bar M}^{3/5}$.

%Notice also that since in this regime $\tilde e_2\gtrsim 1$, this constraint ensures that the unitarity bound is lower that the naive scale $\tilde e_2^{1/2}\bar{\bar M}$ at which $\sigma$ excitation become superluminal and where ghosts would enter in the spectrum~\footnote{Notice that strictly speaking the scale at which the ghosts would become part of the spectrum is controlled by the unitary gauge operator $\ddot\sigma^2/M_g^2$, where the scale $M_g$ can be in principle different from $\bar{\bar M}$.}. Since the wavefunction is different than in the case of a linear dispersion relation, the shape of the four-point function is different in this case. The resulting four-point function can be clearly detectable. Notice that here the situation is different than in the case of single field inflation, where this operator cannot be the leading one for the four-point function for symmetry reasons~\cite{Senatore:2010jy}. Depending on the actual level of the non-Gaussianities, it is possible that also higher derivative terms might be detectable. Because of the non-relativistic scaling and the shift-symmetry, the next leading ones are of scaling dimension $3/2$, they contain only spatial derivatives and schematically are of the form $(\d^2_i\sigma)^2(\d_j\sigma)^2$ or with similar index contractions.

A second way to make the four-point function the leading signal in the case of $\omega\propto k^2$ dispersion relation is by making the interactions associated to the soft-symmetry breaking the leading ones.  In this case, the leading operator is the one in $\sigma^4$. Because of the non-linear dispersion relation, the constraint that the induced mass be smaller than $\sim(n_s-1)^{1/2}H$ becomes
\be
\mu^4\lesssim (n_s-1)  H^2 \Lambda_{S}^2\ .
\ee
Here we are normalizing the fields in the soft symmetry breaking sector by $\Lambda_S$, {\it i.e.} the soft-breaking potential takes the form $\sim \mu^4\cos(\sigma_c/\Lambda_S)$. Because of the non-linear dispersion relation $\Lambda_S$ is not equal to the unitarity bound $\Lambda_{U,S}$ in the soft-breaking sector but  is related to it  $\Lambda_S^4\simeq \Lambda_{U,S}\bar{\bar M}^3\tilde e_2^{3/2}$. As in the case of a linear dispersion relation, interaction terms compatible with the Abelian symmetry do not renormalize the soft breaking terms. However, the soft-breaking terms can generate interactions compatible with the symmetry. In particular the coefficient $\tilde{\tilde M}_1^4$ suppressing the operator $(\d_i\sigma)^4$ has to be smaller than $\tilde{\tilde M}_1^4\lesssim \Lambda_{U,S}^3 \bar{\bar M}^{5}\tilde e_2^{1/2}/\mu^4$. They also generate the $(\d_i\sigma)^2$ operator. Imposing that the generated term to be smaller than the tree-level value amounts to an inequality that it is satisfied for $\Lambda_{
U,S}\gtrsim (n_s-1)^{1/2} H$, a constraint that we will easily be able to  satisfy. The $\sigma^4$ operator gives a local four-point function parametrically of the form
\be
\left.\frac{{\cal L}^{(4)}_{\rm soft-breaking}}{{\cal L}_2}\right|_{E\sim H}N_e\sim \frac{\mu^4}{H^{3/2}\Lambda_{U,S}\bar{\bar M}^{3/2}\tilde e_2^{3/4}}N_e\lesssim N_e(n_s-1)\left(\frac{H}{\Lambda_{U,S}}\right)^{1/2}\ .
\ee
The operator $\dot\sigma\sigma^3$, if present, would give rise to an additional contribution to the local four-point function suppressed by a factor of $H/\Lambda_{S}$ and could be detectable.
The operator $\sigma^3$, if present, would give a dominant and detectable local three-point function. The operator $\sigma(\d_i\sigma)^2$  is subleading with respect to the four-point function but is potentially detectable:
\be\label{eq:cubic-broken-ratio-bis}
\left.\frac{{\cal L}^{(3)}_{\rm soft-breaking}}{{\cal L}^{(4)}_{\rm soft-breaking}}\right|_{E\sim H}\frac{1}{N_e}\sim \frac{\mu^4(\d_i\sigma_c)^2\sigma_c/\Lambda_S^5}{\mu^4 \sigma_c^4/\Lambda_S^4}\frac{1}{N_e}\sim \frac{H^{3/4}}{N_e \Lambda_{U,S}^{1/4}\tilde e_2^{1/4}\bar{\bar M}^{1/2}}\lesssim\frac{1}{N_e} \left(\frac{H}{\Lambda_{U,S}}\right)^{3/4}\ll 1\ ,
\ee
where in the next-to-last passage we have used that $\Lambda_{U,S}\lesssim \tilde e_2^{1/2}\bar{\bar M}$ to avoid superluminality.

It is easy to estimate that  the quartic operator $(\d_i\sigma)^4$ as generated by loops of the soft-symmetry terms gives a four-point function that is suppressed with respect to the one from $\sigma^4$ by a factor of $H^2/(\Lambda_{U,S}^2N_e)$ and would be hard to detect. Instead, the operator $\sigma^2(\d_i\sigma)^2$ generates a four-point function that is suppressed with respect to one of $\sigma^4$ by a factor of $H/(\Lambda_{S}N_e)$ that could allow for a subleading detection.

\subsubsection*{$\bullet$ Non-Abelian case}

The story in the non-Abelian case proceeds in very similar terms for both the possible kind of dispersion relations. For simplicity we will mainly discuss the case of a linear dispersion relation. Given the Abelian treatment we have just studied the extension to the non-Abelian case is straightfroward apart for a small subtlety we will comment about. If we impose an approximate Lorentz invariance for the theory of the fluctuations, we remove all the cubic interactions and we are left with operators of the form
\be\label{eq:quartic_non_Abelian}
\tilde{\tilde c}_{11}\;\d_\mu\sigma_a\d^\mu\sigma_a\d_\nu\sigma_b\d^\nu\sigma_b+\tilde{\tilde c}_{12}\;\d_\mu\sigma_a\d^\mu\sigma_b\d_\nu\sigma_a\d^\nu\sigma_b\ .
\ee
Notice that at quartic order there are other less irrelevant operators (i.e. of dimension six instead of dimension eight) coming from the terms in $D_{\mu}^2$ in (\ref{Spimulti-non-Abelian}) that we wrote down in (\ref{eq:non_Abelian_quartic_interactions}). If the mass scale $F_1$ suppressing these operators were the same as the one suppressing the operators in (\ref{eq:quartic_non_Abelian}) then these operators would be much more important at energy scales of order Hubble, by a factor of order $F_1^2/H^2\gg 1$. However as we noted in eq.~(\ref{eq:conversion_non_Abelian}) through a rotation of the unbroken group, the adiabatic fluctuations are sensitive only to one linear combination of the $\sigma_a$ fields. Without loss of generality it can be chosen to be the $a=1$ component. If are were looking only at adiabatic fluctuations then we would be looking at a four-point function of the $\sigma_a$'s where all the indeces $a$ are equal to one. Because of the antisymmetry of the structure constants $C_{abc}$, the contribution of the dimension six operators vanishes. For adiabatic fluctuations we are therefore left with the interactions in (\ref{eq:quartic_non_Abelian}) that are of the same in form as in the Abelian case. 

As we noted in the paragraph below eq.~(\ref{eq:isocurvature}) isocurvature fluctuations, if present, will in general be sensitive to a different linear combination of the $\sigma_a$'s raising the possibility that the operators in (\ref{eq:non_Abelian_quartic_interactions}) are important in the case of mixed adiabatic-isocurvature four-point functions. We will explore this further in the next subsections. 

The $g_{NL}$ induced by the operators in (\ref{eq:quartic_non_Abelian}) is of the order of
\be
g_{NL}\zeta^2\sim\left.\frac{{{\cal L}}_4}{{\cal L}_2}\right|_{E\sim H}\sim\frac{H^2\sigma^2}{F^4}\sim
\frac{H^4}{F^4}\qquad\Rightarrow
\qquad g_{NL}\sim 10^{10} \frac{H^4}{F^4}\ ,
\ee
and could be detectable.

Operators similar to the ones in (\ref{eq:quartic_non_Abelian}) are included also in terms of the form ${\rm Tr}\left[{\mathscr{D}}_\nu D^{\mu}{\mathscr{D}}^\nu D_{\mu}\right]$ and others with similar index contractions. Notice that naively these terms contain operators that are cubic in the $\sigma$'s and that are not removed by imposing a quasi Lorentz invariance for the theory. However, it is worth to check that they cancel because of the antisymmetry of the structure constants.

Another interesting case in which the four-point function is relevant observationally is when ${\rm Tr}[x_a x_a x_a]=0$, which happens for some groups like a broken $SU(2)$. In this case the three point function for the $\sigma$ fields is suppressed while the four-point function is not.  This is another way in which a four-point function becomes the leading non-Gaussianity that it valid only in the non-Abelian case and that does not involve the requirement for the theory to be quasi Lorentz invariant.

A detectable four-point function can also be associated with the soft-symmetry breaking. For example, the non-Abelian analogous of the operator in (\ref{eq:potential-even}), schematically of the form $\sigma^4$, can give a detectable four-point function for $\mu$ in the interval $10^{-4}\lesssim \mu^4N_e/\Lambda_{U,S}^4\lesssim (n_s-1) N_e H^2/ \Lambda_{U,S}^2$. The non-Abelian analogous of the terms in  (\ref{eq:operator-peculiar}), when present, could also be potentially detectable. The first term, schematically of the form $\sigma^2\dot\sigma$, can give a comparable effect in the three-point function. The second term, schematically of the form $\sigma(\d\sigma)^2$, is suppressed by a factor $H/(\Lambda_{U,S} N_e)$, which could still leave room for a possible detection. It is interesting to point out that it is possible that the lowest dimension operators are absent because of the way the symmetry is broken. In such a case, subleading operators might become the leading ones, and operators that are not detectable if the lower dimension operators are present, such as those of the form $\mu^4\sigma^2(\d\sigma)^2/\Lambda_{U,S}^6$, can become detectable.

Finally we add just one comment on the case the dispersion relation is of the form $\omega\propto k^2$. As we said, in this case everything proceeds as in the Abelian case except that the symmetries of the Lagrangian do not seem to allow to impose the same structure as in single field inflation and that the operators that generate the kinetic terms also induce some self-interactions. A straightforward estimate shows that these operators would renormalize the standard spatial kinetic term $(\d_i\sigma)^2$ unless the unitarity bound is smaller than $\Lambda_U\lesssim {\bar{\bar F}} / \tilde e_2^{1/6}$. Here $\bar{\bar F}$ plays in the non-Abelian case, the same role as $ \bar{\bar M}$ does for the Abelian one. While, as in the Abelian case, we cannot have interactions of the form $(\d_i\sigma)^4$, this inequality can be satisfied by introducing soft-breaking interactions whose unitarity bound $\Lambda_{U,S}$ is smaller than ${\bar{\bar F}}/\tilde e_2^{1/6}$. In other words, in the non-Abelian case with the dispersion relation $\omega\propto k^2$, it is necessary to introduce soft-breaking interactions. This tells us that detection of a three-point function as produced by the operators $\dot\sigma^4,\, (\d^2_i\sigma)^4$ and similar index contractions would rule out the possibility that the additional fields are the Goldstone bosons of a non-Abelian group.

%that the non-Abelian analogous of the operator in (\ref{eq:ghost-dispertion-operator}) that gives a spatial kinetic term induces also some self-interactions that start at quartic order in the Golsdtone bosons. It is straightforward to check that such an interaction does not necessarily destroy the radiative stability under the analogous condition of  $\tilde{\tilde M}_1\lesssim \tilde e_2^{1/8}\bar{\bar M}$ that ensures that the presence of a four-point function with a dispersion relation of the form $\omega\propto k^2$ also in the non-Abelian case.

There are finally other possibilities in which the four-point function is smaller than the three-point function but still detectable that we will discuss in the next subsection.

\subsubsection*{$\bullet$ Supersymmetric case} 

In the case where the lightness of the $\sigma$ fields is partially protected by an approximate supersymmetry, there are two quartic interactions that can be relevant.
The first is the $\lambda^2\sigma^2\sigma^*{}^2$ interaction. This leads to a four-point function of the order
\be\label{eq:susy_local_four_point}
g_{NL}\zeta^2\sim\left.\frac{{{\cal L}}_4}{{\cal L}_2}\right|_{E\sim H}N_e\sim\lambda^2 N_e\qquad\Rightarrow
\qquad g_{NL}\sim 10^{10} \lambda^2 N_e\ ,
\ee
where we have inserted the factor of the number of $e$-foldings $N_e$ coming from the fact that this interaction acts  from horizon crossing to the end of inflation.
Remembering that the lightness of the $\sigma$ field is controlled by $\lambda\ll 1$, this four point function would be detectable in the interval $10^{-4}\lesssim \lambda^2 N_e\ll N_e$. We saw that such a shape for a four-point function is to a very good approximation the local one, which is produced also from non-linearities in the relationship between $\zeta$ and $\sigma$. As we will see in the next subsection, in the supersymmetric case this four-point function is naturally accompanied by a three-point function of the local form whose size is parametrically larger. The fact that we have not seen yet such a local three-point function constraints the four-point function in (\ref{eq:susy_local_four_point}) to be practically undetectable.

If we instead impose an approximate shift-symmetry on the imaginary part of $\sigma$, then a four-point function of the local form induced by the operator $\lambda^2{\rm Im}(\sigma)^4$ can be generated and can be the leading signal.

It is also possible to have a large four-point function induced by the derivative operator  
$\sigma\sigma^*\d_\mu\sigma\d^\mu\sigma^*/{M_{s,2}^2}$ coming from the Kahler potential. By imposing an approximate $\Sigma\rightarrow -\Sigma$ symmetry it is possible to suppress the three-point function, and the resulting $g_{NL}$ is of the order
\be\label{eq:new_susy_four-point}
g_{NL}\zeta^2\sim\left.\frac{{{\cal L}}_4}{{\cal L}_2}\right|_{E\sim H}\sim\frac{H^2}{M^2_{s,2}}\qquad\Rightarrow
\qquad g_{NL}\sim 10^{10} \frac{H^2}{M^2_{s,2}}\ .
\ee
The $\sigma$-mass that this operator induces can be estimated to be of the order of $H^2/M_{s,2}$, where the loop is cutoff at energies of order $H$ because of supersymmetry restoration. So, we should concentrate on the interval $1\ll M^2_{s,2}/H^2\lesssim 10^4$ for detectability. %The resulting shape for the four-point function is unique to this case and therefore it would represent a clear indication of an approximate supersymmetry during inflation. 
However, we saw that supergravity corrections induce an operator of the form $\sigma^4 H^2/M_{s,2}^2$.
The four point function induced by this operator is of the form
\be
g_{NL}\zeta^2\sim\left.\frac{{{\cal L}}_4}{{\cal L}_2}\right|_{E\sim H}N_e\sim\frac{H^2}{M^2_{s,2}}N_e\qquad\Rightarrow
\qquad g_{NL}\sim 10^{10} \frac{H^2}{M^2_{s,2}} N_e\ ,
\ee
where the factor of $N_e$ comes from the fact that this interaction keeps acting for all the time the modes are outside of the horizon. This is detectable in the interval $10^{-4}\lesssim H^2 N_e/M^2_{s,2}\lesssim 1$. The size of this four-point function is a factor of $N_e$ larger then the one in (\ref{eq:new_susy_four-point}) originating from the operator $\sigma^2(\d_\mu\sigma)^2$ and it has a local shape. It is therefore hard that the shape originating from $\sigma^2(\d_\mu\sigma)^2$ can be detected unless we happen to detect a very large $g_{NL}$ of the local kind. We would be tempted to declare that such a detection would be a signature of supersymmetry as an approximate symmetry during inflation, the caveat being that in this case the lightness of the $\sigma$ fields is guaranteed only by a tuning of order $n_s-1$, and it is therefore delicate to make claims about the signatures of these models that are based on using naturalness at the percent level. \\

We conclude this subsection by summarizing that it {\it is possibile} for multifield inflation to generate a large and detactable four-point function even in the absence of a detectable three-point function. Contrary to the case of single field inflation with a linear dispersion relation $\omega\propto k$, where only one shape is possible here there are five shapes generated by five independent operators. In some cases the signal is either peaked on equilateral quadrangular configurations (for the operators involving derivatives on each fluctuation) or in more general configurations (for the operators without derivatives on each fluctuation). We will study in detail the shape of the four point function in a subsequent paper~\cite{mari1}. One unique combination of these operators is Lorentz invariant. A detection of any of the shapes that are not produced by single field inflation would be a {\it clear indication} of multifield inflation, and depending on the particular shapes and a possible parallel  detection of a three-point function we could learn if {\it Lorentz symmetry} or even, with some further luck, {\it supersymmetry} are  approximate symmetries for the Lagrangian of the fluctuations during inflation.
We could also have indication of the fact that the signals might be originating from soft-symmetry breaking terms. An analysis of the four-point function in the CMB data in search for the signal from the above operators is in progress \cite{kensen1}.

\subsection{Three-point function}

In the former subsection we concentrated on the four-point function as multifield inflation can predict an interesting  structure for this signal that is completely new. This obviously does not excludes the possibility  that in different regions of the parameters space we can have a large and detectable three-point function. Let us start with the Abelian case. 

\subsubsection*{$\bullet$ Abelian case}

The action in (\ref{Spimulti3}) has all possible interactions of the form $\pi^3,\; \pi^2
\sigma_I,\;\pi\sigma_I^2$ and $\sigma_I^3$ subject only to the constraint that a derivative must act on each fluctuations and the operator must be rotational invariant. Since the conversion of the $\pi$ and $\sigma_I$'s fluctuations into $\zeta$ in eq.~(\ref{eq:zeta_relation}) is local in real space, this means that each of these operators will give rise to a three-point function whose size will depend on the particular coefficients of (\ref{eq:zeta_relation}) and on the particular operator chosen from (\ref{Spimulti3}), but whose shape will be equal either to the one of $\dot\pi^3$ or to the one of $\dot\pi(\d_i\pi)^2$. These are the two kind of shapes for the three-point function that are produced in single field inflation when the Goldstone boson is protected by an approximate continuos shift symmetry. Exactly the same shape can be produced in multifield inflation. For example in~\cite{Langlois:2008wt} the possibility of these large interactions had been highlighted by studying the case of multifield DBI inflation, though in that case the coefficients of $\dot\sigma^3$ and $\dot\sigma(\d_i\sigma)^2$ are related because of the additional symmetry imposed on the fields by the DBI Lagrangian. In general in the Abelian case they are independent and they give rise to non-Gaussianities with size of order
\be\label{eq:three-point-Abelian}
f_{NL}\zeta\sim\left.\frac{{{\cal L}}_3}{{\cal L}_2}\right|_{E\sim H}\sim\frac{H^2}{\Lambda_U^2}\qquad\Rightarrow
\qquad f_{NL}\sim 10^{5} \frac{H^2}{\Lambda_U^2}\ ,
\ee
where now $\Lambda_U\sim c_s^{1/4}\tilde M_1$ and can potentially be detected with high signal to noise. Here in this subsection we are assuming that the unitarity bound associated to the cubic term is smaller or equal to the one associated to the quartic interactions compatible with the symmetry, so that the leading signal is indeed the three-point function.
%peculiar Lagrangians of the form $P(X^{IJ},\phi^K)$, with $X^{IJ}=\d_\mu\phi^I\d^\mu\phi^J$. 
The linear combination of the induced shapes interpolates continuously from shapes where the signal is dominated by equilateral triangles in  Fourier space, whose amplitude is characterized by $f_{NL}^{\rm equil.}$, to shapes where the signal is dominated both by equilateral triangles and by flat triangles (with opposite signs), whose amplitude is characterized by $f_{NL}^{\rm orthog.}$.  A detailed and optimal analysis of any possible linear combination of these two shapes was performed in \cite{Senatore:2009gt} on the WMAP data finding no detection but constraining the amplitude of the signal of these two shapes to be~\cite{Komatsu:2010fb,Senatore:2009gt}: 
\be
-214<f_{NL}^{\rm equil. }<266 \ \ \ {\rm at} \ 95\% {\rm \ C.L.}\ , \qquad \qquad -410<f_{NL}^{\rm orthog. }<6 \ \ \ {\rm at} \ 95\% {\rm \ C.L.}\ ,
\ee
with $f_{NL}^{\rm orthog. }$ being zero almost excluded at 2$\sigma$ level. 
Using the effective field theory of single field inflation these constraints were then translated into contour plots for the coefficients $M_2^4$ and $M_3^4$ of the single field operators $\dot\pi^3$ and $\dot\pi(\d_i\pi)^2$~\cite{Senatore:2009gt}. Those contour plots could be extended to include all the additional operators we have in the case of multifield inflation, though, given the large degeneracies (many operators give in fact only two shapes) a large fraction of the parameter space would be left unconstrained. We leave this for future work. Here we just mention that among the many different ways in which the three-point function can be large there are the cases in which the $\sigma_I$ fields have a small speed of sound and there is also the case in which the time-kinetic mixing is characterized by having the $\epsilon_{\rm unmix}$ defined in (\ref{eq:time-kinetic-mixing}) much smaller than one. Of these two cases, the second one is an intrinsically multifield effect.

Operators that induce a shape for the three-point function that is different from the ones produceable in single-field inflation are the ones associated to the soft-breaking of the $U(1)$ symmetry of eq.s~(\ref{eq:potential-odd}) and~(\ref{eq:interactions_mixed}). First lets consider the case when the symmetry breaking is happens due to only one spurion, the leading operators are the following. The operator in (\ref{eq:operator-peculiar}), schematically of the form $\mu^4\sigma_c^2\dot\sigma_c/\Lambda_{S}^4\sim\mu^4 H \sigma_c^3/\Lambda_{S}^4$, gives a local three-point function of size
\be
\label{eq:estimate-breaking-3-point}
\left.\frac{{\cal L}^{(3)}_{\rm soft-breaking}}{{\cal L}^{(2)}}\right|_{E\sim H}N_e\sim\frac{\mu^4 H\sigma_c^3/\Lambda_{U,S}^4}{\dot\sigma_c^2}\sim N_2\frac{\mu^4}{\Lambda_{U,S}^4}\lesssim (n_s-1)N_e\left(\frac{H}{\Lambda_{U,S}}\right)^2\sim\left(\frac{H}{\Lambda_{U,S}}\right)^2\ ,
\ee
where we have used that $\mu^4\lesssim (n_s-1)H^2\Lambda_{U,S}^2$, which could be detectable. This is the leading signal. Subleading signals are given by the operators $\mu^8\sigma^3/\Lambda_{S}^7$, whose signal is suppressed by an additional factor $(n_s-1)H/\Lambda_{U,S}$, and $\mu^4\sigma(\d\sigma)^2/\Lambda_{S}^5$, whose signal is suppressed by an additional factor $H/(\Lambda_{U,S}N_e)$. These signals could be marginally detectable.
%\bea\label{eq:estimate-breaking-3-point}
%&&\left.\frac{{\cal L}^{(3)}_{\rm soft-breaking}}{{\cal L}^{(3)}}\right|_{E\sim H}N_e\sim \frac{\mu^4 H\sigma_c^3/\Lambda_U^4}{(\d\sigma_c)^3/\Lambda_U^2}N_e\sim \frac{\mu^4}{\Lambda_U^2 H^2}N_e\lesssim(n_s-1) N_e\sim 1\ , \\ \nonumber
%&&\left.\frac{{\cal L}^{(3)}_{\rm soft-breaking}}{{\cal L}^{(3)}}\right|_{E\sim H}N_e\sim \frac{\mu^8\sigma_c^3/\Lambda_U^7}{(\d\sigma_c)^3/\Lambda_U^2}N_e\sim \frac{\mu^8}{\Lambda_U^5 H^3}N_e\lesssim \frac{H}{\Lambda_U}\frac{((n_s-1) N_e)^2}{N_e}\ll 1\ , \\ \nonumber
%&&\left.\frac{{\cal L}^{(3)}_{\rm soft-breaking}}{{\cal L}^{(3)}}\right|_{E\sim H}\sim \frac{\mu^4(\d\sigma_c)^2\sigma_c/\Lambda_U^5}{(\d\sigma_c)^3/\Lambda_U^2}\sim \frac{\mu^4}{\Lambda_U^3 H}\ll \frac{H}{\Lambda_U}\ll 1\ ,
%\eea
%where $\Lambda_U\sim c_s^{1/4} \tilde M_1$ and $\mu^4\lesssim (n_s-1)H^2\Lambda_U^2$.
Further, as we saw in sec.~(\ref{sec:four-point-multifield}), we can impose an approximate Lorentz invariance symmetry which forbids the cubic operators compatible with the $U(1)$ symmetry as well as the operator schematically of the form $\sigma^2\dot\sigma$ from symmetry breaking. In this case the leading cubic operator has the form $\sigma(\d_\mu\sigma)^2$. This is unique and it induces a shape for the three-point function that cannot be produced as the leading ones in single field inflation, and whose size is of the order
\be
f_{NL}\zeta\sim\left.\frac{{{\cal L}}_3}{{\cal L}_2}\right|_{E\sim H}\sim\frac{\mu^4 \sigma_c}{ \Lambda_{U,S}^5}\qquad\Rightarrow\qquad f_{NL}\sim 10^{5} \frac{\mu^4 H}{ \Lambda_{U,S}^5}\ ,
\ee
which can be detectable in the regime $10^{-4}\lesssim H\mu^4/ \Lambda_{U,S}^4\lesssim (n_s-1) H^3/ \Lambda_{U,S}^3$.
Notice that the non-Gaussianities induced by this operator are expected to be smaller than the ones induced by the operator generating the local four-point function in (\ref{eq:potential-even}), schematically of the form $\sigma^4$, by a factor of $H/( \Lambda_{U,S} N_e)$. Because of this, detectability of such a shape is possible only in the case we happen to detect a large four-point function of the local kind. Notice that while this shape could be produced in single-field inflation as well, it is in general either subleading with respect to different shapes than the local four-point function and/or accompanied by other shapes of comparable size. Therefore the combination local four-point function and shape of the form $\sigma(\d_\mu\sigma)^2$ is a clear indication of multifield inflation with an approximate Lorentz invariance. Further notice that for the interactions of the form $\sigma(\d\sigma)^2$ one of the fluctuations does not have a derivative acting upon it. Therefore the resulting shape of the three-point function will have a non-vanishing squeezed limit and so it will affect the clustering statistics of collapsed objects such as galaxies and there would be constraints in studies similar to  \cite{Slosar:2008hx} in addition to direct searches for a three-point function.

In this same regime where we suppress the leading cubic operators by imposing an approximate Lorentz invariance for the fluctuations, the local three point function induced by the soft-symmetry braking $\sigma^3$ operator in eq.~(\ref{eq:potential-odd}) could be detectable. Its non-Gaussianity is suppressed by a factor of order $\lesssim(n_s-1)H/ \Lambda_{U,S}$, which could still allow for a detection.

In the case the explicit symmetry breaking is performed with two spurions of different charge such  that a large cubic operator of the form $\sigma^3$ is generated, the induced local three-point function is of the size
\be
f_{NL}\zeta\sim N_e\left.\frac{{{\cal L}}_3}{{\cal L}_2}\right|_{E\sim H}\sim N_e\frac{\tilde\mu^4}{ \Lambda_{U,S}^3 H}\qquad\Rightarrow\qquad f_{NL}\sim 10^{5} N_e\frac{\tilde \mu^4}{ \Lambda_{U,S}^3 H}\ ,
\ee
which can be detectable in the interval $10^{-4}\lesssim N_e\tilde \mu^4/(\Lambda_{U,S}^3 H)\lesssim H/\Lambda_{U,S}$ . We remark that it is possible that the soft symmetry breaking might be proportional to the parameter breaking time diffs. (related to the inflationary clock field). In this case, the leading soft-breaking terms might be absent, making the subleading ones even more detectable.
 
As in the former section when we discussed about the four-point function in the Abelian case, we have so far concentrated on the case where the dispersion  relation is linear in $k$, of the form $\omega=c_s k$. As we discussed, it is also possible that the dispersion relation at horizon crossing is of the form $\omega=k^2/(\tilde e_2^{1/2}\bar{\bar M})$. Because of the non-relativistic scaling, in this case the leading operators are of the form $\dot\sigma(\d_i\sigma)^2$ and $(\d_j^2\sigma)(\d_i\sigma)^2$. The level of non-Gaussianity induced by the first operator is of the form
\be\label{3-point-ksq-q}
f_{NL}\zeta\sim\left.\frac{{{\cal L}}_3}{{\cal L}_2}\right|_{E\sim H}\sim\frac{H^{1/4}\bar{\bar M}^{7/4}}{\tilde M^2 \tilde e_2^{5/8}}\sim \left(\frac{H}{\Lambda_U}\right)^{1/4}
\qquad\Rightarrow
\qquad f_{NL}\sim 10^{5} \left(\frac{H}{\Lambda_U}\right)^{1/4}\ ,
\ee
where $\Lambda_U\sim \tilde e_2^{5/2}\tilde M^8/\bar{\bar M}^7$. %\sim \tilde e^{1/2}_2 {\bar{\bar M}}$ and we have taken $\tilde M\sim {\bar{\bar M}}/\tilde e_2^{1/4}$ in order for the unitarity bound to coincide with the bound from the naive scale at which ghost excitations would become part of the spectrum. 
 A similar estimate holds also for the operator $(\d_j^2\sigma)(\d_i\sigma)^2$, and the resulting signal can be detectable. The same signal can be produced also in single field inflation, where optimal limits on the three-point function as induced from these two operators were given in~\cite{Senatore:2009gt} by deriving and interpreting the limits on $f_{NL}^{\rm equil.}$ and $f_{NL}^{\rm orthog.}$. 

In analogy to what happened for  the four-point function  in the absence of a symmetry the operator in $\dot\sigma(\d_i\sigma)^2$ can generate at loop level a kinetic term of the form $(\d_i\sigma)^2$.  Imposing that such a generation is still subleading with respect to the tree-level value of $(\d_i\sigma)^2$ restricts $\tilde M\lesssim\bar{\bar M}/ \tilde e_2^{3/8}$. Since the leading interactions are proportional to ${\tilde M}$ imposing such an inequality is radiatively stable. The constraint $\tilde e_2\gtrsim {\bar{\bar M}}^2/H^2$ forces the unitarity bound $\Lambda_U$ to be smaller than $H$. Given the estimate in (\ref{3-point-ksq-q}) such a situation is not viable. This would naively force us either to forbid the cubic interactions compatible with the symmetry, or
%Notice also that since in this regime $\tilde e_2\gtrsim 1$, this constraint ensures that the unitarity bound is lower that the naive scale $\tilde e_2^{1/2}\bar{\bar M}$ at which $\sigma$ excitations become superluminal and where ghosts would enter in the spectrum.
to impose a symmetry  by which we can make sure that the operator $(\d_i\sigma)^2$ does not get renormalized. We can go back to the case of single-clock inflation in the near de Sitter limit with a dispersion relation of the form $\omega\propto k^2$ where there are three interactions of the form $\dot\pi^3,\ \dot\pi(\d_i\pi)^2$ and $(\d_i\pi)^4$ with some specific coefficients fixed by symmetries. There can also be an additional unrelated interactions proportional to $(\d_j^2\pi)(\d_i\pi)^2$, accompanied by symmetry reasons by higher order interactions. Contrary to the naive estimates loops of these operators cancel each other and do not generate a $(\d_i\pi)^2$ term. This is in fact a result of a symmetry: the $(\d_i\pi)^2$ term has to be proportional to $\dot H\mpl^2$ and such a term breaks the shift symmetry of $\pi$. It can therefore be generated only through operators that explicitly break this symmetry. This ensures that there is a cancellation among the loop-diagrams which might seem remarkable at a diagrammatic level. Coming back to the case of multifield inflation where the $\sigma$ fields have a dispersion relation of the form $\omega\propto k^2$, we see that if we impose that the coefficients of the operators $\dot\sigma^3,\ \dot\sigma(\d_i\sigma)^2$ and $(\d_i\sigma)^4$ to have exactly the same relationship as in the case of single-clock inflation, then we are guaranteed  not to generate the $(\d_i\sigma)^2$ term (and that the relative coefficients between the operator remains unchanged). In this way we are able to eliminate the constraint $\tilde M\lesssim\bar{\bar M}/ \tilde e_2^{3/8}$ and we take $\tilde M\sim \bar{\bar M}/\tilde e_2^{1/4}$ so that the unitarity bound $\Lambda_U\sim \tilde e_2^{1/2}{\bar{\bar M}}$ coincides with the scale at which $\sigma$ excitations become superluminal. The resulting non-Gaussian signal is dominated by a three-point function as in (\ref{3-point-ksq-q}) with a subleading, and undetectable, four-point function as in  (\ref{eq:4-point-k2}).

Let us now consider  the symmetry breaking terms. In the case we break the symmetry with only one spurion, the leading operators are of the form $\mu^4\dot\sigma_c\sigma_c^2/\Lambda_S^4\sim \mu^4 H\sigma_c^3/\Lambda_{S}^4$ and $\mu^4\sigma_c(\d_i\sigma_c)^2/\Lambda_{S}^5$. The first one gives rise to a local three-point function of the form
\be
\left.\frac{{\cal L}^{(3)}_{\rm soft-breaking}}{{\cal L}^{(2)}}\right|_{E\sim H}N_e\sim\frac{\mu^4 }{\Lambda_{S}^4}\frac{H \sigma_c^3}{H^2\sigma_c^2} N_e\sim  \frac{\mu^4}{H^{3/4}\bar{\bar M}^{9/4}\tilde e_2^{9/8}\Lambda_{U,S} }N_e\lesssim (n_s-1)N_e\left(\frac{H}{\Lambda_{U,S}}\right)^{5/4}\ .
\ee
where we have used that ${\bar{\bar M}}\tilde e_2^{1/2}\gtrsim\Lambda_{U,S}$, that $\Lambda_{S}^4\sim \Lambda_{U,S}\bar{\bar M}^3\tilde e_2^{3/2}$ and that $\mu^4\lesssim (n_s-1)H^2\Lambda_S^2$.
The second operator gives a three-point function of the order of
\be
\left.\frac{{\cal L}^{(3)}_{\rm soft-breaking}}{{\cal L}^{(2)}}\right|_{E\sim H}\sim\frac{\mu^4 }{\Lambda_{S}^5}\frac{H {\bar{\bar M}}\tilde e_2^{1/2} H \sigma_c^3}{H^2\sigma_c^2} \sim \frac{\mu^4}{H^{3/4}\bar{\bar M}^{2}\tilde e_2\Lambda_{U,S}^{5/4} }\lesssim (n_s-1)\left(\frac{H}{\Lambda_{U,S}}\right)^{5/4}\ .
\ee
Notice that the three-point function induced by these two operators is smaller than the local four-point function induced by the $\sigma^4$ operator, which scales as $(n_s-1)N_e(H/\Lambda_{U,S})^{1/2}$. Still the signal could be detectable even in the presence of the $\sigma^4$ operator.

If the symmetry is broken with a second spurion $\tilde\mu^4$, then it is possible to have a large local three-point function as induced by the $\tilde\mu^4\sigma_c^3/\Lambda_{S}^3$ operator. The signal scales as 
be
\be
\left.\frac{{\cal L}^{(3)}_{\rm soft-breaking}}{{\cal L}^{(2)}}\right|_{E\sim H}N_e\sim\frac{\tilde\mu^4 }{\Lambda_{S}^3}\frac{\sigma_c^3}{H^2\sigma_c^2} N_e\sim \frac{\tilde\mu^4}{ H^{7/4}\bar{\bar M}^{3/2}\tilde e_2^{3/4}\Lambda_{U,S}^{3/4} }\lesssim (n_s-1)N_e\left(\frac{H}{\Lambda_{U,S}}\right)^{1/4}\ ,
\ee
which can be  detectable and which, if present, would  make the contribution from the other operators negligible. In order for the operators associated to soft-breaking not to renormalize the $(\d_i\sigma)^2$ kinetic term, we have to impose $\Lambda_{U,S}\gtrsim (n_s-1)^{1/2}H$, a constraint that is straightforward to satisfy.

In multifield inflation, there is one additional form of three-point function that does not have an analogue in single field inflation. This is due to the three-point function induced by the quadratic relationship in (\ref{eq:zeta_relation}) between $\zeta$ and $\sigma$. This induces a three-point function of the local form, with $f_{NL}^{\rm loc.}$ as in (\ref{eq:fnlloc}). The optimal analysis of the WMAP data for this kind of signal was performed in \cite{Komatsu:2010fb,Smith:2009jr}, where the constraint were combined with analogous limits from the SDSS data \cite{Slosar:2008hx} to obtain
\be
-5<f_{NL}^{\rm loc.}<59 \ \ \ {\rm at} \ 95\% {\rm \ C.L.}\ .
\ee
This limit can be translated, using (\ref{eq:zeta_relation}), into limits on the parameters that characterize the conversion from $\sigma_I$'s to $\zeta$. Unfortunately, the large degeneracy in the parameter space limits the power of this constraint. 

It is worth stressing that single field inflation cannot produce a value of $f_{NL}^{\rm loc.}$ that is larger than the deviation of the power spectrum from scale invariance, which is generically very small \cite{Maldacena:2002vr,Creminelli:2004yq,Cheung:2007sv}. This means that a detection of such a three-point function in the absence of a detection of a large deviation from scale invariance would {\it rule out} single field inflation.

\subsubsection*{$\bullet$ Non-Abelian case}

The non-Abelian case is a straightforward generalization of the Abelian one and for simplicity we shall treat explicitly only the case of a linear dispersion relation, adding just a brief comment concering the non-linear dispersion relation towards the end. The main difference is that the possible operators, and therefore the possible shapes, can be further constrained depending on the non-Abelian group considered. If ${\rm Tr}[x_a x_b x_c]\neq0$, then we can have the same kind of cubic interactions $\dot\sigma^3$ and $\dot\sigma(\d_i\sigma)^2$ present in the Abelian and in the single-clock cases. Notice that because of the field redefinition (\ref{eq:conversion_non_Abelian}), in the case of adiabatic fluctuations we are actually interested in groups where ${\rm Tr}[x_a x_a x_a]\neq0$. We have cubic interactions of the form 
\be
\dot\sigma^3\ ,\qquad \dot\sigma(\d_i\sigma)^2\ , \quad {\rm only \ if\ }\quad {\rm Tr}[x_a x_a x_a]\neq0 \ .
\ee
The trace is non-zero whenever there is a commuting Abelian subgroup  (which had to be the case as this case includes the Abelian case of the former section). But for other groups such as $SU(2)$ it can be zero.

When the trace is zero we are left with the last three cubic operators in (\ref{eq:non_Abelian_three_point}) and with the terms from soft-symmetry breaking. Notice however that the last three operators in  (\ref{eq:non_Abelian_three_point}) require that both the $\sigma$ fields and the $\pi$ field be important at a comparable level for the adiabatic fluctuations.  The operators from soft-symmetry breaking require a relevant breaking at the time of horizon crossing, which again does not need to be there. We find that this strengthen the relevance of the four-point function in the non-Abelian case.

%We are left with the only three operators in (\ref{eq:non_Abelian_three_point}) plus additional possible ones of the form $\d\sigma\d\sigma_a\d\sigma_a$   that are possible if there is an additional Goldstone boson in  a singlet representation ($\sigma$ in this case). Notice that however these interactions involve always more than one kind of a field: either the non-Abelian Goldstone bosons and $\pi$, or the non-Abelian Goldstone bosons and a singlet. This means that in order for a three-point function of the adiabatic fluctuations to be induced at a relevant level, more than one field in (\ref{eq:zeta_relation}) need to be important at a comparable level. In some sense, this makes the case for a large four-point function in the non-Abelian case even more interesting.

We now come back to the operators associated with the soft-symmetry breaking. The natural suppression of the three-point function for some non-Abelian groups makes the discussion even more interesting than in the Abelian case. If we break the symmetry with only one spurion, the leading operator might be given by the non-Abelian analogous of the cubic one in eq.~(\ref{eq:interactions_mixed}), schematically of the form $\sigma(\d\sigma)^2$. This gives an effect of the order
\be\label{eq:non-Abelian-three-point-operator}
f_{NL}\zeta\sim\left.\frac{{{\cal L}}_3}{{\cal L}_2}\right|_{E\sim H}\sim\frac{\mu^4 H}{ \Lambda_{U,S}^5}\qquad\Rightarrow\qquad f_{NL}\sim 10^{5} \frac{\mu^4 H}{ \Lambda_{U,S}^5}\ .
\ee
In the limit $10^{-4}\lesssim \mu^4/(H \Lambda_{U,S}^5) \ll H^3/\Lambda_{U,S}^3$ this signal is detectable. A simple estimate shows that the non-Gaussianity induced by this operator is subleading by a factor $H/(\Lambda_{U,S} N_e)\ll1$ with respect to the four-point function induced by the non-Abelian analogous of the operator in (\ref{eq:potential-even}), schematically of the form $\sigma^4$. There is however a regime in which both shapes could be detected. Notice that if the nature of the symmetry breaking is such that the first of the operators in (\ref{eq:operator-peculiar}) is present, then there is a local three-point function whose size is comparable to the one of the four-point function. The stronger constraints currently available from the data for the local three-point function~\cite{Senatore:2009gt,Komatsu:2010fb} imply that detection of  the signal from (\ref{eq:non-Abelian-three-point-operator}) would be problematic. However, as we stressed, the presence of any of these operators depends on the characteristics of the symmetry breaking terms and it can be absent. This is for example the case for the well-known chiral $SU(2)\times SU(2)$ symmetry of the QCD Lagrangian spontaneously broken to the diagonal subgroup and explicitly broken by the quark masses. Here there are even more possibilities, as the explicit symmetry breaking could be proportional to the breaking of time diffs. by the inflationary clock, possibly forcing the leading symmetry breaking terms to have derivatives acting on the $\sigma$'s.
%Further, the three-point function induced by the non-Abelian analogous of the operator in (\ref{eq:potential-odd}) is down with respect to the one from (\ref{eq:interactions_mixed}) we just considered by another factor of $\mu^4/(H^2F^2)$, which makes it hard to detect. 
%Notice that the signal resulting from the combination of the $\sigma^4$ operator and the non-Lorentz invariant versions of the $\sigma(\d\sigma)^2$ operator is unique to the non-Abelian case.

Because of the natural suppression of the three-point function that can happen for some non-Abelian groups the local three point function induced by the soft-symmetry breaking $\sigma^3$ operator could be detectable. If the operator in $\dot\sigma \sigma^2$ is present its non-Gaussianity is suppressed by a factor of order $\lesssim(n_s-1)H/\Lambda_{U,S}$ with respect to the one from the $\sigma^4$ operator. This makes it comparable to the non-Gaussianity from the operator $\sigma(\d\sigma)^2$  in (\ref{eq:non-Abelian-three-point-operator}) and it could still allow for a detection. As in the Abelian case this operator can be made much more important and actually the leading one from the symmetry breaking sector if we break the symmetry with the non-Abelian analogous of two spurions with different charge.

We finally add a brief comment on the non-Abelian case when the dispersion relation is of the form $\omega\propto k^2$. We notice that since the non-Abelian analogous of the kinetic operator in~(\ref{eq:ghost-dispertion-operator}) induces automatically higher order terms, imposing the Lagrangian to have the same structure as in single-clock inflation to prevent the generation of the $(\d_i\sigma)^2$ term seems hard.  Non-renormalization of the standard kinetic term forces the unitarity bound to be smaller than $\bar{\bar F}/\tilde e_2^{1/6}$, which is smaller than the scale $\sim \tilde e_2^{1/2}\bar{\bar F}$ at which these interactions become strongly coupled. As before we have to forbid the cubic interactions compatible with the symmetry, and introduce soft-breaking terms with a unitarity bound $\Lambda_{U,S}\lesssim \bar{\bar F}/\tilde e_2^{1/6}$. Again soft-breaking interactions are in this case necessary. This also means that detecting a three-point function coming from the operators $\dot\sigma(\d_i\sigma)^2$ or $\d^2_j\sigma(\d_i\sigma)^2$ with a dispersion relation $\omega\propto k^2$ excludes the possibility that the additional fields are Goldstone bosons of a non-Abelian group.
%In this case we simply impose the non-Abelian analogous of the inequality~$\tilde M\lesssim\bar{\bar M}/ \tilde e_2^{3/8}$ to ensure that radiative corrections do not spoil the model.

\subsubsection*{$\bullet$ Supersymmetric case}

In the supersymmetric case we can have two kind of three-point functions. The first comes from the operator $\lambda m_s(\sigma+\sigma^*)\sigma\sigma^*\sim\lambda H(\sigma+\sigma^*)\sigma\sigma^*$, where we have used that $m_s\sim H$ in order to cancel the mass from supergravity corrections. This operator induces a possibly detectable three-point function of the order of 
\be
f_{NL}\zeta\sim\left.\frac{{{\cal L}}_3}{{\cal L}_2}\right|_{E\sim H}N_e\sim\lambda N_e\qquad\Rightarrow\qquad f_{NL}\sim 10^{5} \lambda N_e\ ,
\ee
which in the interval $10^{-4}\lesssim\lambda  N_e \lesssim N_e$ is detectable. The shape is local and the factor of $N_e$ comes as usual from the fact that this interaction keeps operating after horizon crossing. 

Another operator that can induce a large three-point function is the higher derivative one $(\sigma+\sigma^*)\d_\mu\sigma\d^\mu\sigma^*/M_{s,1}$. This induces a three-point function of order 
\be\label{eq:three-point-susy-kahler}
f_{NL}\zeta\sim\left.\frac{{{\cal L}}_3}{{\cal L}_2}\right|_{E\sim H}\sim\frac{H }{ M_{s,1}}\qquad\Rightarrow\qquad f_{NL}\sim 10^{5} \frac{H}{M_{s,1}}\ ,
\ee
which would be detectable in the interval $1\ll M_{s,1}/H\lesssim 10^{4}$. The lower limit on $M_{s,1}$ comes from imposing that the $\sigma$-mass induced at loop level is much smaller than one (after cutting off the loop at scales of order $H$ where supersymmetry is restored). As we saw, such a term in the Kahler potential induces also a cubic operator of the form $H\sigma^3/M_{s,1}$ from supergravity corrections. The induced three-point function  is of the local form and has a size of order
\be
f_{NL}\zeta\sim\left.\frac{{{\cal L}}_3}{{\cal L}_2}\right|_{E\sim H}N_e\sim\frac{H }{ M_{s,1}}N_e\qquad\Rightarrow\qquad f_{NL}\sim 10^{5} \frac{H}{M_{s,1}}N_e\ .
\ee
This is a factor of $N_e$ larger than the one in (\ref{eq:three-point-susy-kahler}) and is detectable in the interval $10^{-4}\lesssim H N_e/ M_{s,1}\lesssim 1$. Given the current constraints on $f_{NL}^{\rm local}$, the signal in (\ref{eq:three-point-susy-kahler}) will be hard to detect.

\subsection{Isocurvature fluctuations and their non-Gaussianity}

As we anticipated in sec.~\ref{sec:isocurvaute} in the case of multifield inflation isocurvature fluctuations can be generated. In this case $\sigma$ fluctuations would be visible not only through their effect on the adiabatic fluctuations but also through their imprint in the isocurvature ones. Of course this has been known for a long time (see for example \cite{Komatsu:2008hk} and references therein for the two-point function, see for example \cite{Lyth:2005fi} for non-Gaussian correlation functions). Our results build upon the former ones. We believe the framework we give in (\ref{eq:isocurvature}), together with the Lagrangian for the fluctuations, offers a general parametrization for the study of these effects and for the identification of new signatures. Furthermore it offers the most general way of relating experimental constraints (or measurements) to parameters of the theory.

From the linear components of (\ref{eq:isocurvature}) it is straightforward to see that we can have isocurvature two-point functions of  various kinds, as well as cross correlations between any two kinds of isocurvature fluctuations and between adiabatic and isocurvature fluctuations. Cross correlations happen in cases when two different fluctuations depend on the same fields or if they depend of two different fields that are mixed at quadratic level in the Lagrangian. This mixing can only happen between the $\pi$ field and the Goldstone boson of a commuting Abelian subgroup (see sec.~\ref{sec:Abelian_lagrangian}).
 
%Finally, we would like to point out the following new possibility of generating observable non-Gaussianities in multifield inflaton. So far in this section, the only way in which we have allowed the $\sigma$ fluctuations to be visible, is by imposing them to be converted in $\zeta$ fluctuations. This however is not the only possibility. As we have seen in sec.~\ref{sec:isocurvaute}, complementary to the possibility for the $\sigma$ fields to induce a relevant $\zeta$ fluctuation, they can induce isocurvature fluctuations. We have characterized these isocurvature fluctuations by three independent functions in (\ref{eq:isocurvature}). 

Isocurvature fluctuations can be non-Gaussian. As in the adiabatic case there are several mechanism that can lead to non-Gaussianity. The simplest example is of course if the quadratic terms in (\ref{eq:isocurvature}) are relevant. This would induce an isocurvature three-point function of the local type. For the same reason as in the adiabatic case we do not expect a four-point function to be generically induced in this way.

Alternatively, through the linear dependence in the conversion parameterization of (\ref{eq:isocurvature}), non-Gaussianities in the statistics of the $\sigma$ fluctuations can be translated into the isocurvature sector. Here the discussion develops as in the case of adiabatic fluctuations: we can have three-point functions with the same shape as in the adiabatic case. Furthermore by imposing the same kind of symmetries as in the adiabatic case to the multifield Lagrangian we can have a large four-point function without a detectable three-point function in an isocurvature mode. However, there are  important differences with respect to the purely adiabatic case that we are now going to explain.

\subsubsection*{$\bullet$ Abelian case}

For the Abelian case, the possible shapes for the three and four-point function arise from the same kind of operators $(\d\sigma_I)^3$ and $(\d\sigma_I)^4$ where there is a derivative acting on each fluctuation, plus the operators of the form $\sigma(\d_\mu\sigma)^2$, $\sigma^4$, $\dot\sigma\sigma^2$ and $\sigma^3$, if present, associated with soft-breaking of the $U(1)$ symmetry. Further, in the case of isocurvature fluctuations we can have more than one kind of three and four-point functions depending on the kind of correlation being considered. All the correlations  of the form 
\be
\langle\zeta\zeta\,\zeta_{\rm iso}\zeta_{\rm iso}\ldots\rangle\ ,
\ee
are possible, where we have defined $\zeta_{\rm iso}$ as one of the isocurvature flucutations in eq.~(\ref{eq:isocurvature}) and where the number of insertions of $\zeta$ and $\zeta_{\rm iso}$ is in general arbitrary. Since we know observationally that isocurvature fluctuations are smaller then the adiabatic ones, the leading effect comes from inserting the smallest possible number of isocurvature fluctuations in the correlation functions. The $U(1)$-compatible operators give rise to a non-Gaussianity of the order of
\bea\label{eq:iso1}
&&\epsilon_{\rm iso}\left.\frac{{{\cal L}}_3}{{\cal L}_2}\right|_{E\sim H}\sim\epsilon_{\rm iso}\frac{H \sigma_c}{\Lambda_U^2}\sim\epsilon_{\rm iso} \frac{H^2}{\Lambda_U^2}\ ,\qquad \epsilon_{\rm iso}\left.\frac{{{\cal L}}_4}{{\cal L}_2}\right|_{E\sim H}\sim\epsilon_{\rm iso}\frac{H^2 \sigma_c^2}{\Lambda_U^4}\sim\epsilon_{\rm iso} \frac{H^4}{\Lambda_U^4}\ ,
\eea
Notice that when the four-point function is made the leading signal by imposing some symmetry of the $\sigma$ Lagrangian, the scale suppressing the cubic and the quartic operators is different, the scale associated to the quartic operators being actually lower in order for them to be relevant (see former subsections). Here we have for simplicity used the same scale. Here we have also defined $\epsilon_{\rm iso}$ as the suppression of the isocurvature fluctuations with respect to the adiabatic ones and we have inserted only one factor of $\epsilon_{\rm iso}$. WMAP data already constrain the isocurvature component to be subdominant so $\epsilon_{\rm iso} \lesssim 0.1-0.3$, the actual limit depending on the kind of isocurvature fluctuations considered \cite{Komatsu:2010fb,Komatsu:2008hk}. Forecasts for Planck give roughly $\epsilon_{\rm iso}\sim 10^{-2}$~\cite{Bucher:2000hy}.  The $U(1)$-soft-breaking terms of the form $\sigma^2\dot\sigma$, $\sigma(\d_\mu\sigma)^2$ and $\sigma^4$, whenever present, instead give a level on non-Gaussianity of the form respectively
\bea\label{eq:iso2}
&&\epsilon_{\rm iso}\left.\frac{{{\cal L}}_3}{{\cal L}_2}\right|_{E\sim H}N_e\sim\epsilon_{\rm iso}\frac{\mu^4 H \sigma_c}{\Lambda_{U,S}^4 H^2}N_e\sim\epsilon_{\rm iso} \frac{\mu^4 }{\Lambda_{U,S}^4}N_e\ , \\ \nonumber
&&\epsilon_{\rm iso}\left.\frac{{{\cal L}}_3}{{\cal L}_2}\right|_{E\sim H}\sim\epsilon_{\rm iso}\frac{\mu^4 \sigma_c}{\Lambda_{U,S}^5}\sim \epsilon_{\rm iso}\frac{\mu^4 H}{\Lambda_{U,S}^5}\ ,\qquad \epsilon_{\rm iso}\left.\frac{{{\cal L}}_4}{{\cal L}_2}\right|_{E\sim H}N_e\sim\epsilon_{\rm iso}\frac{\mu^4  \sigma_c^2}{ \Lambda_{U,S}^4 H^2}N_e\sim \epsilon_{\rm iso} \frac{\mu^4}{\Lambda_{U,S}^4}N_e\ ,
\eea
where the factor of $N_e$ comes from the fact that the $\sigma^4$ and $\dot\sigma\sigma^2$ interactions keep operating outside of the horizon. The $\sigma^3$ operator, if present, would give a level of non-Gaussianity of order
\be
\epsilon_{\rm iso}\left.\frac{{{\cal L}}_3}{{\cal L}_2}\right|_{E\sim H}N_e\sim\epsilon_{\rm iso}\frac{\tilde \mu^4  \sigma_c}{ \Lambda_{U,S}^3 H}N_e\sim \epsilon_{\rm iso} \frac{\tilde \mu^4}{\Lambda_{U,S}^3 H}N_e\ ,
\ee
which would be the leading one.

The limits on the ratios in for example (\ref{eq:iso1}) and (\ref{eq:iso2}) scale as $1/N_{\rm pix}^{1/2}$, where $N_{\rm pix}$ is the number of signal dominated modes. Clearly if $\epsilon_{\rm iso}$ is not too small non-Gaussianities involving isocurvature fluctuations might be detected.

As in the former subsections, so far we have dealt with the case in which there is a linear dispersion relation for the modes. In the case the dispersion relation is of the form $\omega\propto k^2$, the discussion proceeds in very analogous terms. The analogous of Eq. (\ref{eq:iso1}) becomes
\bea\label{eq:iso1_ksq}
&&\epsilon_{\rm iso}\left.\frac{{{\cal L}}_3}{{\cal L}_2}\right|_{E\sim H}\sim\epsilon_{\rm iso} \left(\frac{H}{\Lambda_U}\right)^{1/4}\ ,%\qquad \epsilon_{\rm iso}\left.\frac{{{\cal L}}_4}{{\cal L}_2}\right|_{E\sim H}\sim\epsilon_{\rm iso} \left(\frac{H}{\Lambda_U}\right)^{1/2}\ ,
\epsilon_{\rm iso}\left.\frac{{{\cal L}}_4}{{\cal L}_2}\right|_{E\sim H}\sim\epsilon_{\rm iso} \left(\frac{H}{\Lambda_U}\right)^{5/2}\ ,
\eea
with the value of $\Lambda_U$ given in the former subsection.

The terms associated to the explicit symmetry breaking in eq.~(\ref{eq:iso2}): $\sigma^2\dot\sigma$, $\sigma(\d_i\sigma)^2$ and $\sigma^4$ give:
 \bea\label{eq:iso2_ksq}
&&\epsilon_{\rm iso}\left.\frac{{{\cal L}}_3}{{\cal L}_2}\right|_{E\sim H}N_e\sim\epsilon_{\rm iso} \frac{\mu^4 }{H^{3/4} \Lambda_{U,S} \bar{\bar M}^{9/4}\tilde e_2^{9/8}}N_e\lesssim \epsilon_{\rm iso}(n_s-1) N_e\left(\frac{H}{\Lambda_U}\right)^{5/4}\ , \\ \nonumber
&&\epsilon_{\rm iso}\left.\frac{{{\cal L}}_3}{{\cal L}_2}\right|_{E\sim H}\sim \epsilon_{\rm iso}\frac{\mu^4}{H^{3/4} \Lambda_{U,S}^{5/4}\bar{\bar M}^2\tilde e_2}\lesssim \epsilon_{\rm iso}(n_s-1)\left(\frac{H}{\Lambda_U}\right)^{5/4}\ , \\ \nonumber
&& \epsilon_{\rm iso}\left.\frac{{{\cal L}}_4}{{\cal L}_2}\right|_{E\sim H}N_e\sim \epsilon_{\rm iso} \frac{\mu^4}{\Lambda_{U,S} H^{3/2}\bar{\bar M}^{3/2}\tilde e_2^{3/4}}N_e\lesssim \epsilon_{\rm iso}N_e(n_s-1)\left(\frac{H}{\Lambda_U}\right)^{1/2}\ .
\eea
where in the last passage we have used that $\tilde e_2^{1/2}\bar{\bar M}\gtrsim \Lambda_{U,S}$.
Given the current constraints on the non-Gaussianities, all of these terms could be detectable. If the term in $\sigma^3$ were to be present with a spurionic coefficient proportional to $\tilde \mu^4$, it would induce a level of non-Gaussianity of order
\be
 \epsilon_{\rm iso}\left.\frac{{{\cal L}}_3}{{\cal L}_2}\right|_{E\sim H}N_e\sim \epsilon_{\rm iso} N_e\frac{\tilde\mu^4}{\Lambda_{U,S}^{3/4} H^{7/4}\bar{\bar M}^{3/2}\tilde e_2^{3/4}}\lesssim \epsilon_{\rm iso}(n_s-1)N_e\left(\frac{H}{\Lambda_U}\right)^{1/4}
\ee
which could be  detectable as a local three-point function.

\subsubsection*{$\bullet$ Non-Abelian and Supersymmetric cases}

Everything in the former discussion generalizes to the non-Abelian case and the supersymmetric case. There are however some  important differences  concerning the three and the four-point functions in the non-Abelian case that we now discuss. For the three-point function we noticed in the adiabatic case that there are additional cases  where the three-point function was naturally suppressed when only one field was relevant for the adiabatic fluctuations. This was because we required that ${\rm Tr}[x_ax_ax_a]\neq0$. Observation of a three-point function in this case required that the explicit symmetry breaking was large enough. In the case of isocurvature fluctuations it is somewhat more natural to expect that more than one field is relevant for observations, and that might alleviate the suppression of the three-point function. 

A more interesting interaction is the one in (\ref{3-pt-iso-non-Abelian}) of the form $\sigma(\d\sigma)^2$ that can appear in the non-Abelian case when there is explicit symmetry breaking and one of the Goldstone bosons gets a time dependent vev. In this case the level of non-Gaussianity for a three-point function 
 \be
\langle\zeta\zeta\,\zeta_{\rm iso}\rangle\
\ee
is approximately of the order of~\footnote{See footnote~(\ref{footnote:mass scales}) for an explanation on how the fields are normalized in this case.}
\be\label{eq:iso-non-Abelian-three-point}
\epsilon_{\rm iso}\left.\frac{{{\cal L}}_3}{{\cal L}_2}\right|_{E\sim H}\sim\epsilon_{\rm iso} \frac{\mu^4 \sigma_c}{\Lambda_{U,S}^4H}\sim \epsilon_{\rm iso}\frac{\mu^4 }{ \Lambda_{U,S}^4}\ .
\ee
This combination of isocurvature and adiabatic fluctuations is chosen in order to minimize the suppression from inserting isocurvature fluctuations.
This signal is expected to be dominated by at least an adiabatic four-point function of the local form, that is larger by a factor of $N_e/\epsilon_{\rm iso}$, and so could be detectable only if we happen to detect a large four-point function. If the operator associated to soft breaking $\sigma^2\dot\sigma$ is also present, it induces a local three-point function in the adiabatic sector comparable to the four-point function. Given the more stringent constraints we have on the local three-point function, in this case detection of the signal in (\ref{eq:iso-non-Abelian-three-point}) would become problematic. As we stressed, the presence of the various symmetry-breaking operators in the non-Abelian sector depends on the spurionic transformation properties of the terms explicitly breaking the symmetry. It might be that some operators are absent in specific cases.%It is also possible that for some groups the term in (\ref{eq:iso-non-Abelian-three-point}) vanishes.
 
Concerning the four-point function, we saw in eq.~(\ref{eq:non_Abelian_quartic_interactions}) in sec.~(\ref{sec:non-Abelian-case}) that there are dimension six operators of the form $\sigma^2(\d\sigma)^2$ which could not contribute to the four-point function of adiabatic fluctuations because of the antisymmetry of the structure constants and to the fact that adiabatic fluctuations could depend only on one linear combination of the $\sigma_a$'s (see~eq.~(\ref{eq:redefinition})). For the adiabatic case this forced the leading operator contributing to the four-point function to be of dimension eight. As we pointed out in the paragraph below eq.~(\ref{eq:isocurvature}), the linear combination of fields that sources isocurvature fluctuations is in general different from the one sourcing the adiabatic fluctuations. This implies that the operators in eq.~(\ref{eq:non_Abelian_quartic_interactions}), schematically of the form $\sigma^2(\d\sigma)^2$, can give a non-zero effect in the case of isocurvature fluctuations. Of course in this case the effect is suppressed because we know that observationally isocurvature fluctuations are smaller than the adiabatic ones. For this reason the leading contribution will come from mixed adiabatic-isocurvature four-point functions of the form
\be
\langle\zeta\zeta\,\zeta_{\rm iso}\zeta_{\rm iso}\rangle\ . 
\ee
which is non-zero and has the smallest suppression from the smallness of the isocurvature fluctuations. The amount of non-Gaussianity induced by the operators in~(\ref{eq:non_Abelian_quartic_interactions})   schematically of the form $\sigma^2(\d\sigma)^2$ is of order
\be\label{eq:quartic-iso-non-ab}
\epsilon_{\rm iso}^2\left.\frac{{{\cal L}}_4}{{\cal L}_2}\right|_{E\sim H}\sim\epsilon_{\rm iso}^2\frac{\sigma_c^2}{\Lambda_U^2}\sim\epsilon_{\rm iso}^2\frac{H^2}{\Lambda_U^2}\ .
\ee
Notice that in the adiabatic case, the level of non-Gaussianity scales as
\be
\left.\frac{{{\cal L}}_4}{{\cal L}_2}\right|_{E\sim H}\sim\frac{H^2\sigma_c^2}{\Lambda_U^4}\sim\frac{H^4}{\Lambda_U^4}\ .
\ee
The ratio of these two four-point functions scales as $( \epsilon_{\rm iso}\Lambda_U/H)^2$ which, if isocurvature fluctuations are not too small,  might lead to the leading detectable signal. Notice that these dimension six operators are absent in the Abelian case. Detection of such a kind of mixed four-point function with such a shape in the absence of a large non-Gaussianity in the adiabatic fluctuations is therefore possible and it would be a clear indication of a non-Abelian sector of Goldstone bosons active during inflation. 

The same shape can also appear in a correlation function of the form
\be
\langle\zeta\zeta\,\zeta\zeta_{\rm iso}\rangle\ .
\ee
This is possible when there is explicit symmetry breaking and  the non-Abelian Goldstone bosons develops a time-dependent vev. In this case the operator in (\ref{4-pt-iso-non-Abelian}) again of the form $\sigma^2(\d\sigma)^2$ gives a non-Gaussianity of order
\be\label{eq:quartic-bis-iso-non-ab}
\epsilon_{\rm iso}\left.\frac{{{\cal L}}_4}{{\cal L}_2}\right|_{E\sim H}\sim\epsilon_{\rm iso}\frac{\mu^4\sigma_c^2}{H \Lambda_{U,S}^5}\sim\epsilon_{\rm iso}\frac{\mu^4 H}{\Lambda_{U,S}^5}\ ,
\ee
which, if $\mu^4$ happens to be close to its upper bound $(n_s-1)H^2\Lambda_{U,S}^2$ and $\epsilon_{\rm iso}$ not too small, could be detectable. 
This same shape as induced by an operator of the form $\sigma^2(\d\sigma)^2$ can also be generated in the supersymmetric case. However in this case we would not expect a suppression of the non-Gaussianities in the adiabatic sector and, as we have seen, we would expect an even larger three-point function of the local form.

Another interesting shape proper of the non-Abelian case that can appear in 
$\langle\zeta\zeta\,\zeta\zeta_{\rm iso}\rangle$
is given by the operators in (\ref{eq:non-Abelian-four-mixed}), schematically of the form $\sigma(\d\sigma)^3$. They induce a four-point function that scales as
\be\label{eq:non-ab-quartic-new-iso}
\epsilon_{\rm iso}\left.\frac{{{\cal L}}_4}{{\cal L}_2}\right|_{E\sim H}\sim\epsilon_{\rm iso}\frac{H\sigma_c^2}{ \Lambda_U^3}\sim\epsilon_{\rm iso}\frac{H^3}{\Lambda_U^3}\ .
\ee
Since one of the fluctuations does not have a derivative this shape will have a non-vanishing squeezed limit and might be detectable by studying the clustering statistic of collapsed objects.  Because of the non-Abelian symmetry, these interactions have to be dominated by cubic ones of the form $(\d\sigma)^3$ by a factor of $F_1/(\epsilon_{\rm iso}H)$. In turn they give a level of non-Gaussianity that scales as
\be
\left.\frac{{{\cal L}}_3}{{\cal L}_2}\right|_{E\sim H}\sim\frac{H\sigma}{ F_1^2}\sim\frac{H^2}{F_1^2}\ .
\ee
If we detect a large level of non-Gaussianity in the three-point function and if $\epsilon_{\rm iso}$ is not too small then we might marginally detect this four- point function. Such a correlation of a signal between a three-point and a four-point function would be a clear indication of the non-Abelian nature of the symmetry group.

The situation  when  the dispersion relation is  $\omega\propto k^2$ is also very interesting. In the three-point function the effect of the operator considered in (\ref{eq:iso-non-Abelian-three-point}) schematically of the form $\sigma(\d_i\sigma)^2$ becomes
\be\label{eq:iso-non-Abelian-three-point-ksq}
\epsilon_{\rm iso}\left.\frac{{{\cal L}}_3}{{\cal L}_2}\right|_{E\sim H}\sim \epsilon_{\rm iso}\frac{\mu^4 }{H^{7/4} \Lambda_{U,S} \bar{\bar M}^{5/4}\tilde e_2^{5/8}}\lesssim \epsilon_{\rm iso}(n_s-1)\left(\frac{H}{\Lambda_{U,S}}\right)^{1/4}\ .
\ee
Though in the non-Abelian case one might not have  all the naively allowed  symmetry breaking terms, it is worth noticing that if an adiabatic interaction of the form $\sigma^4$ is present, then the induced signal would compare to the above one by a factor of $N_e(H/\Lambda_{U,S})^{1/4}$ (with possibly an additional factor of $1/\epsilon_{\rm iso}$ in case the $\sigma^4$ operator contributes in the purely adiabatic sector), which could be either dominant or subleading depending on parameters. The same signal would be suppressed by a factor of $N_e$ (modulo an additional possible factors of $1/\epsilon_{\rm iso}$ as before) if the operator of the form $\sigma^3$ were to be present with coefficient $\tilde\mu^4$. %As we said, a detection of such a shape as induced by the $\sigma(\d_i\sigma)^2$ operator  would be a clear indication of the non-Abelian nature of the symmetry group.

For the operators considered in (\ref{eq:quartic-iso-non-ab}) schematically of the form $\sigma^2(\d\sigma)^2$, we notice that the leading term is the one with two time derivatives $\sigma^2\dot\sigma^2$. While naively, because of  the non-linear dispersion relation of the form $\omega\propto k^2$, the term in $\sigma^2(\d_i\sigma)^2$ would be much more relevant (it would actually be a dangerous relevant operator), the structure of the Lagrangian forces its coefficient to be suppressed by a factor of $1/\tilde e_2\lesssim (H/\Lambda_U)^2$ with respect to the coefficient of $\sigma^2\dot\sigma^2$ which therefore becomes more relevant. Its effect scales as
\be\label{eq:quartic-iso-non-ab-ksq}
\epsilon_{\rm iso}^2\left.\frac{{{\cal L}}_4}{{\cal L}_2}\right|_{E\sim H}\sim\epsilon_{\rm iso}^2\left(\frac{H}{\tilde e_2^{1/2}\bar{\bar F}}\right)^{1/2}\lesssim \epsilon_{\rm iso}^2\left(\frac{H}{\Lambda_{U,S}}\right)^{4/3}\ .
\ee
with $\bar{\bar F}$ playing the same role in the non-Abelian case as $\bar{\bar M}$ in the Abelian case and we have used that ${\bar{\bar F}}\gtrsim \Lambda_{U,S}\tilde e_2^{1/6}$.

Considering now the operator in (\ref{eq:quartic-bis-iso-non-ab}), schematically of the form $\sigma^2(\d_i\sigma)^2$, its effect scales as 
\be\label{eq:quartic-bis-iso-non-ab-ksq}
\epsilon_{\rm iso}\left.\frac{{{\cal L}}_4}{{\cal L}_2}\right|_{E\sim H}\sim\epsilon_{\rm iso}\frac{\mu^4 }{H^{3/2}\Lambda_{U,S}^{5/4}\bar{\bar M}^{5/4}\tilde e_2^{5/8}}\lesssim\epsilon_{\rm iso}(n_s-1)\frac{H^{1/2}\bar{\bar M}^{1/4}\tilde e_2^{1/8}}{\Lambda_{U,S}^{3/4}} \ .
\ee
This compares to the local four-point function by the factor $\bar{\bar M}^{1/4}\tilde e_2^{1/8}/(\Lambda_{U,S}^{1/4}N_e)$ which could be either larger or smaller than one. Detection of such a shape would be another stricking signature of the non-Abelian symmetry group.

The spatial-kinetic operator in the non-Abelian case, schematically of the form $({\cal D}_iD^i\sigma)^2$, induces interactions of the form $\sigma^2(\d_i^2\sigma)^2$ where one of the fluctuations has to be isocurvature. The induced four-point function is of order
\be
\epsilon_{\rm iso}\left.\frac{{{\cal L}}_4}{{\cal L}_2}\right|_{E\sim H}\sim\epsilon_{\rm iso}\left(\frac{H}{\tilde e_2^{1/2} \bar{\bar M}}\right)^{1/2}\lesssim \epsilon_{\rm iso}^2\left(\frac{H}{\Lambda_{U,S}}\right)^{4/3}\ ,
\ee
which is subleading with respect to the leading ones. 

Finally the operator considered in (\ref{eq:non-ab-quartic-new-iso}), schematically of the form $\sigma(\d\sigma)^3$, would give a four-point function of the form
\be\label{eq:non-ab-quartic-new-iso-ksq}
\epsilon_{\rm iso}\left.\frac{{{\cal L}}_4}{{\cal L}_2}\right|_{E\sim H}\sim\epsilon_{\rm iso}\left(\frac{H}{\Lambda_U}\right)^{1/2}\ .
\ee
This is subleading with respect to the three-point function that has to be present at the same time by a factor of order $\epsilon_{\rm iso}(H/\Lambda_U)^{1/4}$, which would make it very hard to detect.

\section{Conclusions}

Given the ongoing experimental effort to test inflation and the proliferation of different models, it is
quite important to characterize the most general theory of inflation. In this paper we have extended the recently developed effective theory of single-clock inflation \cite{Cheung:2007st,Cheung:2007sv,Senatore:2009gt} to include additional light scalar fields that are in their vacuum at horizon exit. In a subsequent paper~\cite{senatore2} we shall include non-scalar light fields and fields that are not in their vacuum state at horizon exit. By concentrating on the theory of the fluctuations, we have been able to write down the most general Lagrangian coupling the Goldstone boson  of time-translation~($\pi$) with additional light scalar fields. We have assumed the additional scalar fields to be either the Goldstone boson associated with the spontaneous breaking of some symmetry, or to be protected by an approximate supersymmetry. These two mechanisms provide a natural way to make scalar fields light.

Contrary to the case of single-clock inflation, the Lagrangian for the fluctuations around the time during which the observable modes cross the horizon is not sufficient to make full contact with observations. In principle, knowledge of the full Lagrangian in field space is necessary to determine how fluctuations  in the additional light fields get converted into perturbations of cosmological interest. However, most of this conversion happens when gradients are negligible and this  allowed us to parametrize the general conversion mechanism with a few parameters, without significantly affecting the predictive power of the effective theory of multifield inflation.

We have been able to identify many observational signatures that can be produced both in single-clock inflation and in multifield inflation, such as a large three-point function of the density fluctuations. However, we have found a very rich structure of possible additional signatures specific to multifield inflation, such as some specific shapes of  the three-point function and more interestingly, the realization that it is quite generic for multifield inflation to produce a large and detectable four-point function with a larger signal-to-noise ratio than the three-point function. This opens up a new observational window into inflation that has so far been largely unexplored. We have also shown that there are specific signatures associated to the Abelian or the non-Abelian nature of the symmetry group of which the additional light scalar fields might be the Goldstone bosons off, as well as signatures associated to a possible approximate supersymmetry of the Lagrangian during inflation. Depending on the specific signatures we might detect, we could be able to distinguish between these three mechanisms for keeping the scalar fields light. Our estimates for the size and the characteristics of the signals we have identified have been approximate and valid at the order of magnitude level; clearly a more detailed study on the lines of the one performed in \cite{Senatore:2009gt} for single-clock inflation should be performed in the future. 

Concluding, we can summarize the two main results of our paper by saying that we have developed a universal framework where all multifield inflation models can be studied at the same time and their signatures analyzed, and we have realized the importance of a new, largely unexplored possible signature of inflation:  the four-point function. A detection of such a signature would teach us a lot about the dynamics during inflation and even possibly about the fundamental symmetries of Nature.

%To this point, it should be stressed that the amount of information contained in the four-point function is unprecedented, much larger than the information contained in the three-point function (already very large on its own) and 

\subsubsection*{Acknowledgments}

We thank Nima Arkani-Hamed, Cliff Burgess, Richard Holman, Lam Hui, Shamit Kachru, Zohar Komargodski, Juan Maldacena, Michele Papucci, Massimo Porrati, David Shih, Eva Silverstein, Kendrick Smith, Yuji Tachikawa, Giovanni Villadoro and Jay Wacker for interesting conversations. L.S.~is supported in part by the National Science Foundation under PHY-0503584.
M.Z. is supported by the National Science Foundation under PHY-
0855425 and AST-0907969 and by the David and Lucile 
Packard Foundation and the John D. and Catherine~T.~MacArthur~Foundation.

 \begingroup\raggedright\endgroup

%\bibliography{trapbib}
\end{document}